\documentclass[aip,jcp,amsmath,amssymb, reprint]{revtex4-1}

\usepackage{graphicx}
\usepackage{dcolumn}
\usepackage{amsfonts}
\usepackage{braket}
\usepackage{multirow}
\usepackage{threeparttable}
\usepackage{xspace}
\usepackage{verbatim}
\usepackage{mhchem}
\usepackage{soul}
\usepackage{ulem}
\usepackage{siunitx}
\usepackage[colorlinks = true,
            linkcolor = blue,
            urlcolor  = black,
            citecolor = blue,
            anchorcolor = black]{hyperref}

\usepackage{amsmath}
\usepackage{mathtools}
\usepackage{xcolor}
\usepackage{xspace}
\usepackage{ifthen}

\newcommand*{\Eh}{$E_{\rm h}$\xspace}
\newcommand*{\Ecorr}{$E_{\rm{corr}}$\xspace}
\newcommand*{\RCMnorm}{$\norm{\pmb{\lambda}_{2}}^{2}_\mathrm{F}$\xspace}

\newcommand*{\nfci}{N_\mathcal{H}}
\newcommand*{\ncomp}{\mathcal{V}_X}

\providecommand{\norm}[1]{\lVert#1\rVert}

\usepackage{newtxtext,newtxmath}
\usepackage[scaled=1.0]{helvet}
\usepackage[compact]{titlesec}
\usepackage{xcolor}
\usepackage{notes2bib}
\bibnotesetup{ note-name = , use-sort-key = false}
\titleformat{\section}
  {\normalfont\sffamily\bfseries}
  {\thesection.}{0.5 em}{\MakeTextUppercase}
\titlespacing{\section}{0pt}{12pt}{12pt}

\titleformat{\subsection}[block]
  {\normalfont\sffamily\bfseries}
  {\thesubsection.}{0.5 em}{}
\titlespacing{\subsection}{0pt}{12pt}{8pt}

\titleformat{\subsubsection}[block]
  {\normalfont\itshape\sffamily\bfseries\raggedright}
  {\arabic{subsubsection}.}{0.5 em}{}
\titlespacing{\subsubsection}{0pt}{8pt}{8pt}

\begin{document}

\title{Exploring Hilbert space on a budget: Novel benchmark set and performance metric for testing electronic structure methods in the regime of strong correlation}
\author{Nicholas H. Stair}
\email{nstair@emory.edu}
\author{Francesco A. Evangelista}
\email{francesco.evangelista@emory.edu}
\affiliation{Department of Chemistry and Cherry Emerson Center for Scientific Computation, Emory University, Atlanta, GA, 30322}

\begin{abstract}
This work explores the ability of classical electronic structure methods to efficiently represent (compress) the information content of full configuration interaction (FCI) wave functions.
We introduce a benchmark set of four hydrogen model systems of different dimensionality and distinctive electronic structures: a 1D chain, a 1D ring, a 2D triangular lattice, and a 3D close-packed pyramid.
To assess the ability of a computational method to produce accurate and compact wave functions, we introduce the accuracy volume, a metric that measures the number of variational parameters necessary to achieve a target energy error.
Using this metric and the hydrogen models, we examine the performance of three classical deterministic methods: i) selected configuration interaction (sCI) realized both via an \textit{a posteriori} (ap-sCI) and variational selection of the most important determinants, ii) an \textit{a posteriori} singular value decomposition of the FCI tensor (SVD-FCI), and iii) the matrix product state representation obtained via the density matrix renormalization group (DMRG).
We find that DMRG generally gives the most efficient wave function representation for all systems, particularly in the 1D chain with a localized basis.
For the 2D and 3D systems, all methods (except DMRG) perform best with a delocalized basis, and the efficiency of sCI and SVD-FCI is closer to that of DMRG.
For larger analogs of the models, DMRG consistently requires the fewest parameters, but still scales exponentially in 2D and 3D systems, and the performance of SVD-FCI is essentially equivalent that of ap-sCI.
\end{abstract}

\maketitle

\section{\label{sec:intro}Introduction}

An outstanding challenge in modern electronic structure theory is solving the many-body Schr{\"o}dinger equation for strongly correlated electrons.
The availability of accurate computational methods for strongly correlated systems is imperative to study a myriad of important phenomena such as bond breaking and photochemical processes,\cite{Mok1996DynamicalAnd,Roca2012MulticonfigurationSecond} molecular magnetism,\cite{malrieu2014magnetic} high-temperature superconductivity,\cite{Lee2007HighTemp} and many others.\cite{Imada1998MetalInsulator, Salamon2001PhysicsOf, Tokura2006CriticalFeatures, murthy2003hamiltonian}
In brief, strong correlation arises when the cost of promoting electrons to higher energy orbitals is small in comparison to the electron pairing energy (Coulombic repulsion).
Consequently, strongly correlated electrons cannot be qualitatively described by a mean-field picture because the wave function may contain nontrivial contributions from many Slater determinants.\cite{Tew2007ElectronCorrelation, kutzelnigg2003theory}
In this situation, electronic structure methods that build upon a mean-field reference cannot effectively approximate the wave function with a polynomial number of parameters and, therefore, often yield inaccurate energies and molecular properties.

The full configuration interaction (FCI) expansion captures all correlation effects for $N$ electrons in $L$ spatial orbitals.
Restricting FCI to a complete active space (CAS), with (CASSCF)\cite{Roos:1980wj} or without (CASCI) orbital optimization is also a common strategy when strong correlation effects are limited to few orbitals.
However, the size of the FCI (or CASSCF) determinant space scales like a binomial coefficient in $N$ and $L$, making these methods intractable for most systems of chemical interest\cite{Laughlin2000TheTheory} containing more than approximately 18 electrons in 18 orbitals (18e,18o)---although massively-parallel computations have recently managed to push this figure to (22e,22o).\cite{vogiatzis2017pushing}

Fortunately, for many ground and low-lying states, the complexity of the wave function is reduced by symmetry restrictions, sparsity bread by non-interacting determinants, and regular structure resulting from the local nature of Coulombic correlation.
Much work has thus been devoted to the development of methods that can exploit sparsity or use decomposition techniques to compactly approximate the wave function.  
However, it is not generally known what approaches are the most efficient (i.e., which ones can reach a target accuracy using the fewest parameters) given the physical dimension of the system, the degree of correlation strength, and choice of molecular orbitals.

Understanding the degree to which different wave functions may be compressed is important to guide future development of both classical and quantum computational methods.\cite{mcardle2020quantum}
In particular, there is a growing need for benchmark sets that may be used to compare classical and new quantum algorithms in various regimes of electron correlation.
Since many classes of emerging quantum algorithms---such as variational quantum eigensolvers\cite{Peruzzo:2014kca, yung2014transistor, McClean:2016bs, grimsley2019adaptive} and quantum subspace diagonalization techniques\cite{mcclean2017hybrid, motta2019determining, Parrish:2019tc, Stair_2020}---use parameterized ans\"{a}tze, one way to compare them to classical algorithms is to quantify their efficiency in terms of classical resources needed to achieve a target energy accuracy.
Such characterization is also useful in answering whether or not a quantum algorithm has an advantage over a purely classical approach.

The goal of this work is to examine how to best compress the FCI wave function of strongly correlated systems using classical methods.
To this end, we introduce a benchmark set and a simple metric to analyze the performance of a method.
We consider three families of deterministic methods that systematically approach FCI in a near-continuous fashion: i) selected CI (sCI),\cite{Huron1973IterativePerturbation, Buenker1974IndividualizedConfiguration, Buenker1975EnergyExtrapolation, Evangelisti1983ConvergenceOf} ii) singular value decomposition FCI (SVD-FCI) (related to the methods discussed in Refs.~\citenum{Koch1992AVariational,Taylor2013LosslessCompression,Fales2018LargeScale}), and iii) the density matrix renormalization group (DMRG).\cite{White1992DensityMatrix}
These exemplify different strategies to approximate the exact wave function; however, they all converge to the exact energy in the limit of no truncation, and their accuracy can be controlled by a single parameter.

Selected CI schemes approximate the FCI solution using a subset of the full determinant space. Therefore, they are most efficient when the exact wave function has a sparse structure.
Contrary to other forms of truncated CI, selected CI methods identify an optimal determinant basis using an iterative selection procedure that gradually expands the determinant space.
Although selected CI \cite{Huron1973IterativePerturbation, Buenker1974IndividualizedConfiguration, Buenker1975EnergyExtrapolation, Evangelisti1983ConvergenceOf} was proposed decades ago, in recent years it has received renewed attention with new deterministic,\cite{Garcia1995AnIterative, Neese2003ASpectroscopy, Nakatsuji2005IterativeCI, Abrams2005ImportantConfigurations, Bytautas2009APriori, Roth2009ImportanceTruncation, Evangelista2014Adaptive, Knowles2015CompressiveSampling, Liu2016iCI, Schriber2016Adaptive, Holmes2016HeatBath, Schriber2017Adaptive} stochastic,\cite{Greer1995Estimating, Greer1998MonteCarlo, Coe2014ApplyingMonte, Coe2013StateAveraged, Coe2012DevelopmentOf, Gyorffy2008MonteCarlo} and semistochastic\cite{Sharma2017SemistochasticHeat, Holmes2017ExcitedStates, Chien2018ExcitedStates, li2018fast} variants being proposed.
The closely related family of determinant-based Monte Carlo methods\cite{Booth2009FermionMonte, Cleland2010SurvivalOf, Booth2011BreakingThe, Cleland2011AStudy, Cleland2012TamingThe, Thomas2014SymmetryBreaking, Booth2014LinearScaling, Li2016CombiningThe} has also been explored.

The singular value decomposition finds use in several areas of quantum chemistry as a way to achieve low-loss compression of data.\cite{Kinoshita2003SingularValue, Hino2004SingularValue,Lewis2016ClusteredLow,Lowdin1956NaturalOrbitals, Bischoff2011LowOrder, Malmqvist2012TheBinatural, Beran2004ExtracrtingPair, Mayer2007UsingSingular}
The SVD can be applied to compress the FCI state once the coefficient vector is reshaped as a matrix, a representation naturally suggested by string-based CI algorithms.\cite{Knowles:1984ud}
Taylor has proposed to reduce the memory requirements of FCI by performing a SVD decomposition at each iteration of the Davidson procedure.\cite{Taylor2013LosslessCompression}
Another method that employs a compressed representation of the FCI vector is rank-reduced  FCI (RR-FCI), originally proposed by Koch\cite{Koch1992AVariational} and recently extended by Fales and co-workers.\cite{Fales2018LargeScale}
RR-FCI approximates the FCI solution with a polar decomposition of the FCI vector (represented as a matrix) combined with variational minimization of the energy.
We note that both Taylor's ``gzip'' approach and RR-FCI approaches cannot be justified on the basis of a symmetry or a physical principle, although a variant of RR-FCI that exploits locality has been proposed (see Ref.~\citenum{Weinstein2011ReducingMemory}).

Tensor network states (TNSs) represent a broad family of methods that approximate the FCI coefficients (viewed as a tensor) with a collection of tensors connected by contractions.
The simplest type of TNS is a matrix product state (MPS), the underlying ansatz\cite{ostlund1995thermodynamic, dukelsky1998equivalence, schollwock2011density} of the density matrix renormalization group (DMRG).\cite{White1992DensityMatrix}
MPSs are able to maximally exploit local orbital entanglement, that is, for states that satisfy an area law for the entanglement entropy, MPSs can yield near-exact results in 1D and quasi-1D systems.
\cite{Eisert2009AreaLaws,Evenbly2011TensorNetwork}
The generalization of the MPS ansatz to two- (2D) and three-dimensional (3D) TNS using high-order tensor factorizations is also an active area of research.\cite{Evenbly2011TensorNetwork, Murg2010SimulatingStrongly, Nakatani2013EfficientTree} In practice, the variational optimization of TNSs suffers from very high scaling and is less efficient relative to MPSs.
DMRG (as applied to quantum chemistry),\cite{Chan2002HighlyCorrelated, chan2011density, Moritz2005RelitivisticDMRG, Kurashige2009HighPerformance, Olivares2015TheAbinitio} has been tremendously successful in describing the ground states of quasi-linear molecular systems.
For example, DMRG has enabled the investigation of long hydrogen chains,\cite{Chan2002HighlyCorrelated, Hachmann2006MultireferenceCorrelation, Mitrushchenkov2012OnTheImportance, Wouters2012LongitudinalStatic, Ma2013AssessmentOf} oligoacenes,\cite{Hachmann2007TheRadical, Raghu2002StructuralAnd, Raghu2002DensityMatrix} and large biochemically-relevant transition metal complexes with up to 100 orbitals.\cite{Kurashige2013EntangledQuantum, Sharma2014LowEnergy}

To test the performance of electronic structure theories in the strongly correlated regime, we have introduced a benchmark set of one, two, and three-dimensional (3D) hydrogen systems.
These systems model strongly correlated electrons in significantly different regimes and dimensionalities, and allow us to explore the physics of Mott insulators and spin frustrated systems in 2- and 3D.
1D hydrogen systems have recently been the subject of comprehensive benchmark studies aimed at treating strong correlation in real materials.\cite{Sinitskiy2010StrongCorrelation,Motta2017TowardsThe, motta2019ground}
Hydrogen lattices with localized spins are also related to the more fundamental Heisenberg and Hubbard models, exhibiting similar spin correlation patterns and band structures.
Our benchmark set contains four \ce{H10} models: the well-investigated 1D chain and ring, as well as a 2D triangular lattice (referred to as ``sheet'' throughout the paper), and a 3D close-packed pyramid.
For each model, we consider both the effect of the \ce{H}--\ce{H} distance on the strength of correlation, and the use of different molecular orbital bases (delocalized/localized).
We characterize these models by computing various metrics of correlation, including the norm of the two-body cumulant, the total quantum information, and spin-spin correlation functions.
Additionally, to investigate the compression efficiency as a function of system size, we also consider \ce{H12}, \ce{H14} , and \ce{H16} analogs of the four models.

Since the methods considered here play an important role as substitutes for FCI in multireference treatments of electron correlation,\cite{Yanai:2010kf,Saitow:2013ij,Kurashige:2014bq,Guo:2016fu,Wouters:2016fb,schriber2018combined} we are particularly interested in assessing their performance when applied only to valence orbitals.
To simulate this scenario, our computations employ a minimal basis set.
Note, that this treatment may be considered equivalent to diagonalizing a valence effective Hamiltonian\cite{Freed:1974vo} with interactions modified by dynamical correlation effects.
It is important to point out that since we only consider zero-temperature quantum chemistry approaches, we focus in particular on regimes of electron correlation that range from weak to medium/strong.
We intentionally avoid the limit of infinite \ce{H}--\ce{H} separation, because all the models considered here develop a massively degenerate ground state containing $2^{10}$ states.
At large separation, it is ludicrous to characterize a ground state, and one should instead seek to compute thermal averages employing a finite-temperature approach.\cite{welden2016exploring, white2018time, harsha2019thermofield}

To compare the performance of each method, we evaluate the error in the energy and the two-body density cumulant as a function of the number of variational parameters.
These errors measure how well electronic correlation effects are preserved in the compression, and therefore can indicate the quality of an approximate wave function and its properties.
From this information, we extract a single metric, the \textit{accuracy volume}, which measures the number of variational parameters necessary to achieve a target energy error.
Although the accuracy volume does not take into account the actual cost of a computation, this metric serves as a proxy for the computational resources required by each method, independently of implementation details.
We also compare the energy errors produced with Hartree--Fock theory, second-order M{\o}ller--Plesset many-body perturbation theory (MP2), coupled cluster theory with singles and doubles (CCSD),\cite{Purvis1982FullCoupled} CCSD with perturbative triples [CCSD(T)],\cite{Raghavachari1989AFifth} the completely renormalized CC approach with perturbative triples [CR-CC(2,3)],\cite{piecuch2005renormalized} and the variational two-particle reduced density matrix (V2RDM) method.\cite{Colmenero1993ApproximatingQorder, Nakatsuji1996DirectDetermination, Mazziotti1998ContractedSchro,Mazziotti2011TwoElectron,Fosso2016LargeScale}
We have collected the data generated in this study in an online repository\cite{HstudyRepo2020} in the hope that it will be useful in future studies.

The remainder of this article is organized as follows: Sec.~\ref{sec:theory} defines the accuracy volume,  summarizes the three methods compared in this study, and defines the metrics used to assess correlation strength in the hydrogen model systems.
Section~\ref{sec:compu_details} provides the computational details of our study.
Numerical results are reported in Sec.~\ref{sec:results}, and Sec.~\ref{sec:conclusions} summarizes our findings and discusses their relevance in the context of classical and quantum algorithms for strongly correlated systems. 

\section{THEORY}
\label{sec:theory}

\subsection{Definition of the accuracy volume}
For a systematically improvable method $X$ we indicate the energy computed using $N_{\rm{par}}$  parameters as $E_X(N_{\rm{par}})$. We then define the accuracy volume, $\ncomp(\alpha)$, to be the smallest number of parameters such that the error per electron with respect to the FCI energy ($E_\mathrm{FCI}$) is less than or equal to $10^{-\alpha}$:
\begin{equation}
\label{eq:Vx}
\ncomp(\alpha) = N_{\rm{par}} : \frac{ |E_X(N_{\rm{par}}) - E_\mathrm{FCI}|}{N} \leq  10^{-\alpha}  \; E_\mathrm{h}.
\end{equation}
For convenience, in the rest of the paper we always assume the target energy error is 1 m\Eh for the \ce{H10} systems, which corresponds to a 0.1 m\Eh error per electron  ($\alpha = 4$) and use the more compact symbol $\ncomp$ instead.
For methods that exploit the sparsity of the FCI wave function (e.g., selected CI), the accuracy volume is a measure of the number of Slater determinants or configuration state functions (equal to the number of parameters).
This literal interpretation of the accuracy volume does not extend to approximation schemes based on tensor decomposition, in which case it only reflects the total number of parameters employed. 
We intend the accuracy volume to be used as a performance metric of a method, since it approximately measures the computational resources (memory and CPU) necessary to achieve a target accuracy.
Because the accuracy volume can be equally applied to purely classical and hybrid quantum-classical methods, it provides a straightforward way to compare the two on more equal footing.
Our definition of  $\ncomp$ [Eq.~\eqref{eq:Vx}] considers the energy error per electron to allow the comparison of systems with different numbers of electrons.
This approach is consistent with the fact that approximate methods that are size consistent, when applied to noninteracting fragments, give an error that is additive in the error of each fragment.
We also choose to define $\ncomp$ as the absolute number of parameters, as opposed to the fraction of the total Hilbert space, since the former is proportional to the computational resources required by a method.
In contrast, a comparison based on the fraction of Hilbert space parameters employed by a method would be dependent on the exploitable symmetries for the orbitals that are chosen (e.g. symmetry adapted delocalized vs. localized orbitals) making comparisons of different computations less indicative of actual computational resources.

\subsection{Overview of the computational methods}
Given a basis of $K$ spin orbitals $\{\psi_p\}$ with $p = 1,\ldots,K$, we indicate a generic $N$-electron determinant $\ket{\psi_{i_1}\cdots\psi_{i_N}}$ using the notation $\ket{\Phi_I}$ where the multindex $I = (i_1,\ldots,i_N)$ represents an ordered list of indices ($i_1 < i_2 < \ldots < i_N$).
The set of $N$-electron determinants ($\mathcal{H}_N$) forms a Hilbert space of dimension $|\mathcal{H}_N| = \nfci$.
Using this notation, the FCI wave function is written as a linear combination of determinants, each parameterized by a coefficient ($C_{i_1,\ldots,i_N} \equiv C_{I} $)
\begin{equation}
\ket{\Psi_{\text{FCI}}} = \sum_{i_1 < i_2 < \ldots < i_N}^{K} C_{i_1 \cdots i_N} \ket{\Phi_{i_1 \cdots i_N}}  = \sum_{I}^{\nfci} C_{I} \ket{\Phi_{I}} 
\label{eq:fci_def1}.
\end{equation}
An equivalent way to express the FCI wave function employs occupation vectors.
In this representation, each determinant $\ket{\Phi_I}$ is associated with a vector of length $K$, $\ket{\mathbf{n}} =\ket{n_1, n_2, \ldots, n_K}$, where $n_i \in \{0,1\}$ is the occupation number of spin orbital $\psi_i$.
The FCI wave function represented in the occupation vector form is given by
\begin{equation}
\ket{\Psi_{\text{FCI}}}  = \sum_{\{  n_{i}\}} C_{n_{1} \ldots n_{K}} \ket{n_{1} \ldots n_{K}}
= \sum_{\mathbf{n}} C_{\mathbf{n}} \ket{\mathbf{n}}
\end{equation}
where the sum over all occupation vectors ($\{  n_{i}\} \equiv \mathbf{n}$) is restricted to $N$-electron determinants ($\sum_i n_j = N$) of given spin and spatial symmetry.

\subsubsection{Selected CI}
\label{sec:SCI}

Selected CI methods approximate the FCI wave function using a subset $\mathcal{M}$ (model space) of the full determinant space
\begin{equation}
\ket{\Psi_\mathrm{sCI}} = \sum_{I \in \mathcal{M}} \tilde{C}_I \ket{\Phi_I}.
\end{equation}
All flavors of selected CI aim to approximate the FCI vector with the smallest number of elements and differ primarily in the way they determine the set $\mathcal{M}$.
For sCI methods, we report $N_{\rm{par}}$ as the size of the space $\mathcal{M}$ (or equivalently, the size of the vector $\tilde{C}_I$).

The first approach we consider consists of an \textit{a posteriori} selected CI (ap-sCI) compression of the exact FCI wave function.
This compressed representation is obtained by sorting the determinants according to their weight $w_I = |C_I|^2$, and discarding elements with the smallest weight while satisfying the condition
\begin{equation}
\sum_{I \notin \mathcal{M}} |C_I|^{2} < \tau_{\text{sCI}}.
\end{equation}
The compressed ap-sCI vector $\tilde{C}_I$ is then normalized and the energy is computed as the expectation value of the Hamiltonian.
Even though this compression scheme does not yield a variationally optimal solution, the error in the ap-sCI energy is quadratic in the wave function error.
Still, this ideal (albeit impractical) version of selected CI is useful in assessing the error introduced by the different selection schemes used in practical sCI approaches.

The second approach we consider, the adaptive configuration interaction (ACI),\cite{Schriber2016Adaptive, Schriber2017Adaptive} identifies the space $\mathcal{M}$ via an iterative procedure that seeks to control the energy error.
ACI is unique in the regime of selected CI methods as it aims to approximate the FCI energy within a user-specified error tolerance $\sigma$
\begin{equation}
E_{\text{ACI}}(\sigma) - E_{\text{FCI}} \approx \sigma,
\end{equation}
where $E_{\text{ACI}}(\sigma)$ is the ACI energy.
In ACI, the model space is divided into two spaces $\mathcal{M} = \mathcal{P} \cup \mathcal{Q}$, where $\mathcal{P}$ contains the most important determinants and $\mathcal{Q}$ contains singly and doubly excited determinants spawned from $\mathcal{P}$.
New candidate determinants ($\Phi_{A}$) for the model space are selected from the singly and doubly excited determinants generated from the current $\mathcal{P}$ space. Each candidate determinant is ranked by its energy contribution, $\epsilon(\Phi_{A})$, a quantity estimated by diagonalizing the Hamiltonian in the basis of the ACI wave function at the current iteration and $\Phi_{A}$.
To determine an improved model space, the candidate determinants are sorted according to $|\epsilon(\Phi_{A})|$ and unimportant elements are removed until the sum of their estimated energy is less than or equal to $\sigma$
\begin{equation}
\label{eq:aci_selection}
\sum_{\Phi_{I} \notin \mathcal{M}} | \epsilon(\Phi_{I}) | \leq \sigma.
\end{equation}
Optionally, additional determinants are included in $\mathcal{M}$ at each iteration to ensure spin completeness.

After adding these determinants, the Hamiltonian is diagonalized and a new $\mathcal{P}$ space is formed by coarse graining $\mathcal{M}$ according to their weight using a cumulative metric similar to Eq.~\eqref{eq:aci_selection}.
The course graining step increases the overall efficiency of the procedure and reduces the dependency of the final solution on the initial guess (usually the HF determinant or a small CASCI).

The final ACI energy is computed by diagonalization of the Hamiltonian in the model space basis. However, during the selection process it is possible to accumulate the estimate of the energy contributions from the discarded determinants ($E_\mathrm{PT2}$) and this quantity can be added to the ACI energy to obtain an improved energy ($E_\mathrm{ACI+PT2}$).

\subsubsection{\label{sec:SVD-FCI}Singular value decomposition FCI}

In this work we consider an \textit{a posteriori} rank reduction of the FCI tensor obtained via a singular value decomposition (SVD-FCI). 
Our approach is essentially identical to the ``gzip'' treatment used by Taylor (Ref.~\citenum{Taylor2013LosslessCompression}) with the caveat that we only perform SVD of the final converged wave function rather than at each FCI iteration.
The SVD-FCI approach is also inspired by the the rank-reduced FCI method (RR-FCI).\cite{Koch1992AVariational,Fales2018LargeScale} 
However, because we do not variationally optimize the SVD-FCI wave function, RR-FCI would yield lower energies than SVD-FCI for a specified rank (particularly at low ranks).

SVD-FCI starts from a string-based representation of the FCI wave function,\cite{Handy1980MultiRoot}
in which each determinant is labeled by separate multi-indices (strings) for alpha and beta electrons ($I_{\alpha}$ and $I_{\beta}$), and the determinant  $\ket{\Phi_I} = \ket{\Phi_{I_\alpha} \Phi_{I_\beta}}$ factorizes into products of alpha ($\Phi_{I_\alpha}$) and beta ($\Phi_{I_\beta}$) spin orbitals.
Consequently, the FCI vector $C_I$ is represented as a matrix  indexed by string configurations ($I_{\alpha}$/$I_{\beta}$), $(\mathbf{C})_{I_{\alpha}I_{\beta}} = C_{I_{\alpha}I_{\beta}}$, and the wave function is written as 
\begin{equation}
\ket{\Psi_{\text{FCI}}} =\sum_{I_{\alpha}}^{N_{\alpha}} \sum_{I_{\beta}}^{N_{\beta}}  C_{I_{\alpha}I_{\beta}} \ket{\Phi_{I_\alpha} \Phi_{I_\beta}},
\label{eq:fci_mat}
\end{equation}
where $N_\alpha$ and $N_\beta$ are the number of alpha and beta strings, respectively.
While the original RR-FCI algorithm is based on variational minimization of the energy, in this work we consider only an \textit{a posteriori} compression.
To this end we perform the singular value decomposition of the FCI coefficient matrix,  $\mathbf{C} = \mathbf{U} \mathbf{S} \mathbf{V}$, where we assume that the entries of $\mathbf{C}$ are real.
To find the most compact reduced-rank approximations of $\mathbf{C}$ we reconstruct an approximate matrix $\widetilde{\mathbf{C}}^{\text{SVD}}$ defined as
\begin{equation}
\widetilde{\mathbf{C}}^{\text{SVD}} = \mathbf{U} \widetilde{\mathbf{S}} \mathbf{V},
\end{equation}
where $\widetilde{\mathbf{S}}$ is a truncated version of $\mathbf{S}$.
Assuming the singular values $s_i = S_{ii}$ are sorted in decreasing order, we keep in $\widetilde{\mathbf{S}}$ the diagonals $s_1, \ldots, s_R$ such that the sum of the square of the elements excluded is less than a user-provided threshold ($\tau_{\text{SVD}}$)
\begin{equation}
\sum_{i=R + 1} s_{i}^{2} < \tau_{\text{SVD}}.
\end{equation}
Therefore, $R$ represents the rank of $\widetilde{\mathbf{C}}^{\text{SVD}} $ and the error in the FCI wave function is given by
\begin{equation}
\|\mathbf{C} - \widetilde{\mathbf{C}}^{\text{SVD}}\|_\mathrm{F} < \sqrt{\tau_{\text{SVD}}},
\end{equation}
where $\|\cdot\|_\mathrm{F}$ is the Frobenius norm.
The SVD-FCI energy is computed as
\begin{equation}
E^{\text{SVD-FCI}} =(\widetilde{ \mathbf{C} }^{\text{SVD}})^{\dagger} \mathbf{H} \widetilde{ \mathbf{C} }^{\text{SVD}} ,
\end{equation}
and although it does not correspond to the optimal energy for a wave function of rank $R$, this estimate deviates from the variational energy by a quadratic term.
For SVD-FCI, we calculate the number of parameters as $N_{\rm{par}} = R (N_{\alpha} + N_{\beta})$, where we have assumed that the singular values $\widetilde{\mathbf{S}}$ are folded into either $\mathbf{U}$ or $\mathbf{V}$.
Note that with no truncation, the SVD-FCI requires twice the number of parameters as the size of the Hilbert space.
We also point out that since the FCI wave function is invariant with respect to unitary rotations of the orbitals, the rank $R$ SVD approximation yields the same approximate wave function in any orbital basis.
However, the number of parameters may differ from one orbital basis to another if symmetry is employed and the SVD is applied only to the non-zero blocks of $\mathbf{C}$.

\subsubsection{Density-matrix renormalization group}

The matrix product state representation at the basis of the DMRG is a conceptually different form of compression that aims to exploit the local character of entanglement. 
A MPS decomposition of the FCI tensor in the occupation number representation is given by
\begin{equation}
\
C_{n_{1} \ldots n_{K}} \approx C_{n_{1} \ldots n_{K}} ^\mathrm{DMRG}= 
\mathbf{A}^{n_1}_1 \mathbf{A}^{n_2}_2 \cdots \mathbf{A}^{n_K}_K,
\end{equation}
where, for a given value of the occupation number $n_j$, a generic term $\mathbf{A}^{n_j}_j$ is a $M \times M$ matrix, except for the first and last terms which are a row and a column vector of size $M$, respectively.
Given an occupation number pattern ($\mathbf{n}$), the corresponding tensor element $C_{n_{1} \ldots n_{K}}$ is approximated by the product of all the $\mathbf{A}^{n_j}_j$ matrices.
Quantum chemistry implementations of DMRG exploit the symmetry group of the Hamiltonian (particle number, spin, point group) to induce a block-sparse structure in the MPS tensors $\mathbf{A}^{n_j}_j$, with consequent reduction in computational and storage costs.
We calculate $N_{\rm{par}}$ for DMRG as the sum of the number of parameters in each site tensor $\mathbf{A}^{n_j}_j$ in the converged MPS, taking into account the block structure induced by symmetries (assuming at most abelian point groups).

Formally, the MPS representation can be derived by performing a series of successive SVDs on the FCI tensor (appropriately reshaped), at each step retaining only $M$ terms.
Therefore, it is exact in the limit of $M \rightarrow \nfci$.
In practice, the DMRG method directly builds the MPS representations via a sweep algorithm using a fixed value of $M$ specified by the user.
For chemical applications, the quality of the MPS as a function of $M$ is controlled by two choices: the type (localized vs. delocalized) and ordering of the orbitals. These aspects present a challenge for practical calculations since different orbital types and orderings can dramatically affect the final outcome of a calculation. Although there are rules of thumb for specific cases---such as choosing localized orbitals ordered to be spatially adjacent for elongated molecules\cite{Wouters2014TheDensity}---the choice of these parameters is generally a non-trivial problem beyond 1D. 
Various approaches to ordering delocalized orbitals have also been explored.\cite{legeza2003controlling,moritz2005convergence,Legeza2003OptemizingThe,Rissler2006MeasuringOrbital}

\subsection{\label{metrics_of_correlation}Metrics of strong electronic correlation}

\subsubsection{Metrics based on mean-field and coupled cluster wave functions}

In computational quantum chemistry, the prevailing measure of electronic correlation is the correlation energy. This metric dates back to the work of L{\"o}wdin \cite{lowdin1958correlation} and is defined as the difference between the FCI and mean-field ($E_{\text{MF}}$) energy
\begin{equation}
E_{\text{corr}} = E_{\text{FCI}} - E_{\text{MF}}.
\label{eq:cor_energy}
\end{equation}
The correlation energy may be further partitioned into dynamical and non-dynamical contributions, as proposed by Sinano{\v g}lu and others.\cite{Sinanoglu1964QuantumTheory,bartlett1994applications}

One can similarly estimate correlation effects from the magnitude of the overlap of the Hartree--Fock determinant with the \textit{normalized} FCI wave function, $|C_\mathrm{HF}| = |\braket{\Phi_{\mathrm{HF}}|\Psi}|$.
This metric has been discussed as a diagnostic tool for determining the quality of single-reference electron correlation methods.\cite{lee1989diagnostic}
However, for infinite systems $|C_\mathrm{HF}| \rightarrow 0$, so this metric is probably suited only for comparing systems with the same number of electrons.

In the context of coupled cluster theory, several diagnostics have been introduced.
The $D_1$ diagnostic captures deficiencies in the reference, and is defined as the 2-norm of the matrix of singles cluster amplitudes $(\mathbf{T})_{ia} = t_i^a$, where the indices $i$ and $a$ span the occupied and virtual orbitals, respectively.
This metric is defined as
\begin{equation}
D_1 = \| \mathbf{T} \|_2 = \sqrt{\lambda_{\max} (\mathbf{TT}^T)},
\end{equation}
where $\lambda_{\max} (\mathbf{TT}^T)$ indicates the largest eigenvalue of the matrix $\mathbf{TT}^T$.
The $D_2$ diagnostic is a measure of correlation, and it is similarly defined using doubles amplitudes ($t_{ij}^{ab}$) with the above equation modified to make this metric orbital invariant.\cite{nielsen1999double}

\subsubsection{Measures based on the two-body density cumulant}

The norm of the two-body cumulant ($\pmb{ \lambda}_{2}$) has become a well established metric of correlation.\cite{Luzanov2005IrreducibleCharge,  Huang2006Entanglement, Juhasz2006TheCumulant, Luzanova2007HighOrder, Alcoba2010OnThe}
This quantity is the portion of the two-body density matrix $\pmb{ \gamma}_{2}$ that is not separable into one-body contributions, and it is defined as
\begin{equation}
\lambda_{pq}^{rs} = \gamma_{pq}^{rs}
- \gamma_{p}^{r} \gamma_{q}^{s}
+ \gamma_{p}^{r} \gamma_{s}^{q},
\end{equation}
where $\gamma^{p}_{q} $ and $\gamma_{pq}^{rs}$ are the one- and two-body reduced density matrices:
\begin{align}
\gamma^{p}_{q} = \bra{\Psi} a_{p}^{\dagger} a_{q} \ket{\Psi}, \quad 
\gamma^{pq}_{rs} = \bra{\Psi} a_{p}^{\dagger} a_{q}^{\dagger} a_{s}  a_{r} \ket{\Psi}.
\end{align}
The information contained in $\pmb{ \lambda}_{2}$ can be distilled down to a single value metric via its Frobenius norm:
\begin{equation}
\norm{\pmb{ \lambda}_{2}}_\mathrm{F} = \sqrt{\sum_{pqrs} |\lambda_{pq}^{rs}|^{2}},
\end{equation}
which captures both spin entanglement and Coulombic correlation effects,\cite{Juhasz2006TheCumulant, Alcoba2010OnThe} and is null for a single determinant.
The two-body density cumulant also has a direct connection to the number of effectively unpaired electrons, which itself has been used as a metric of correlation.\cite{Bochicchio1998OnSpin, Lain2009ADecomposition, Alcoba2006OnTheDefinition}
For two non-interacting fragments A and B with no interfragment spin entanglement, the \textit{square} Frobenius norm is additive,\cite{Juhasz2006TheCumulant,Alcoba2010OnThe} that is $\norm{\pmb{ \lambda}_{2}(\mathrm{A})}^2_\mathrm{F} +\norm{\pmb{ \lambda}_{2}(\mathrm{B})}^2_\mathrm{F} = \norm{\pmb{ \lambda}_{2}(\mathrm{A}\cdots\mathrm{B})}^2_\mathrm{F}$, where ``$\mathrm{A}\cdots\mathrm{B}$'' indicates A and B at infinite separation. Therefore, in our comparison of the models we report the square Frobenius norm.

Moreover, the two body cumulant is directly related to the definition of the intrinsic correlation energy (ICE) proposed by Kutzelnigg.\cite{kutzelnigg2003theory}
By expressing the energy in terms of 1- and 2-RDMs and expanding the latter in terms of the two-body cumulants, one may rewrite the two-body contribution to the total energy as a sum of Coulombic, exchange, and correlation contributions, $E_{2} = E_{\text{Coul}} + E_{\text{ex}} + E_{\text{ICE}}$.
Here, $E_{\text{ICE}}$ is a pure two-body potential energy term which may be expressed using the two-body cumulant represented in coordinate space [$\lambda_{2}(1,2;1,2)$] as
\begin{equation}
E_{\text{ICE}} =\frac{1}{4} \sum_{pqrs} \lambda^{pq}_{rs} \langle{rs \| pq} \rangle = \frac{1}{2}\int \frac{\lambda_{2}(1,2;1,2)}{r_{12}} d\tau_{1}d\tau_{2} 
.
\end{equation}
This intrinsic correlation energy has the advantage of being defined irrespective of a reference mean-field wave function.

\subsubsection{Spin correlation metrics}

We also characterize electronic states using various metrics based on the spin-spin correlation function as they are helpful n diagnosing spin frustration.
The spin-spin correlation function ($A_{ij}$), defined as 
\begin{equation}
A_{ij} = \langle \mathbf{\hat{S}}_{i} \cdot \mathbf{\hat{S}}_{j} \rangle - \langle \mathbf{\hat{S}}_{i} \rangle \cdot \langle \mathbf{\hat{S}}_{j} \rangle,
\end{equation}
measures the irreducible correlation of total spin ($ \mathbf{\hat{S}}_{i}$) for two localized spatial orbitals $\phi_i$ and $\phi_j$.
In this work we employ Pipek--Mezey localized orbitals\cite{Pipek1989FastIntrinsic} to define spin-spin correlation metrics.
We also compute the spin-spin correlation density $A_{i}(\mathbf{r})$, which can be used to graphically represent the spatial correlations of spin with respect to a localized orbital $\phi_i$.
For well localized atomic orbitals, $A_{i}(\mathbf{r})$ can be approximated as
\begin{equation}
A_{i}(\mathbf{r}) = \langle \mathbf{\hat{S}}_{i} \cdot \mathbf{\hat{S}}(\mathbf{r})\rangle -   \langle \mathbf{\hat{S}}_{i} \rangle \cdot \langle \mathbf{\hat{S}}(\mathbf{r}) \rangle\approx \sum_{j} A_{ij} |\phi_{j}(r)|^{2},
\end{equation}
where $\mathbf{\hat{S}}(\mathbf{r})$ is the total spin operator in real space, $|\phi_{j}(r)|^{2}$ is the spatial density of the $j$-th orbital, and $A_{ij}$ are elements of the spin-spin correlation function.

Additionally, we consider three scalar metrics introduced in previous molecular spin frustration studies: i) the sum of the absolute value of the spin-spin correlations $\langle S^2 \rangle_{\rm{abs}}$,\cite{Hoyos2014PolyradicalCharacter}
\begin{equation}
\label{eq:S_sq_abs}
\langle S^2 \rangle_{\rm{abs}} = \sum_{ij} |  \langle \mathbf{\hat{S}}_{i} \cdot \mathbf{\hat{S}}_{j} \rangle |,
\end{equation}
ii) the sum of the absolute value of the long range spin-spin correlations $\langle S^2 \rangle_{\rm{abs,lr}}$, 
\begin{equation}
\label{eq:S_sq_abs_lr}
\langle S^2 \rangle_{\rm{abs,lr}} = \langle S^2 \rangle_{\rm{abs}} - \sum_{i} |  \langle \mathbf{\hat{S}}_{i} \cdot \mathbf{\hat{S}}_{i} \rangle | - 2\sum_{\langle kl \rangle} |  \langle \mathbf{\hat{S}}_{k} \cdot \mathbf{\hat{S}}_{l} \rangle |,
\end{equation} 
and iii) the sum of the nearest-neighbor spin-spin interactions $\langle S^2 \rangle_{\rm{nn}}$, 
\begin{equation}
\label{eq:S_sq_nn}
\langle S^2 \rangle_{\rm{nn}} = \sum_{\langle kl \rangle}  \langle \mathbf{\hat{S}}_{k} \cdot \mathbf{\hat{S}}_{l} \rangle,
\end{equation}
where $i$ and $j$ index all orbital sites, and $\langle kl \rangle$ is a double sum over nearest neighbor orbital sites.

\subsubsection{Metrics based on quantum information theory}

Metrics inspired by quantum information theory have also been recently used to investigate various phenomena related to strong correlation and entanglement,\cite{Boguslawski2015OrbtitalEntanglement}
and find several applications in computational chemistry.\cite{Boguslawski2013OrbitalEntanglement,stein2017automated,fertitta2014investigation,Legeza2003OptemizingThe, Rissler2006MeasuringOrbital} 

We consider two quantities, the single-orbital entanglement entropy (SOEE) and the total quantum information ($I_\mathrm{tot}$), both of which can be derived from the 1- and 2-RDMs.
The SOEE describes the entanglement of a spatial orbital $\phi_i$ with the remaining \textit{bath} orbitals.
For a given spatial orbital $\phi_i$, we can write four occupation patterns for the corresponding $\alpha$ and $\beta$ spin orbitals $\ket{p} \equiv \ket{n_{i_\alpha}n_{i_\beta}} \in \{\ket{00},\ket{01},\ket{10}, \ket{11}\}$, which we label with the index $p = 1,2,3,4$.
The reduced density matrix $\rho^i_{pq} = \mathrm{Tr}_\mathrm{bath} [\braket{p|\Psi}\braket{\Psi|q}]$ is computed by projecting the wave function onto single-orbital configurations $\ket{q}$ and $\ket{p}$ of orbital $\phi_i$ and tracing out all other degrees of freedom. For states with fixed number of electrons, this matrix is diagonal with elements given by
\begin{align}
\rho_{11}^i &= 1 - \gamma_{i^{\alpha}}^{i^{\alpha}} - \gamma_{i^{\beta}}^{i^{\beta}} + \gamma_{i^{\alpha} i^{\beta}}^{i^{\alpha} i^{\beta}}, \\
\rho_{22}^i &= \gamma_{i^{\alpha}}^{i^{\alpha}} - \gamma_{i^{\alpha} i^{\beta}}^{i^{\alpha} i^{\beta}}, \\
\rho_{33}^i &= \gamma_{i^{\beta} }^{i^{\beta} } - \gamma_{i^{\alpha} i^{\beta}}^{i^{\alpha} i^{\beta}}, \\
\rho_{44}^i &=\gamma_{i^{\alpha} i^{\beta} }^{i^{\alpha} i^{\beta}}.
\end{align}
The SOEE of orbital $\phi_i$ is then computed as the Shannon entropy with respect to the four occupations
\begin{equation}
S_{i} = -\sum_{p=1}^{4} \rho^{i}_{pp} \ln(\rho^{i}_{pp}).
\label{eq:soee}
\end{equation}

The total quantum information ($I_\mathrm{tot}$) is given as the sum of the SOEEs for all spatial orbitals
\begin{equation}
I_\mathrm{tot} = \sum_{i=1}^{L} S_{i}.
\end{equation}
Large values of $I_\mathrm{tot}$ indicate departure from integer orbital occupations and are associated with strong correlation effects.\cite{murg2015tree}
We note, however, that the value of $I_\mathrm{tot}$ is not invariant with respect to unitary rotations of the orbitals, and therefore, will depend on the type of orbital basis employed in a computation.

\section{\label{sec:compu_details}Computational Details}

The ground-state singlet energies and two-body density cumulants of the model systems were calculated using FCI, ACI, and DMRG.
The ap-sCI and SVD-FCI wave functions were obtained from FCI wave functions as described in Secs.~\ref{sec:SCI} and \ref{sec:SVD-FCI}, respectively. 
All computations employed self consistent field (SCF) orbitals obtained with the open-source quantum chemistry package \textsc{Psi4}\cite{Parrish2017Psi4, smith2020psi4} and used a STO-6G basis set.\cite{Hehre1969ASelf}
Canonical (delocalized) orbitals were computed using restricted Hartree--Fock (RHF).
Localized orbitals were obtained by first performing a restricted open-shell Hartree-Fock  (ROHF) computation using maximum multiplicity (e.g., $S = 5$ for \ce{H10}) and then localizing the orbitals with the Pipek--Mezey (PM) \cite{Pipek1989FastIntrinsic} procedure (allowing rotations among all orbitals).

Computations based on canonical RHF orbitals were run in $D_{\rm{2h}}$ symmetry for the \ce{H10} chain, ring, and sheet and in $C_{2\rm{v}}$ symmetry for the \ce{H10} pyramid.
The \ce{H12}, \ce{H14}, and \ce{H16} analogs of the four systems were run with the same symmetry as their \ce{H10} counterparts with the exception of the \ce{H14} pyramid, which used $D_{\rm{2h}}$ symmetry.   
All computations using localized orbitals were performed in $C_{\rm{1}}$ symmetry.
The ranges of threshold parameters used for each method are given in the Supplementary Material.

MP2, CCSD, and CCSD(T) computations were performed using the  \textsc{Psi4}, while V2RDM calculations employed the open source \textsc{v2rdm-CASSCF} plugin. \cite{Fosso2016LargeScale}  
CR-CC(2,3) computations were performed using GAMESS.\cite{barca2020recent}
FCI and ACI computations were performed using our open-source code \textsc{Forte}.\cite{Evangelista2019Forte}  
All ACI computations included additional determinants to ensure spin completeness of the $\mathcal{P}$ and $\mathcal{Q}$ spaces.
The rank-reduction procedure used for SVD-FCI and the \textit{a posteriori} determinant screening procedure for ap-sCI were  implemented in a development version of \textsc{Forte}.

Density matrix renormalization group calculations were performed with \textsc{CheMPS2}.\cite{Wouters2014Chemps2}
DMRG calculations associated with a particular final value of $M$ were preceded by three preliminary computations with smaller bond dimension and added noise. 
This procedure has been shown to make the overall DMRG calculation converge more rapidly and produce more accurate results.\cite{Chan2002HighlyCorrelated, Moritz2006ConstructionOf}   
In the first two preliminary computations $M$ is set to 150, 500, 500, and 500 (for \ce{H10}, \ce{H12}, \ce{H14}, and \ce{H16} respectively) to build an initialization for the last two instructions with a larger value of $M$. 
In cases where the final value of $M$ is less than the values specified above, the same value of $M$ is used for the three preliminary calculations and for the final calculation.
As mentioned already, due to the block structure of the DMRG tensors induced by symmetries, the final MPS in general does not correspond to a set of dense matrices of dimension $M^2$.
For DMRG calculations using a localized basis, orbitals for the 1D chain and ring were ordered to be spatially consecutive.  
Localized orbitals for the 2D sheet and 3D pyramid systems used a Fiedler vector ordering derived from the two electron integrals to account fo physical proximity and orbital overlap.\cite{Olivares2015TheAbinitio}
Plots of the localized orbitals and the site orderings are reported in the Supplementary Material.
For canonical MOs, orbitals were grouped into blocks by irreducible representation and (within each irreducible representation) ordered energetically.
For calculations using $D_{\rm{2h}}$ symmetry, the irreducible representation blocks were ordered as $A_{\rm g}$, $B_{\rm  1u}$, $B_{\rm  3u}$, $B_{\rm  2g}$, $B_{\rm  2u}$, $B_{\rm  3g}$,  $B_{\rm 1g}$, $A_{\rm  u}$ such that blocks corresponding to bonding and anti-bonding orbitals were adjacent on the DMRG lattice.  
This strategy has been shown to be successful for several DMRG studies\cite{Yanai2010MultireferenceQuantum, Kurashige2009HighPerformance, Barcza2011QuantumInfromation, Boguslawski2012EntanglementMeasures} and is rationalized by quantum information principles.\cite{Rissler2006MeasuringOrbital} 
For calculations using symmetries other than $D_{\rm{2h}}$, the ordering of the irreducible representations followed Cotton's ordering.

\section{RESULTS}
\label{sec:results}

\begin{figure}[bt!]
\centering
\includegraphics[width=2.25in]{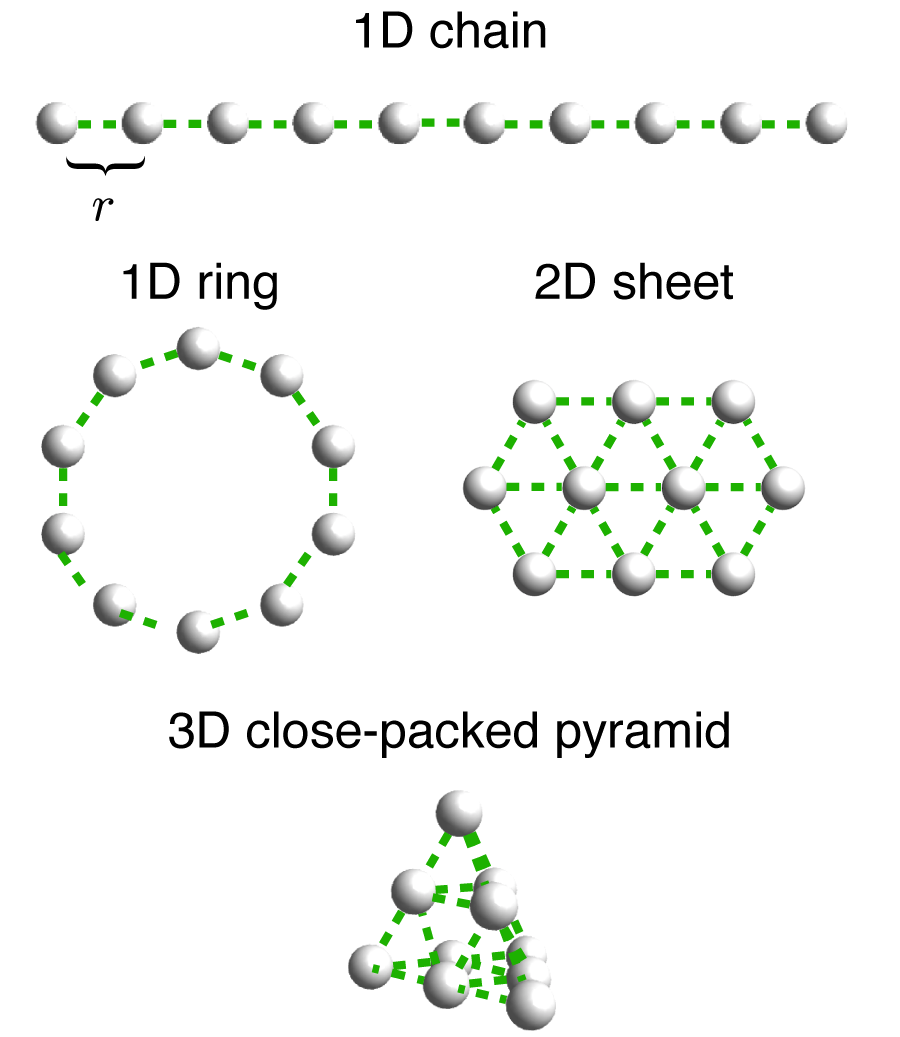}
\caption{Structure of the \ce{H10} model systems studied in this work. Geometries are parameterized by the nearest-neighbor \ce{H}--\ce{H} distance ($r$), indicated by green dashed lines.}
\label{fig:systems}
\end{figure}

In this section we analyze the results of our study for the \ce{H10} models. 
Fig.~\ref{fig:systems} shows the structure of the four \ce{H10} model systems. 
The geometry of each model is controlled by a parameter $r$ which determines the nearest neighbor \ce{H}--\ce{H}  distance (in {\AA}).
The geometries of all models, raw data for the potential energy curves, and energy errors are collected in a GitHub repository.\cite{HstudyRepo2020}

The $r$ values considered here (0.75--2.0 {\AA}) cover both the weak and strong electron correlation regimes of each model.
This point can be quantified by estimating the $U/t$ ratio of the Hubbard Hamiltonian:
\begin{equation}
\hat {H} =-t\sum _{i,\sigma }\left({\hat {a}}_{i,\sigma }^{\dagger }{\hat {a}}_{i+1,\sigma }+{\hat {a}}_{i+1,\sigma }^{\dagger }{\hat {a}}_{i,\sigma }\right)+U\sum _{i}{\hat {n}}_{i\uparrow }{\hat {n}}_{i\downarrow },
\end{equation}
where $t$ and $U$ are obtained by fitting the excitations energies (for singlet and triplet states) of the Hubbard dimer to those of the \ce{H2} molecule with bond length $r$.
Using this approach, we find that $U/t$ ranges from about 0.94 at $r = 0.75$ {\AA} to 8.55 at $r = 2.0$ {\AA}.

\subsection{\label{sec:low_lying_states}Ground and low-lying electronic states}

\begin{table*}[tb]
\begin{threeparttable}
\centering
\renewcommand{\arraystretch}{1.1}
\scriptsize 
\caption{Properties of the singlet ground state of the four \ce{H10} systems at different values of the \ce{H}--\ce{H}  distance ($r$). Summary of correlation metrics: correlation energy ($E_{\text{corr}}$), the squared Frobenius norm of the two-body density cumulant ($\norm{\pmb{\lambda}_{2}}^{2}_\mathrm{F}$), coupled-cluster amplitude diagnostics ($D_1$ and $D_2$), magnitude of the Hartree Fock coefficient in the normalized FCI expansion ($|C_{\text{HF}}|$), and total quantum information in a RHF canonical basis ($I_{\text{tot}}^\mathrm{d}$) and a localized basis ($I_{\text{tot}}^\mathrm{l}$). For the \ce{H10} pyramid at $r=2.0$~{\AA}, the data reported correspond to an excited state adiabatically connected to the ground state at smaller values of $r$. See Supplemental Material for details.}
\begin{tabular*}{6in}{@{\extracolsep{\stretch{1.0}}}*{2}{l}*{9}{r}@{}}
    \hline

    \hline
    System & $r$ / {\AA}  & $E_{\rm{FCI}}$ / \textit{E}$_{\text{h}}$ & $E_{\text{corr}}$ / \textit{E}$_{\text{h}}$ & $E_{\text{ICE}}$ / \textit{E}$_{\text{h}}$ & $ \norm{\pmb{\lambda}_{2}}^{2}_\mathrm{F}$ &  $D_1$ & $D_2$ & $|C_{\text{HF}}|$ & $I_{\text{tot}}^\mathrm{d}$ & $I_{\text{tot}}^\mathrm{l}$ \\
    \hline
    \multirow{4}{*}{\ce{H10} 1D Chain}

		&0.75&  $-$5.228560       &$-$0.1082 & $-$0.2628	&   0.61 & 0.018 & 0.202 	& 0.96	&1.24 & 13.74 \\
		&1.00&  $-$5.415393       &$-$0.1678 & $-$0.4351	&   1.46 & 0.015 & 0.302 	& 0.91	&2.57 & 13.52 \\
		&1.50&  $-$5.036293       &$-$0.4038 & $-$1.0662	&   6.11 & 0.010 & 0.696  & 0.67	&7.42 & 11.99\\
		&2.00&  $-$4.790989       &$-$0.7912 & $-$1.6754	& 13.27 & -           & -         & 0.37	&11.78 & 9.22 \\[6pt]

    \multirow{4}{*}{\ce{H10} 1D Ring}

		&0.75&  $-$5.151378	&$-$0.1026 & $-$0.2323	&   0.43 & 0.000 & 0.122 	& 0.97	&1.01 & 13.81\\
		&1.00&  $-$5.422958	&$-$0.1475 & $-$0.3650	&   1.02 & 0.000 & 0.189 	& 0.94	&2.05 & 13.67\\
		&1.50&  $-$5.048052	&$-$0.3616 & $-$1.0197	&   5.96 & 0.000 & 0.643 	& 0.67	&7.28 & 12.24\\
		&2.00&  $-$4.794398	&$-$0.7678 & $-$1.6659	& 13.64 & -           & -        & 0.32	&11.87 & 9.35 \\[6pt]

    \multirow{4}{*}{\ce{H10} 2D Sheet}

		&0.75&  $-$3.917633	&$-$0.1040 & $-$0.2325	& 0.35 & 0.008 & 0.107 	& 0.98	&  0.85 & 13.65\\
		&1.00&  $-$4.891538	&$-$0.1393 & $-$0.3262	& 0.71 & 0.014 & 0.159 	& 0.95	&  1.58 & 13.56\\
		&1.50&  $-$4.903192	&$-$0.2868 & $-$0.7820	& 2.85 & 0.038 & 0.337 	& 0.79	&  5.47 & 12.92\\
		&2.00&  $-$4.739235	&$-$0.6886 & $-$1.6949	& 9.22 & -           & -	         & 0.21	&12.36 &  9.44\\[6pt]

    \multirow{4}{*}{\ce{H10} 3D Pyramid}

		& 0.75 & $-$2.853673     &$-$0.1737 & $-$0.4151     & 1.13 & 0.015 & 0.320 	& 0.93	& 1.77 & 13.54\\
		& 1.00 & $-$4.269379     &$-$0.2397 & $-$0.5811	& 2.28 & 0.031 & 0.486	 & 0.84	& 3.13 & 13.40\\
		& 1.50 & $-$4.733459     &$-$0.4051 & $-$0.9765	& 3.54 & 0.067 & 0.635 	& 0.62	& 6.88 & 12.67\\
		& 2.00*& $-$4.694062     &$-$0.7480 & $-$1.7252	& 3.59 & 0.093 & 0.685 	& 0.25	& 12.62 & 9.48\\
    \hline

    \hline
\end{tabular*}
\label{tab:cor_props}
\end{threeparttable}
\end{table*}

We have found a variety of interesting characteristics in the ground and low lying excited states of the \ce{H10} model systems.
Metrics of correlation for the ground state of the four \ce{H10} systems as a function of the $r$ are reported in Tab.~\ref{tab:cor_props}. 
As expected, the numbers show an increase in correlation as $r$ increases across all four systems. 
However, when comparing different systems, there are interesting discrepancies between the metrics. 
For example, at $r=1.5$~{\AA} the 1D  chain has the second largest absolute value of \Ecorr  ($0.4038$ \Eh), the largest absolute value of intrinsic correlation energy ($1.066$ \Eh), a high \RCMnorm value (6.11), and the largest $D_2$ value (0.70); however, this system unexpectedly displays a relatively large weight of the Hartree--Fock determinant  ($|C_\mathrm{HF}| = 0.67$).
A comparison of the ring with the chain, shows that the former is slightly less correlated than the latter.
In the case of the 2D sheet at $r = 1.5$ {\AA}, all metrics of correlation indicate that this system has the smallest degree of electron correlation.
In contrast, the 3D pyramid displays the strongest correlation effects, yielding the largest absolute value of \Ecorr  ($0.4051$ \Eh), a large intrinsic correlation energy ($0.9765$ \Eh), and the smallest HF determinant  weight ( $|C_\mathrm{HF}| = 0.62$).
However, strong correlation in the 3D system is not reflected in the value of \RCMnorm (3.54), which is smaller than that of both 1D systems ($\approx 6$).
The quantum information metric (in a delocalized basis) $I_{\text{tot}}^\mathrm{d}$ paints a similar picture: the pyramid total information lies in between that of the 1D systems and the less correlated 2D system. 
However, in a localized basis, the same metric $I_{\text{tot}}^\mathrm{l}$ decreases for all systems as a function of $r$.
This behavior is interesting as it suggests that quantum information metrics could potentially be useful for choosing orbitals to use with various approximate methods.
As discussed in more detail in Sec.~\ref{sec:spin_correlation}, the low value of \RCMnorm observed for the 3D pyramid is likely a consequence of spin frustration, which results in a rapid decay of spin correlation functions.
We also note that after $r = 1.5$ {\AA} the ground state of the 3D pyramid crosses several low-lying singlet states and by $r = 2.0$ {\AA} it corresponds to the third excited state of $A_g$ symmetry.
See the Supplementary Material for a plot of the low-lying states of the 3D pyramid in the range $r = $ 1.5--2.0 {\AA}.

The small discrepancies observed in the various metrics can be owed to the fact that they measure different aspects of correlation. 
While \Ecorr and  $|C_\mathrm{HF}|$ quantify the deficiency of the mean-field treatment (measured in both energetic and wave function terms), quantities like \RCMnorm and $I_\mathrm{tot}$ capture only statistical aspects of correlation.
The intrinsic correlation energy ($E_\mathrm{ICE}$) appears to offer a good compromise between the mean-field and statical measures of correlation; nevertheless, its value is significantly larger than the \Ecorr values and captures contributions due to Coulomb repulsion (i.e., absent Coulomb repulsion, $E_\mathrm{ICE}$ is zero even for a correlated state).
The $D_1$ and $D_2$ metrics measure the importance of orbital rotations ($D_1$) and correlation effects ($D_2$) in the CCSD wave function.
In particular, since $D_1$ is not directly related to electron correlation, its behavior is very different from that of $D_2$, with the latter growing with $r$ in all models.
In contrast, $D_1$ decreases in the 1D chain, it is identically zero in the 1D ring due to the different symmetry of singly excited determinants, and it grows with $r$ in the 2D and 3D models.

Another common approach to diagnose the onset of strong correlation is symmetry breaking of the Hartree--Fock solution.
The Coulson--Fischer point (here defined in terms of the restricted $\rightarrow$ unrestricted symmetry breaking) of the chain and ring models is found at $r=0.85$~{\AA} and 1.05~{\AA}, respectively.
Consistent with the lower degree of correlation in the 2D sheet, the corresponding UHF solution exhibits spin-contamination at a point farther out in the dissociation curve (1.35~{\AA}).
Instead,  the 3D pyramid exhibits symmetry breaking at the smallest distance (0.70~{\AA}) compared  to the other three systems.

\begin{figure}[bth]
\centering
\includegraphics[width=3.35in]{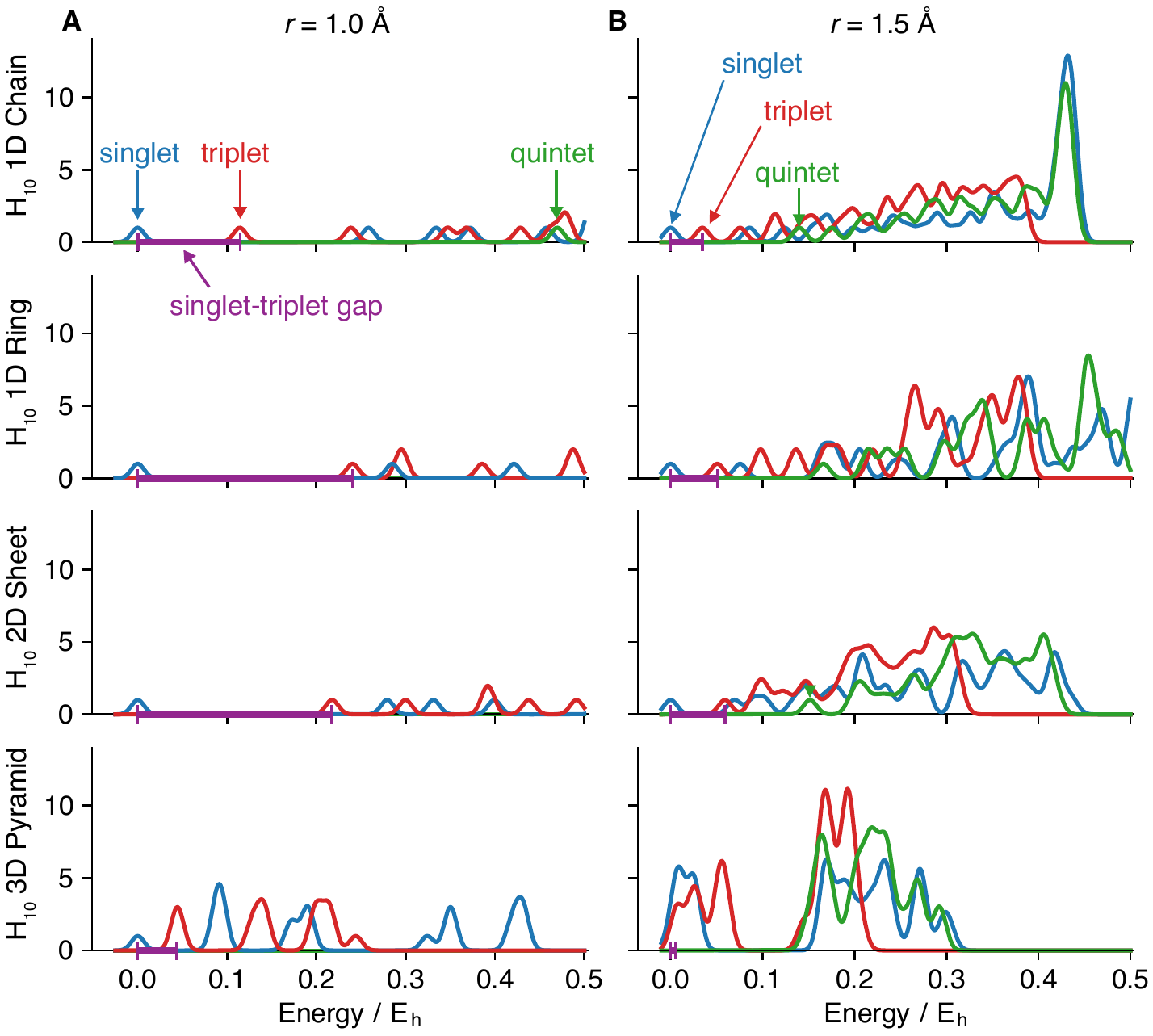}
\caption{Density of states with fixed electron number convoluted with a  Gaussian function [$D_g(E)$, defined in Eq.~\eqref{eq:dos}] computed from the 50 lowest singlet, triplet, and quintet states of the \ce{H10} systems at an H--H distance (A) $r$ = 1 {\AA} and (B) $r$ = 1.5 {\AA}.
The density of states was convoluted with a Gaussian function of exponent $\alpha=10^5$~$E_\mathrm{h}^{-2}$.
}
\label{fig:spectra}
\end{figure} 

Lastly, we characterize the strength of correlations by computing the density of states (DOS) with fixed particle number [$D(E)$].
For convenience, we convolute the density of states with a Gaussian function of exponent $\alpha$ and shift the energies by the ground state energy ($E_0$).
This convoluted DOS is expressed in terms of excitation energies  $\Delta E_i = E_i- E_0 $, where $E_i$ are energies of singlet, triplet, and quintet electronic excited states, and it is given by
\begin{equation}
\label{eq:dos}
D_g(E) = \sum_i \exp(-\alpha (\Delta E_i - E)^2).
\end{equation}
Note that this quantity is different from the DOS computed for electron attached/detached states.

Figure~\ref{fig:spectra} shows the energy spectra in the range 0--0.5 \Eh (0--13.6 eV) relative to the ground state from computations of the lowest 50 singlet, triplet, and quintet states of the \ce{H10} systems. At shorter bond lengths ($r = 1.0$~{\AA}), the 1D and 2D systems show large gaps between the ground state and the lowest triplet state.
However, this gap closes significantly in the 3D pyramid to ca.  0.044~\Eh.
At longer bond lengths ($r = 1.5$~{\AA}), the singlet-triplet gap decreases for all systems. Interestingly, the 3D pyramid shows an almost zero gap (ca. 0.007~\Eh) and several singlet near-degenerate states accumulate near the ground state.

\subsection{\label{sec:spin_correlation}Spin correlation and frustration}

\begin{figure*}[tbh]
\centering
\includegraphics[width=6in]{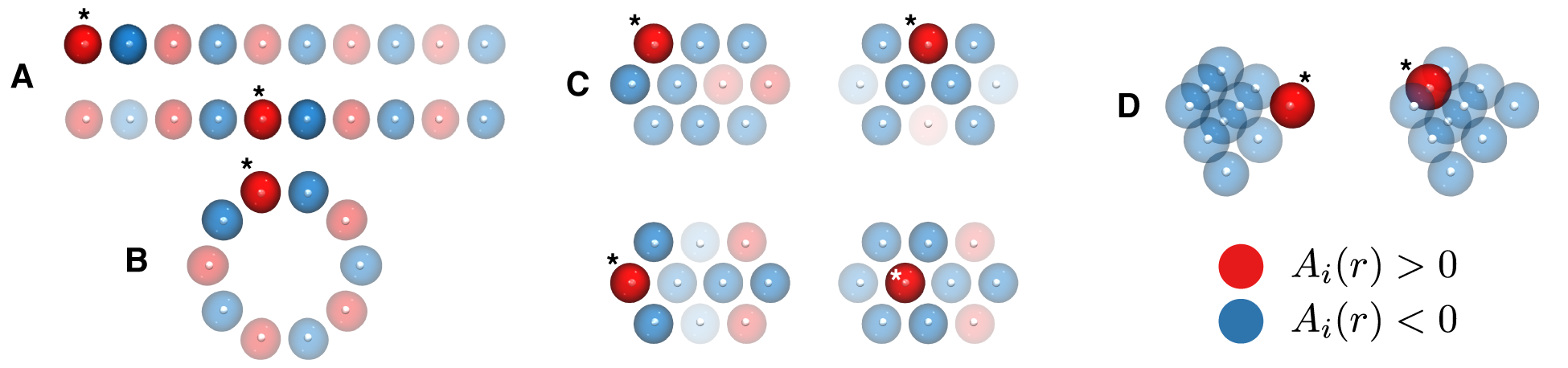}
\caption{Spin correlation density $A_{i}(\mathbf{r})$ of the \ce{H10} models at an H--H distance $r=1.25$~{\AA} plotted for (A) the edge and central localized MO sites of the hydrogen chain, (B) the symmetry unique site of the hydrogen ring, (C) the four symmetry unique sites of the \ce{H10} sheet, and (D) the two symmetry unique sites of the \ce{H10} pyramid. Positive and negative values of $A_{i}(\mathbf{r})$ are indicated in red and blue, respectively, and in each plot the localized orbital $\phi_i$ is denoted by an asterisk.}
\label{fig:spin_correlation}
\end{figure*} 

We have found that there are signs of spin frustration in the 2D sheet and 3D pyramid models.
Frustration is indicated by the inability to satisfy antiferromagnetic interactions---a condition which is not mathematically rigorous but that has, nonetheless, been used to define systems as spin frustrated\cite{Baker2012ClassificationOf}---and the lack of long range antiferromagnetic ordering beyond nearest-neighbor interactions.
 
The spin-spin correlation densities shown in Fig.~\ref{fig:spin_correlation} A and B indicate clear  antiferromagnetic ordering beyond nearest neighbors in the 1D chain and ring.
Each localized spin is anti-correlated with its nearest neighbor, as depicted by the adjacent red and blue shading. 
Fig.~\ref{fig:spin_correlation} C shows spin-spin correlation density for the four symmetry unique sites in the 2D sheet. 
In contrast to the 1D models, it can be seen that there is no way to simultaneously satisfy all antiferromagnetic interactions for the 2D sheet, and consequentially, spin correlations decay more rapidly. 
This is also the case for the 3D pyramid (see Fig.~\ref{fig:spin_correlation} D), for which each site is anti-correlated with all other sites, suggesting no antiferromagnetic ordering beyond nearest neighbors.
Tab.~\ref{tab:spin_cor_props} summarizes spin correlation properties for the \ce{H10}, \ce{H12}, and \ce{H14} systems at $r=1.5$~{\AA}. 
As is the case with the 2-body cumulant norm, the \ce{H10} 1D chain and ring systems have larger absolute spin correlation $\langle S^2 \rangle_{\rm{abs}}$ (17.42 and 18.66, respectively) than the 2D or 3D systems (11.55 and 10.86, respectively).
The short-range nature of spin correlation of the 2D and 3D  \ce{H10} systems is also indicated by their smaller value of  $\langle S^2 \rangle_{\rm{abs,lr}}$ (2.46 and 3.04), compared to the 1D systems (5.25 and 6.51, for the chain and ring respectively).
These results are consistent with the spin correlation density analysis in Figs.~\ref{fig:spin_correlation}.
We note that the scaling of $\langle S^2 \rangle_{\rm{abs}}$ and $\langle S^2 \rangle_{\rm{abs,lr}}$ with $n$ for the sheet and pyramid systems is (in most cases) linear or super-linear, which is not expected for systems absent of long range spin ordering. 
However, this is likely because the lattice sizes considered are still relatively small, and the addition of two hydrogens at a time does not extend the lattices in a completely uniform manner, thus altering (possibly greatly) the frustrated character. 
\begin{table}[!ht]
\begin{threeparttable}
\centering
\renewcommand{\arraystretch}{0.9}
\caption{Ground-state of the four \ce{H}$_{n}$ systems at an H--H distance $r=1.50$~{\AA}. Sum of the absolute value of the spin-spin correlations ($\langle S^2 \rangle_{\rm{abs}}$), the sum of the absolute value of the long range spin-spin correlations ($\langle S^2 \rangle_{\rm{abs,lr}}$), and the sum of the nearest neighbor spin-spin interactions ($\langle S^2 \rangle_{\rm{nn}}$). See Eqs.~\eqref{eq:S_sq_abs}--\eqref{eq:S_sq_nn} for the definition of these metrics.}
\begin{tabular*}{\columnwidth}{@{\extracolsep{\stretch{1.0}}}*{2}{l}*{4}{r}@{}}
    \hline

    \hline
    System & $n$  & $\langle S^2 \rangle_{\rm{abs}}$ & $\langle S^2 \rangle_{\rm{abs,lr}}$ & $\langle S^2 \rangle_{\rm{nn}}$  \\
    \hline
    \multirow{3}{*}{\ce{H}$_{n}$ Chain}

		&10	  &17.42	&5.25	&$-$3.10 	\\
		&12	  &21.77	&7.18	&$-$3.72 	\\
		&14	  &26.29	&9.27	&$-$4.35	\\[6pt]
		
    \multirow{3}{*}{\ce{H}$_{n}$ Ring}

		&10	&18.66	&6.51	&$-$3.16 	 \\
		&12	&24.13	&9.39	&$-$3.84 	 \\
		&14	&28.95     &11.91	&$-$4.42 	 \\[6pt]		

    \multirow{3}{*}{\ce{H$_{n}$} Sheet}

		&10	&11.55	&2.46	&$-$1.94 	 \\
		&12	&14.31	&2.59	&$-$2.66 	 \\
		&14	&17.16	&3.63	&$-$3.06 	 \\[6pt]		

    \multirow{3}{*}{\ce{H$_{n}$} Pyramid}

		& 10  &10.86	&3.04	&$-$1.19 	 \\
		& 12  &18.06	&4.02	&$-$3.63 	 \\
		& 14  &18.30	&5.94	&$-$2.40 	 \\

    \hline

    \hline
\end{tabular*}
\label{tab:spin_cor_props}
\end{threeparttable}
\end{table}
It is also possible to observe a lack of long-range correlation and ordering for the 2D sheet and 3D pyramid by considering the radial distribution of spin-spin correlations and absolute spin-spin correlations reported in the Supplementary Material.

It is evident from the various metrics of correlation, the DOS plots, and our analysis of spin correlation, that the \ce{H10} lattices display a broad range of correlation regimes.
Therefore, we believe it is important to consider the 2D and 3D models in future benchmarks of electronic structure methods because they capture some aspect of the physics of spin frustration that are not displayed by 1D hydrogen models.

\subsection{\label{H10compression} Performance of sCI, SVD-FCI, and DMRG}
\begin{figure*}[ht]
\centering
\includegraphics[width=0.95\textwidth]{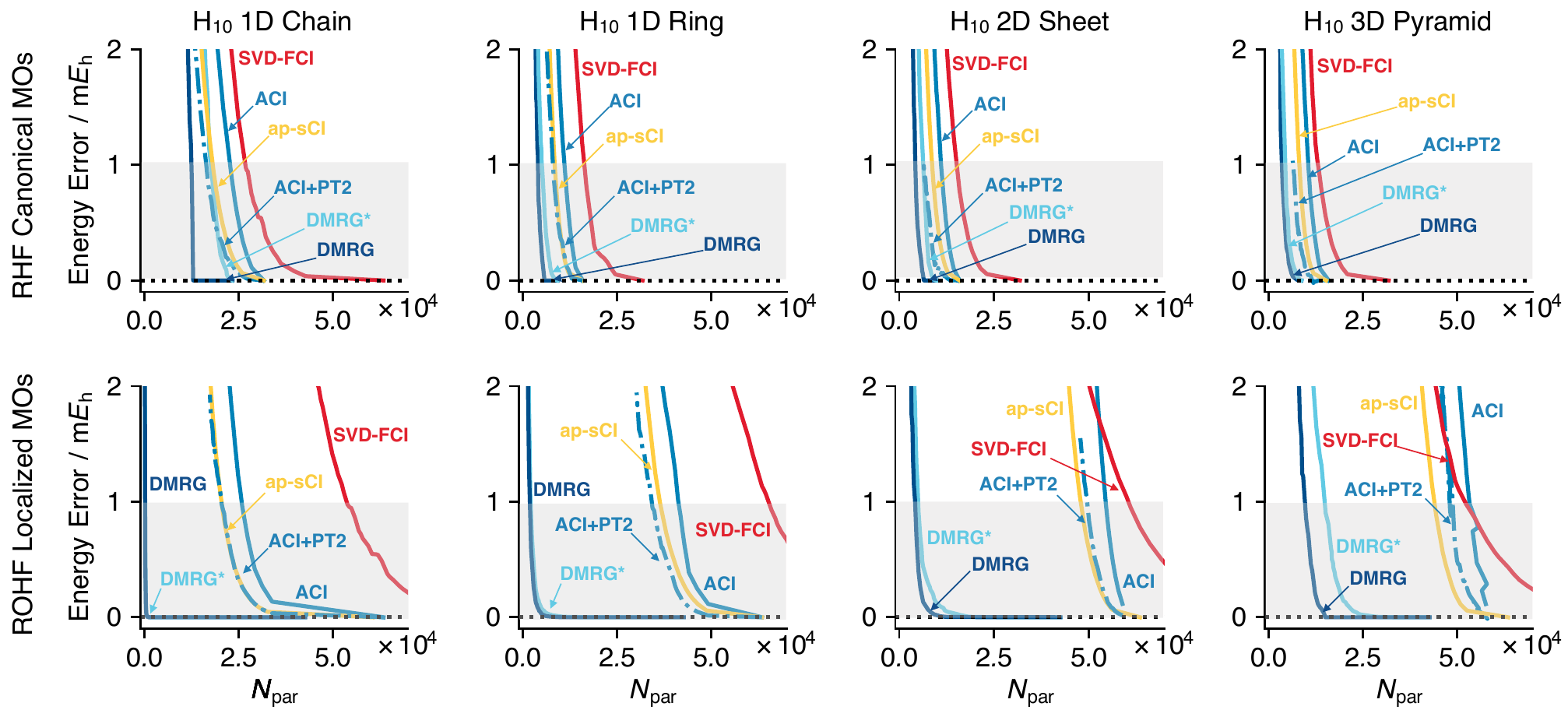}
\caption{Ground-state of the four \ce{H10} models at an H--H distance $r = 1.5$~{\AA}. Energy error with respect to FCI vs. number of parameters ($N_\mathrm{par}$) of approximate methods. The gray shaded region represents chemically accurate energies (error less than 1m\Eh). }
\label{fig:h10compression_energy}
\end{figure*}

Having characterized the nature of the ground state of the \ce{H10} models we now proceed to analyze the efficiency with which sCI, SVD-FCI, and DMRG approximate the wave functions of these systems. 

\begin{figure*}[ht]
\centering
\includegraphics[width=0.95\textwidth]{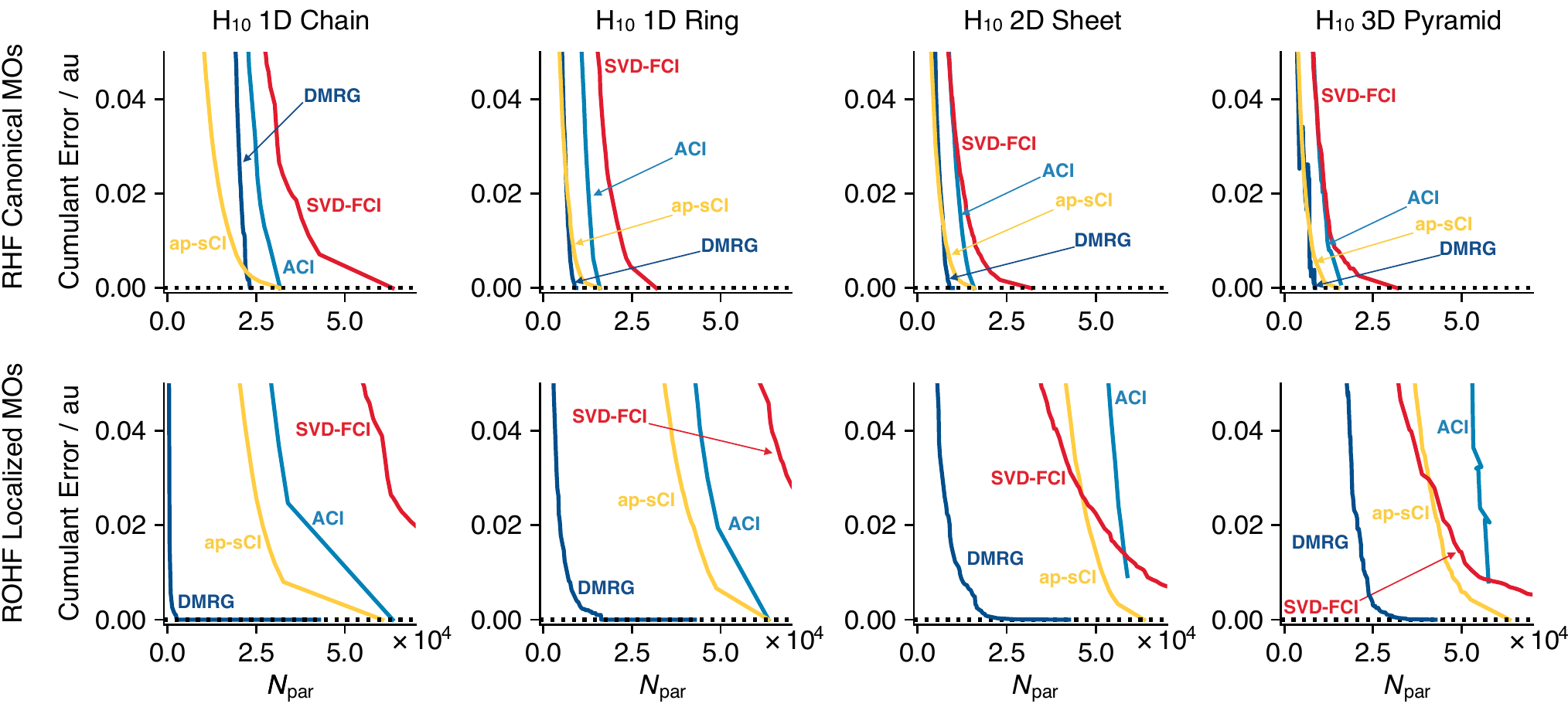}
\caption{Ground-state of the four \ce{H10} models at an H--H distance $r = 1.5$~{\AA}. Density cumulant error $ \norm{\Delta\pmb{\lambda}_{2}}_\mathrm{F}$ with respect to FCI vs. number of parameters ($N_\mathrm{par}$) of approximate methods.}
\label{fig:h10compression_cumulant}
\end{figure*}

In Fig.~\ref{fig:h10compression_energy} we plot the energy error [$E_X(N_{\rm{par}}) - E_\mathrm{FCI}$] as a function of the number of variational parameters for the \ce{H10} systems in the regime of strong electron correlation ($r=1.5$~{\AA}).
The accuracy volume may be obtained from these plots by finding the number of parameters corresponding to a 1 m\Eh error.
When using canonical orbitals, we see that DMRG affords the most compact representation, although $\mathcal{V}_{\text{ap-sCI}}$ and $\mathcal{V}_{\text{ACI+PT2}}$   are within a factor of 1.5--2 of $\mathcal{V}_{\text{DMRG}}$.
ACI without the PT2 correction always requires more variational parameters to match the accuracy of ap-sCI and ACI+PT2, and $\mathcal{V}_{\text{ACI}}$ is 2--3 times $\mathcal{V}_{\text{DMRG}}$.
For all four \ce{H10} systems, SVD-FCI exhibits the worst efficiency, although only by a small margin, such that $\mathcal{V}_{\text{SVD-FCI}}$ is 2--4 times greater than $\mathcal{V}_{\text{DMRG}}$.
Note that we include two sets of results for DMRG: the lowest energy eigenvalue found during all DMRG sweep optimizations (labeled DMRG, see Ref. \citenum{Wouters2014Chemps2} for details), and the energy obtained from the reduced density matrices of the final MPS (indicated with DMRG*). 
When a large bond number $M$ is used, the two energy values are nearly identical, but for smaller values of $M$, the DMRG* value may be slightly higher than the DMRG one.

When using localized orbitals, we see that DMRG again produces the most compact representation and by a much larger margin for all four \ce{H10} systems.  
In particular, for the 1D chain $\mathcal{V}_{\text{DMRG}}$ is two orders of magnitude lower than all other methods.
Comparing the accuracy volume of DMRG with different orbital bases, one notices that the localized basis is more efficient in the 1D systems, while the delocalized basis leads to smaller $\mathcal{V}_{\text{DMRG}}$ for the 3D model. For the 2D model, the localized and canonical basis yield comparable $\mathcal{V}_{\text{DMRG}}$ values.
The advantage of using canonical orbitals is inconsistent with previous findings,\cite{Olivares2015TheAbinitio} and it is likely due to the size of the systems considered here.
In this case, the compression afforded by using a localized basis is outweighed by the advantages of using point group symmetry.
When comparing results across canonical and localized orbitals for the other methods, we find that the accuracy volume is always smaller in the delocalized basis, so there is no advantage to orbital localization.
This is in agreement with past observations\cite{schriber2018combined} that localization is beneficial for sCI methods only after a certain system size is reached.
It is interesting to observe that the accuracy volume for DMRG and sCI mirrors the behavior of the total quantum information for both delocalized and localized bases (see Tab.~\ref{tab:cor_props}), suggesting that this metric may be useful for determining the best orbital basis to use at a given geometry.

We note that for the sheet and pyramid, there are a few values of $\sigma$ for which the ACI results do not converge monotonically and lead to small bumps.
We have also encountered cases where the iterative ACI algorithm finds the first excited state due to near-degeneracies, an issue that may be resolved using a state-averaged version of the method.\cite{Schriber2017Adaptive}
These incorrect energies were not included in Fig.~\ref{fig:h10compression_energy}.

\begin{table*}[t]
\begin{threeparttable}
\centering
\scriptsize
\renewcommand{\arraystretch}{1.1}
\caption{Accuracy volume ($\ncomp$) computed for the ground state of the \ce{H10} models for various methods. Values are reported for both localized and delocalized molecular orbital bases. The Hilbert space size ($|\mathcal{H}_N|$) for all models in $C_1$ symmetry is 63504. Hilbert space sizes with the largest abelian symmetry exploited in the computations are reported in the table. ACI$+$PT2 values with a $<$ sign indicate that the energy error with the reported number of parameters is significantly lower than 1.0~m\Eh. Finding more precise values of $\ncomp$ for ACI$+$PT2 is challenging as the energy error is not monotonic as a function of $\sigma$.}
\begin{tabular*}{5.8in}{@{\extracolsep{3pt}}*{3}{l}*{10}{r}@{}}
    \hline
    
    \hline
       & & & \multicolumn{5}{c}{Delocalized (RHF Canonical)} 
    &  \multicolumn{5}{c}{Localized (Pipek--Mezey)} \\ 
      \cline{4-8} \cline{9-13}
    System & $|\mathcal{H}_N|$ & $r$ / {\AA}  & ap-sCI  & ACI & ACI$+$PT2 & SVD-FCI & DMRG &  ap-sCI & ACI & ACI$+$PT2 & SVD-FCI & DMRG \\[1pt]
  \hline
     \multirow{4}{*}{1D Chain} & 
    \multirow{4}{24 pt}{31752 \\ ($D_{2\rm h}$)} 
                  &0.75 & 1491   & 2066   &   $<$335  & 5292   &  2600  & 41872 & 46882 & 45052 & 10584 &  468 \\
 		& &1.00 & 5122   & 6978   & $<$2156  & 10584 &  4896  & 35962 & 42332 & 39510 & 21168 &  388 \\
 		& &1.25 & 11201 &14231  &        7347  & 17136 &  9598  & 29148 & 35306 & 30732 & 34272 &  376 \\ 
 		& &1.50 & 18176 & 22989 &      16356 & 26964 & 12674 & 20424 & 26008 & 20564 & 53928 &  176 \\[6pt] 

    \multirow{4}{*}{1D Ring} &
    \multirow{4}{24 pt}{15912 \\ ($D_{2\rm h}$)}

		& 0.75 & 577   & 873     & $<$181 & 2784   &  740  & 53358 & 56244 &53448& 11088 & 3359 \\ 
		& & 1.00 & 2019 & 2803   &     663   & 5328   & 1522 & 49982 & 53364 &50084& 21168 & 3164 \\
		& & 1.25 & 4791 & 6384   &     2701 & 9492   & 2663 & 45452 & 49537 &43486& 37800 & 2688 \\
		& & 1.50 & 8520 & 11056 &     7895 & 16296 & 4034 & 36450 & 41254 &34134& 65016 & 1884 \\[6pt]
		
    \multirow{4}{*}{2D Sheet} & 
    \multirow{4}{24 pt}{15912 \\ ($D_{2\rm h}$)}
    
		& 0.75 & 766   &1102   & $<$218  & 2532   & 1117  & 53252  & 59470 & 58050 & 10080 &   4649\\
		& & 1.00 & 1899 & 2809  & $<$718  & 4296   & 1853  & 51822 & 58256 & 56252 & 17136 &  4071\\
		& & 1.25 & 4139 & 5283  &     3478  & 7848   & 2626  & 50318 & 57036 & 53852 & 31248 &   4113\\
		& & 1.50 & 8667 & 11122 &     6468 & 15156 & 4218  & 47916 & 54466 & 49540 & 60480 &  4192\\[6pt]

    \multirow{4}{*}{3D Pyramid} &
    \multirow{4}{24 pt}{15912 \\ ($C_{2\rm v}$)}    

                 & 0.75 & 1478 & 2115 &  $<$787 &  4044 &  1630 & 44062 & 56232 & 55452 & 16128 & 10832\\ 
                 & & 1.00 & 2755 & 3605 &  $<$1607 &  7056 &  2250 & 45812 & 55986 & 55078 & 28224 &12556\\ 
                 & & 1.25 & 4997 & 6530 &  2869 &  9864 &  2927 & 45844 & 56348 & 52728 & 39312 & 10998\\
                 & & 1.50 & 8097 & 10519 &  6457 & 13152 &  3495 & 43932 & 53280 & 48580 & 52416 &  9489\\
		
    \hline

    \hline
\end{tabular*}
\label{tab:h10compression_r}
\end{threeparttable}
\end{table*}           

Fig.~\ref{fig:h10compression_cumulant} shows plots of $N_{\rm{par}}$ vs. the two-body cumulant error $\norm{\Delta\pmb{\lambda}_{2}}_\mathrm{F}$ for the four \ce{H10} systems at $r=1.5$~{\AA}.
These plots do not include ACI+PT2 results since second-order corrections to the ACI 1- and 2-RDMs were not available.
We find similar trends for the efficiency to represent $\pmb{\lambda}_{2}$ as we do for the energy, with the caveat that in a canonical basis, ap-sCI generally gives the best compression efficiency.
It can be seen that with canonical orbitals, ap-sCI actually preserves the accuracy of the two-body density cumulant after compression better than DMRG does for the 1D chain, and similarly to DMRG for the other three systems.
There is also a larger disparity in the performance of ap-sCI and ACI for cumulant compression performance, which can be attributed to two reasons. 
First, ACI adds additional determinants at each iteration to ensure spin completeness (the compressed ap-sCI wave function is not guaranteed to be an eigenfunction of spin). 
Second, ACI selects determinants according to their energetic contribution, and not explicitly their contribution to the wave function.
SVD-FCI is the least efficient in compressing the wave function for the 1D systems, but does nearly as well as ACI for the 2D sheet and 3D pyramid.
However, it possible that if variational optimization is used for SVD-FCI the cumulant error may increase, similarly to the behavior observed for ACI.
When using localized orbitals, it can be seen that DMRG likewise shows the best compression efficiency with respect to $\norm{\Delta\pmb{\lambda}_{2}}_\mathrm{F}$, especially for the 1D chain and ring systems.

As shown in Tab.~\ref{tab:cor_props} of Sec.~\ref{sec:low_lying_states}, the degree of correlation for all models increases as the \ce{H}--\ce{H}  distance $r$ becomes larger. 
In Tab.~\ref{tab:h10compression_r} we can see that in a delocalized basis the complexity of the wave function, as gauged by $\ncomp$, also increases as $r$ becomes larger, such that all methods require a larger number of parameters to achieve chemical accuracy.
In a localized basis $\mathcal{V}_{\text{DMRG}}$, $\mathcal{V}_{\text{ap-sCI}}$, and $\mathcal{V}_{\text{ACI}}$ decrease with increasing $r$, suggesting that these methods can exploit the local character of correlation, although in most cases not enough to outweigh the benefits of symmetry-adapted delocalized orbitals.
In a delocalized basis, we note that for small values of $r$ the ACI+PT2 produces very accurate results with very few parameters, outperforming DMRG using just a few hundred determinants.
It is interesting to note that compression efficiency for SVD-FCI decreases dramatically as $r$ increases, suggesting that the method is not able to take advantage of local correlation.  
Additionally, it can be seen that at more contracted geometries (smaller values of $r$), there is less of a disparity between the compression performance of the various approaches.

\subsection{\label{sec:trad_methods}Comparison with other electronic structure methods}

\begin{figure}[h!]
\centering
\includegraphics[width=3.2in]{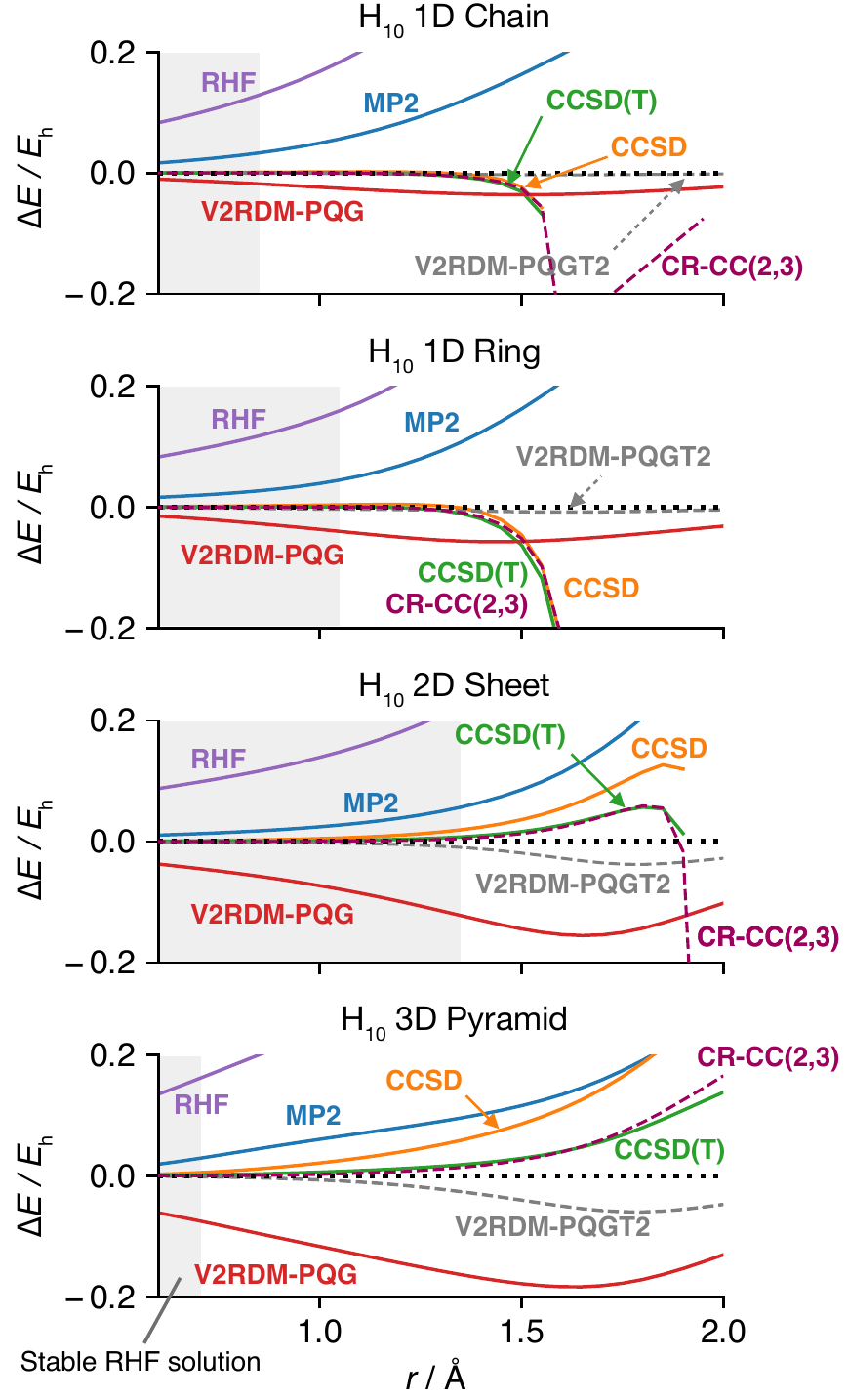}
\caption{Ground-state potential energy curves of the four \ce{H10} models. Energy error ($\Delta E$) with respect to FCI for various electronic structure methods as a function of the H--H distance ($r$). The gray shaded region indicates the range of $r$ for which the restricted Hartree--Fock solution is stable.}
\label{fig:comon_method_errors}
\end{figure}

It is interesting to use the \ce{H10} models to benchmark the robustness and accuracy of conventional  methods that employ a fixed number of parameters.
Fig.~\ref{fig:comon_method_errors} compares the energy errors relative to FCI for RHF, MP2, CCSD, CCSD(T), CR-CC(2,3), V2RDM with the two-body positive-semidefinite P, Q, and G conditions (V2RDM-PQG), and V2RDM-PQG with additional three-body positive semidefinite T$_2$ conditions (V2RDM-PQGT2).
RHF deviates significantly from FCI for all four systems, even near the \ce{H2} equilibrium geometry ($r_e=0.74$~{\AA}), where it gives errors of approximately 80--100~m\Eh. 
MP2 reduces the energy error near $r_e$ to about 10~m\Eh.   
While the RHF and MP2 energies do not diverge, they do not capture the dissociation of the \ce{H10} systems even qualitatively, giving energy errors well over 100--200~m\Eh for $r\geq1.6$~{\AA}. 

The three coupled cluster variants---CCSD, CCSD(T), and CR-CC(2,3)--- achieve chemical accuracy for the 1D chain and ring systems for $r\leq1.0$~{\AA}, and diverge beginning around $r\geq1.5$~{\AA}, past the Coulson--Fisher point. 
Performance for CCSD, CCSD(T), and CR-CC(2,3) is slightly worse for the 2D sheet and 3D pyramid, where chemical accuracy is only achieved for $r\leq0.75$~{\AA}, and divergence is seen once again at larger values of $r$.   
For all four systems, when $r>1.25$~{\AA}, the magnitude of the HF coefficient $|C_{\rm{HF}}|$ in the FCI wave function is less (or significantly less) than 0.9.
It is worth mentioning here, however, that a handful of hydrogen systems have been investigated with variants of CC that provide stable results relative to the examples in Fig. ~\ref{fig:comon_method_errors}. Namely, the paired coupled cluster doubles (pCCD) \cite{Limacher2013NewMean} and the singlet pCCD (CCD0). \cite{Bulik2015CanSingle}
           
The V2RDM approaches achieve the best descriptions of the potential energy surfaces compared to the other methods used in this section. Enforcing the PQG conditions during the optimization gives a good qualitative description of the dissociations, but still produces large quantitative errors in the range of 10--50~m\Eh for the 1D chain and ring systems and 50--200~m\Eh errors for the 2D sheet and 3D pyramid.
Enforcing the additional T2 condition improves the V2RDM results significantly, such that energy errors for the chain and ring systems at $r=1.50$~{\AA} are 3.0~m\Eh and 7.4~m\Eh, respectively.
It can be seen, however, that for the 2D sheet and 3D pyramid, V2RDM-PQGT2 fails to produce chemically accurate results by a large margin, with errors of the order of 10--50~m\Eh at stretched geometries.  
Interestingly, the performance of V2RDM is far less sensitive to $r$ than RHF, MP2, or CC as indicated by smaller values of nonparallelism error (the maximum error minus the minimum error over the entire range of $r$). 
Additionally, the error for V2RDM has a maximum in the re-coupling region ($r\approx1.5$~{\AA}), while all other methods generally decrease in accuracy with increasing $r$.

\section{\label{sec:comparison}Scaling of the accuracy volume and size consistency}
In this section we discuss some of the formal properties of the methods and present numerical results concerning scaling of the accuracy volume and size consistency.
We begin by comparing the scaling with respect to system size.
sCI may be considered a zero-dimensional ansatz, in the sense that it is particularly efficient in the description of few electrons in many virtual orbitals, especially due to the PT2 correction.
If one demands that the sCI energy is size consistent for a set non-interacting fragments $A\cdots B \cdots C\cdots$, one concludes that the number of parameters grows as $N_A N_B N_C \ldots$.
In other words, the sCI accuracy volume (per electron) grows exponentially with the number of electrons $N$ (albeit with a smaller prefactor than FCI), $\mathcal{V}_{\text{sCI}} \propto \exp(N)$.
In practice we find this to be the case for ACI using localized orbitals.
To achieve an accuracy of approximately $10^{-4}$~\Eh per electron for a system of five non-interacting \ce{H2} molecules requires 2380 parameters, rather than the 20 parameters required by a product state built from solutions for each \ce{H2} molecule.

Our analysis also suggests that the SVD-FCI approach suffers from exponential growth of the accuracy volume (independently from  dimensionality), although, to the best of our knowledge a formal analysis has not been reported. 
Even in the best case scenario, the number of parameters for a rank 1 SVD-FCI approximation scales as $N_{\text{SVD-FCI}} \propto 2 \sqrt{N_{\mathcal{H}}}$, which implies $\mathcal{V}_{\text{SVD-FCI}} \geq 2 \sqrt{N_{\mathcal{H}}}$.
We likewise observe that for the same system of dissociated hydrogens, SVD-FCI requires 14616 parameters to achieve approximately $10^{-4}$~\Eh per electron.
Like the result for sCI, this indicates that SVD-FCI with a fixed number of parameters is not size consistent.

In the case of a DMRG, a MPS with bond dimension $M$ can describe a system with entanglement entropy $S$ bound by the condition $S \leq \log_2 M$,\cite{Evenbly2011TensorNetwork} or equivalently, $\exp(S) \leq C M$, with $C$ a constant.
For gapped systems of dimensionality $D$ that satisfy an area law, the entanglement entropy is expected to scale as $S \propto L^{D-1}$ (plus  logarithmic corrections for non-gapped systems), where $L$ is the length scale of the system.\cite{hastings2004lieb}
Therefore, in DMRG the bond dimension $M$ scales at most as $\exp(\gamma L^{D-1}) \propto \exp(\gamma N^{(D-1)/D})$.
Similarly, we estimate that the accuracy volume of DMRG scales as $\mathcal{V}_{\text{DMRG}} \propto N M^2 =  N \exp(2 \gamma N^{(D-1)/D})$.
For one dimensional systems ($D = 1$) $\mathcal{V}_{\text{DMRG}}$ is independent of system size and the ground state can be well approximated by a finite bond dimension.
Beyond one dimension, this analysis suggests that the DMRG bond dimension grows exponentially.
However, for $D = 2$, DMRG is already exponentially more efficient than sCI since  $\mathcal{V}_{\text{DMRG}}\propto N \exp(2 \gamma N^{1/2})$ .
This in practice implies that DMRG is still applicable without exponential cost to both one-dimensional and ``thin'' two-dimensional problems.\cite{stoudenmire2012studying}

DMRG is formally size consistent, giving additively separable energies for non-interacting fragments $A\cdots B$, so long as the orbitals are localized on either $A$ or $B$.\cite{chan2011density, Wouters2014TheDensity} 
If the orbitals in the DMRG lattice are ordered by subsystem ($[A\cdots B]$) then the wave function for noninteracting fragments becomes a product state of the MPS on $A$ and $B$.
Then a MPS obtained by concatenating the MPS of $A$ and $B$ with a bond of dimension one ($M=1$) is sufficient to represent the product state and satisfy size consistency.
This implies that the $\mathcal{V}^{A+B}_{\text{DMRG}}$ for a system of non-interacting fragments is approximately equal to $\mathcal{V}^{A}_{\text{DMRG}} + \mathcal{V}^{B}_{\text{DMRG}}$. In practice we have found this to be nearly true, a product state for a system of five non-interacting \ce{H2} molecules has 20 parameters, and DMRG can reproduce the FCI energy exactly with a bond dimension as small as $M=3$ (with 60 parameters). 
In principle DMRG should be able to achieve this result already with $M=2$, but we find that instead it converges to a local minimum rather than the FCI energy.

\begin{figure}[h!]
\centering
\includegraphics[width=3.2in]{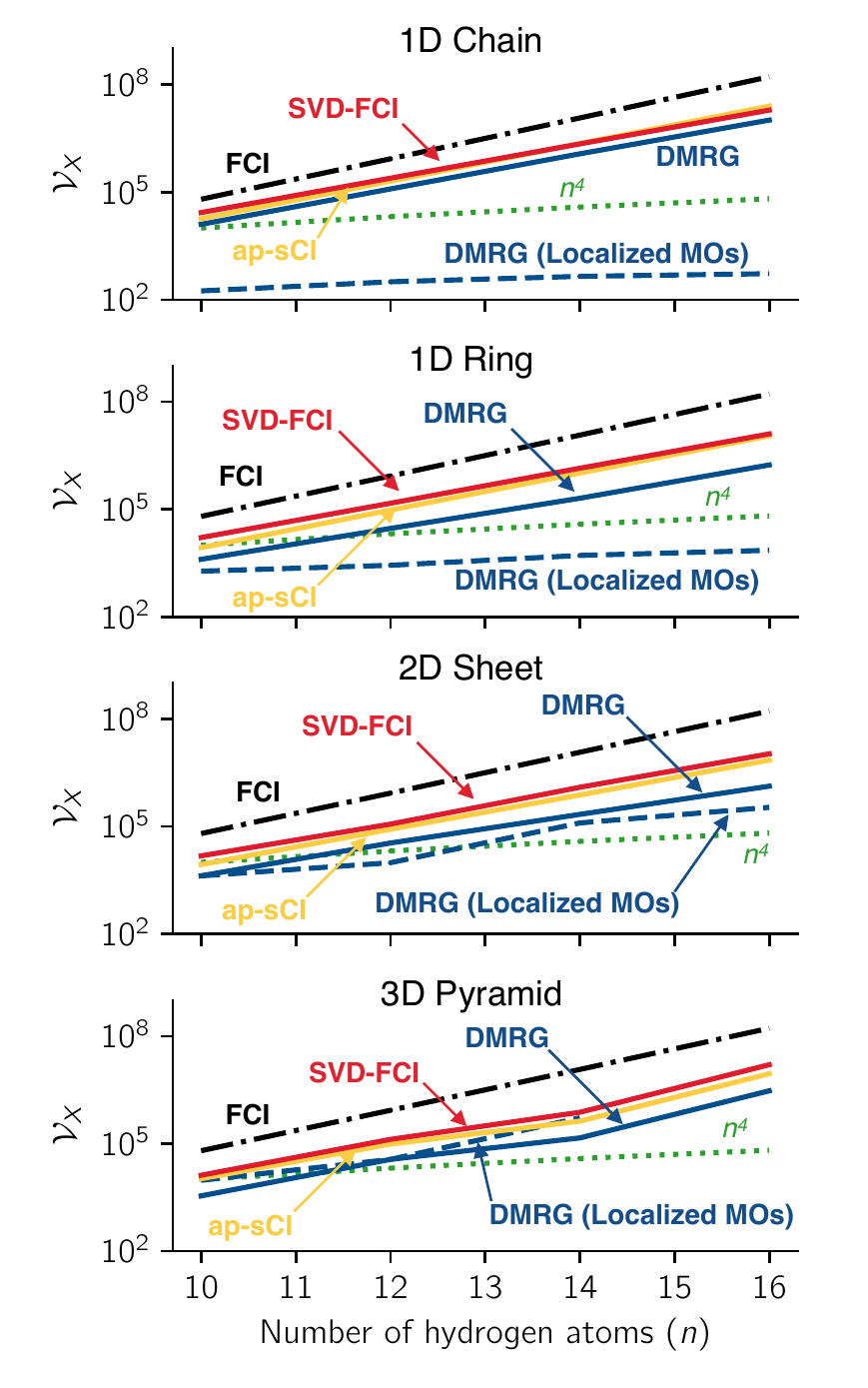}
\caption{Accuracy volume ($\ncomp$) for various approximate methods as a function of the number of number of hydrogen atoms ($n$) for the four \ce{H_n} models.
For comparison we also report the number of FCI determinants (in $C_1$ symmetry) and the curve $n^4$.
The 12, 14, and 16 hydrogen chains, rings, sheets, and pyramids are extensions of the \ce{H10} models in that the additional hydrogens are placed within the same lattice structure.
Unless otherwise noted, all results employ canonical RHF orbitals.}
\label{fig:h10compression_n}
\end{figure}

Finally, we present numerical results for the scaling with system size of the accuracy volume for analogous of our four model systems with upt to up to sixteen hydrogens.
In Fig.~\ref{fig:h10compression_n}, we plot the accuracy volume for $n$ = 10, 12, 14, 16 at $r=1.5$~{\AA}, corresponding to absolute energy errors of 1.0, 1.2, 1.4, and 1.6~m\Eh, respectively.
For comparison, we have also included the size of the FCI space (in $C_1$ symmetry) and a curve with $n^4$ scaling, which is proportional to the number of Hamiltonian matrix elements.
We note that the ground states of the \ce{H12} ring, \ce{H16} ring, and \ce{H16} sheet have symmetries ($B_{1\rm g}$), ($B_{1\rm g}$), and ($B_{3\rm u}$), respectivly, different from that of all other systems ($A_{\rm g}$).
Additionally, we note a small dip in the curve for the 3D systems at 14 hydrogens, which we attribute to the different symmetry used for that lattice, $D_{2\rm h}$ as opposed to $C_{2\rm v}$.
It can be seen that DMRG again provides the best compression of the wave function as measured by the accuracy of the energy for different systems sizes.
In a localized basis, a polynomial fit of $\mathcal{V}_\text{DMRG}$  as a function of the number of hydrogens ($n$) gives a scaling proportional to $n^{2.1}$ for the chain and $n^{3.3}$ for the ring, demonstrating the advantage of this methods for one-dimensional systems.
It is also worth pointing out that for the larger systems (\ce{H12}--\ce{H16}), it is advantageous (though still exponentially scaling) to use localized orbitals with DMRG even for the 2D systems. This result is consistent with other DMRG studies comparing localized vs. canonical orbitals for finite 2D arene systems.\cite{Olivares2015TheAbinitio}
For all the other methods, $\mathcal{V}_{X}$ in a delocalized basis appears to scale exponentially with a prefactor smaller than that of FCI.

It may be helpful to the reader to note that for all the systems considered here we find that in many cases the FCI computations are still faster than ACI and DMRG even up to 16 electrons.
On a single node, FCI computations run in about 1~second up to 3~hours for the \ce{H10}--\ce{H16} systems, whereas the implementation of DMRG used in this work can take up to 1--2 days on a single node for the more challenging 2D and 3D systems. We observe the most striking difference in the case of DMRG applied to the 1D systems, where even a very accurate computation in a localized basis can take on the order of 1~second even for \ce{H16}.
    
\section{\label{sec:conclusions}CONCLUSIONS AND FUTURE WORK}

This work accomplishes two main goals.
Firstly, we propose a series of benchmark hydrogen models with a tunable degree of correlation that cover a wide range of electronic structures.
These include 1D hydrogen chains and rings with antiferromagnetic ground states, a 2D triangular lattice (sheet)  with spin frustrated interactions, and a 3D pyramid system that displays both spin frustration and a vanishing energy gap (dense manifold of near-degenerate states).
We analyze these systems with various correlation metrics and by computing their low-energy spectra and spin-spin correlation functions.
The models are found to have drastically different electronic structures depending on the physical dimension. 
In particular, since 2D and 3D systems exhibit some of the fingerprints of spin frustration and they are not efficiently approximated with MPS, they nicely complement benchmark sets based on 1D lattices.
Our comparison of different metrics of correlation also highlights the importance of using multiple descriptors to characterize electronic states, as our results clearly show that they measure different aspects of correlation.

Secondly, using the hydrogen models, we compare the performance of selected CI, SVD-FCI, and DMRG in various regimes of strong electron correlation.
We focus in particular on determining the ability of each method to efficiently compress the information content of the FCI wave function.
To quantify this property, we introduce a new metric, the accuracy volume ($\ncomp$), which corresponds to the minimum number of variational parameters necessary to achieve a target energy error (in our case, defined as 1m\Eh).
As expected, DMRG affords the most efficient representation for the 1D \ce{H10} chain and ring, using at least an order of magnitude fewer parameters to achieve the same level of energy or two-body cumulant accuracy compared to the other methods.
Nevertheless, this efficiency is gradually lost when going from 1D to higher-dimensional systems.
In contrast, all flavors of sCI perform best in a delocalized basis but are generally less efficient than DMRG.
The SVD-FCI, which we use as a proxy for rank-reduced FCI, is generally found to be the most inefficient approach to approximate the \ce{H10} wave functions.
However, as mentioned previously in Sec.~\ref{sec:SVD-FCI}, the variational optimization in RR-FCI would yield lower accuracy volumes than SVD-FCI, likely making RR-FCI more competitive with sCI and DMRG.
We have similarly analyzed the ability of each method to accurately represent electron distributions, namely the cumulant of the two-body density matrix.
In this case, the trends are similar to those observed for the energy, with the difference that sCI shows better performance for the 2D and 3D systems in a delocalized basis.

In analyzing the compressibility of the wave functions for \ce{H12}, \ce{H14}, and \ce{H16} analogs of the four \ce{H10} models, we have determined that DMRG consistently shows the smallest accuracy volume, and that the performance of SVD-FCI is more on par with that of ap-sCI for the larger systems, suggesting that future developments of RR-FCI methods such as those in Refs. \citenum{Koch1992AVariational} and \citenum{Fales2018LargeScale} are certainly worthwhile, especially for systems larger than those considered in this study. 
Despite the significant reduction in the number of parameters relative to the FCI wave function afforded by selected CI, SVD-FCI, and DMRG, none of these methods bring a reduction in scaling from exponential to polynomial in the general case.
Alternative methods, such as higher-dimensional tensor network states, quantum Monte Carlo, and quantum computational algorithms, may be required to circumvent storage cost of an exponentially scaling wave function.

We note that while the accuracy volume is a generally applicable metric for determining the performance of a method, the benchmark set considered here uses a minimal basis, is restricted to small systems amenable to FCI computations, and does not include atoms with more complex electronic structures.
Therefore, one should be cautious in extrapolating the relative performance of the methods in the case of more complex systems.
In future studies it might be also interesting to investigate the advantages of employing other orbitals bases, like natural orbitals and split-localized orbitals.
In addition, our work has focused only two tensor decomposition methods.
It would be interesting to examine the accuracy volume of projected entangled-pair states,\cite{verstraete2004renormalization} the multi-scale entanglement renormalization ansatz,\cite{vidal2007entanglement} tree tensor network states,\cite{Murg2010SimulatingStrongly, Nakatani2013EfficientTree, murg2015tree} and other more general tensor network states.
With appropriate modifications, the accuracy volume is also applicable to stochastic methods,\cite{needs2009continuum, motta2018ab} both in real and determinant space, and could provide a way to compare these approaches to deterministic methods like the ones considered in this work.

Our work does raise a few important questions as we (potentially) approach an era of quantum advantage for molecular computations.
Although quantum computational algorithms are able to avoid the explicit storage of the wave function, they still suffer from non-trivial classical computational overhead.
For example, the quantum phase estimation\cite{Abrams:1997ha,Abrams:1999ur}  (QPE) algorithm relies on time evolution of the Hamiltonian, which implies a computational scaling and storage costs (ignoring the cost of state preparation) at least proportional to $K^4$ in a delocalized basis, although more efficient representations have been recently proposed.\cite{babbush2018low,mcclean2019discontinuous}
For the purpose of comparing the resource cost of classical and quantum algorithms, in Fig.~\ref{fig:h10compression_n} we have also reported an estimate of the resources needed by quantum algorithms computed as $n^4$, where $n$ is the number of hydrogen atoms (equal to the number of spatial orbitals).
This plot shows that classical compression approaches use more than $n^4$ parameters even with systems as small as 12 electrons.
While this prefatory comparison highlights the importance of quantum algorithm development even for modestly sized systems, it also suggests a threshold for the maximum number of classical parameters a quantum algorithm should employ.
In other words, a successful quantum algorithm should achieve a $\ncomp$ smaller (and with lower $n$-scaling) than state-of-the art classical methods such as selected CI and DMRG for a given level of accuracy.
The competitiveness of any quantum algorithm could be tested for various regimes of correlation by comparing the computational resources (classical variational parameters) required to achieve a 1 m\Eh energy error with those reported in Table~\ref{tab:h10compression_r}.

In summary, this study has explored the limits of classical state-of-the-art electronic structure methodologies as applied to strongly correlated electrons.
The hydrogen benchmark set and the accuracy volume metric are two new tools that will be useful in guiding the development of the next generation of classical and hybrid quantum-classical methods for strongly correlated systems.
An important open problem in electronic structure theory is identifying the practical limits of classical methods and knowing under what circumstances quantum algorithms can overcome these limits.
This work approaches this problem from a computational perspective and sheds some light on the first aspect; in future work we plan to investigate the ability of various quantum algorithms to go beyond the limits of classical methods.

\section*{\label{sec:conclusions}SUPPLEMENTARY MATERIAL}
Ranges of threshold parameters used for each systematically improvable method, examples of the localized orbitals with corresponding DMRG ordering, spin-spin radial distribution function for each \ce{H10} model, and potential energy curves for the low-lying states of the \ce{H10} pyramid are included in the supplementary material.

\begin{acknowledgments}
The authors would like to thank Sebastian Wouters for helpful conversations regarding DMRG calculations, Garnet Chan for discussions on the scaling of DMRG, and Mario Motta for his insights on spin frustrated lattices. 
This work was supported by the U.S. Department of Energy under Award No. DE-SC0019374 and a Camille Dreyfus Teacher-Scholar Award (TC-18-045).
N.H.S. was supported by a fellowship from The Molecular Sciences Software Institute under NSF grant ACI-1547580.
\end{acknowledgments}

\section*{\label{sec:conclusions}DATA AVAILABILITY }
The geometries of the hydrogen models, energy errors at each value of $N_{\rm{par}}$ for ap-sCI, ACI, SVD-FCI and DMRG, and raw data for the potential energy curves of the \ce{H10} models using RHF, MP2, CCSD, CCSD(T), CR-CC(2,3), V2RDM-PQG, V2RDM-PQGT2, and FCI are openly available on GitHub, reference number~\citenum{HstudyRepo2020}.

\bibliography{bibliography,extra}

\begin{thebibliography}{162}%
\makeatletter
\providecommand \@ifxundefined [1]{%
 \@ifx{#1\undefined}
}%
\providecommand \@ifnum [1]{%
 \ifnum #1\expandafter \@firstoftwo
 \else \expandafter \@secondoftwo
 \fi
}%
\providecommand \@ifx [1]{%
 \ifx #1\expandafter \@firstoftwo
 \else \expandafter \@secondoftwo
 \fi
}%
\providecommand \natexlab [1]{#1}%
\providecommand \enquote  [1]{``#1''}%
\providecommand \bibnamefont  [1]{#1}%
\providecommand \bibfnamefont [1]{#1}%
\providecommand \citenamefont [1]{#1}%
\providecommand \href@noop [0]{\@secondoftwo}%
\providecommand \href [0]{\begingroup \@sanitize@url \@href}%
\providecommand \@href[1]{\@@startlink{#1}\@@href}%
\providecommand \@@href[1]{\endgroup#1\@@endlink}%
\providecommand \@sanitize@url [0]{\catcode `\\12\catcode `\$12\catcode
  `\&12\catcode `\#12\catcode `\^12\catcode `\_12\catcode `\%12\relax}%
\providecommand \@@startlink[1]{}%
\providecommand \@@endlink[0]{}%
\providecommand \url  [0]{\begingroup\@sanitize@url \@url }%
\providecommand \@url [1]{\endgroup\@href {#1}{\urlprefix }}%
\providecommand \urlprefix  [0]{URL }%
\providecommand \Eprint [0]{\href }%
\providecommand \doibase [0]{http://dx.doi.org/}%
\providecommand \selectlanguage [0]{\@gobble}%
\providecommand \bibinfo  [0]{\@secondoftwo}%
\providecommand \bibfield  [0]{\@secondoftwo}%
\providecommand \translation [1]{[#1]}%
\providecommand \BibitemOpen [0]{}%
\providecommand \bibitemStop [0]{}%
\providecommand \bibitemNoStop [0]{.\EOS\space}%
\providecommand \EOS [0]{\spacefactor3000\relax}%
\providecommand \BibitemShut  [1]{\csname bibitem#1\endcsname}%
\let\auto@bib@innerbib\@empty
\bibitem [{\citenamefont {Mok}, \citenamefont {Neumann},\ and\ \citenamefont
  {Handy}(1996)}]{Mok1996DynamicalAnd}%
  \BibitemOpen
  \bibfield  {author} {\bibinfo {author} {\bibfnamefont {D.~K.}\ \bibnamefont
  {Mok}}, \bibinfo {author} {\bibfnamefont {R.}~\bibnamefont {Neumann}}, \ and\
  \bibinfo {author} {\bibfnamefont {N.~C.}\ \bibnamefont {Handy}},\ }\href@noop
  {} {\bibfield  {journal} {\bibinfo  {journal} {J. Phys. Chem.}\ }\textbf
  {\bibinfo {volume} {100}},\ \bibinfo {pages} {6225} (\bibinfo {year}
  {1996})}\BibitemShut {NoStop}%
\bibitem [{\citenamefont {Roca-Sanju{\'a}n}, \citenamefont {Aquilante},\ and\
  \citenamefont {Lindh}(2012)}]{Roca2012MulticonfigurationSecond}%
  \BibitemOpen
  \bibfield  {author} {\bibinfo {author} {\bibfnamefont {D.}~\bibnamefont
  {Roca-Sanju{\'a}n}}, \bibinfo {author} {\bibfnamefont {F.}~\bibnamefont
  {Aquilante}}, \ and\ \bibinfo {author} {\bibfnamefont {R.}~\bibnamefont
  {Lindh}},\ }\href@noop {} {\bibfield  {journal} {\bibinfo  {journal} {Wiley
  Interdiscip. Rev. Comput. Mol. Sci.}\ }\textbf {\bibinfo {volume} {2}},\
  \bibinfo {pages} {585} (\bibinfo {year} {2012})}\BibitemShut {NoStop}%
\bibitem [{\citenamefont {Malrieu}\ \emph {et~al.}(2014)\citenamefont
  {Malrieu}, \citenamefont {Caballol}, \citenamefont {Calzado}, \citenamefont
  {de~Graaf},\ and\ \citenamefont {Guihery}}]{malrieu2014magnetic}%
  \BibitemOpen
  \bibfield  {author} {\bibinfo {author} {\bibfnamefont {J.~P.}\ \bibnamefont
  {Malrieu}}, \bibinfo {author} {\bibfnamefont {R.}~\bibnamefont {Caballol}},
  \bibinfo {author} {\bibfnamefont {C.~J.}\ \bibnamefont {Calzado}}, \bibinfo
  {author} {\bibfnamefont {C.}~\bibnamefont {de~Graaf}}, \ and\ \bibinfo
  {author} {\bibfnamefont {N.}~\bibnamefont {Guihery}},\ }\href@noop {}
  {\bibfield  {journal} {\bibinfo  {journal} {Chem. Rev.}\ }\textbf {\bibinfo
  {volume} {114}},\ \bibinfo {pages} {429} (\bibinfo {year}
  {2014})}\BibitemShut {NoStop}%
\bibitem [{\citenamefont {Lee}(2007)}]{Lee2007HighTemp}%
  \BibitemOpen
  \bibfield  {author} {\bibinfo {author} {\bibfnamefont {P.~A.}\ \bibnamefont
  {Lee}},\ }\href@noop {} {\bibfield  {journal} {\bibinfo  {journal} {Rep.
  Prog. Phys.}\ }\textbf {\bibinfo {volume} {71}},\ \bibinfo {pages} {012501}
  (\bibinfo {year} {2007})}\BibitemShut {NoStop}%
\bibitem [{\citenamefont {Imada}, \citenamefont {Fujimori},\ and\ \citenamefont
  {Tokura}(1998)}]{Imada1998MetalInsulator}%
  \BibitemOpen
  \bibfield  {author} {\bibinfo {author} {\bibfnamefont {M.}~\bibnamefont
  {Imada}}, \bibinfo {author} {\bibfnamefont {A.}~\bibnamefont {Fujimori}}, \
  and\ \bibinfo {author} {\bibfnamefont {Y.}~\bibnamefont {Tokura}},\
  }\href@noop {} {\bibfield  {journal} {\bibinfo  {journal} {Rev. Mod. Phys.}\
  }\textbf {\bibinfo {volume} {70}},\ \bibinfo {pages} {1039} (\bibinfo {year}
  {1998})}\BibitemShut {NoStop}%
\bibitem [{\citenamefont {Salamon}\ and\ \citenamefont
  {Jaime}(2001)}]{Salamon2001PhysicsOf}%
  \BibitemOpen
  \bibfield  {author} {\bibinfo {author} {\bibfnamefont {M.~B.}\ \bibnamefont
  {Salamon}}\ and\ \bibinfo {author} {\bibfnamefont {M.}~\bibnamefont
  {Jaime}},\ }\href@noop {} {\bibfield  {journal} {\bibinfo  {journal} {Rev.
  Mod. Phys.}\ }\textbf {\bibinfo {volume} {73}},\ \bibinfo {pages} {583}
  (\bibinfo {year} {2001})}\BibitemShut {NoStop}%
\bibitem [{\citenamefont {Tokura}(2006)}]{Tokura2006CriticalFeatures}%
  \BibitemOpen
  \bibfield  {author} {\bibinfo {author} {\bibfnamefont {Y.}~\bibnamefont
  {Tokura}},\ }\href@noop {} {\bibfield  {journal} {\bibinfo  {journal} {Rep.
  Prog. Phys.}\ }\textbf {\bibinfo {volume} {69}},\ \bibinfo {pages} {797}
  (\bibinfo {year} {2006})}\BibitemShut {NoStop}%
\bibitem [{\citenamefont {Murthy}\ and\ \citenamefont
  {Shankar}(2003)}]{murthy2003hamiltonian}%
  \BibitemOpen
  \bibfield  {author} {\bibinfo {author} {\bibfnamefont {G.}~\bibnamefont
  {Murthy}}\ and\ \bibinfo {author} {\bibfnamefont {R.}~\bibnamefont
  {Shankar}},\ }\href@noop {} {\bibfield  {journal} {\bibinfo  {journal} {Rev.
  Mod. Phys.}\ }\textbf {\bibinfo {volume} {75}},\ \bibinfo {pages} {1101}
  (\bibinfo {year} {2003})}\BibitemShut {NoStop}%
\bibitem [{\citenamefont {Tew}, \citenamefont {Klopper},\ and\ \citenamefont
  {Helgaker}(2007)}]{Tew2007ElectronCorrelation}%
  \BibitemOpen
  \bibfield  {author} {\bibinfo {author} {\bibfnamefont {D.~P.}\ \bibnamefont
  {Tew}}, \bibinfo {author} {\bibfnamefont {W.}~\bibnamefont {Klopper}}, \ and\
  \bibinfo {author} {\bibfnamefont {T.}~\bibnamefont {Helgaker}},\ }\href@noop
  {} {\bibfield  {journal} {\bibinfo  {journal} {J. Comput. Chem.}\ }\textbf
  {\bibinfo {volume} {28}},\ \bibinfo {pages} {1307} (\bibinfo {year}
  {2007})}\BibitemShut {NoStop}%
\bibitem [{\citenamefont {Kutzelnigg}(2003)}]{kutzelnigg2003theory}%
  \BibitemOpen
  \bibfield  {author} {\bibinfo {author} {\bibfnamefont {W.}~\bibnamefont
  {Kutzelnigg}},\ }in\ \href@noop {} {\emph {\bibinfo {booktitle} {Explicitly
  Correlated Wave Functions in Chemistry and Physics}}},\ \bibinfo {editor}
  {edited by\ \bibinfo {editor} {\bibfnamefont {J.}~\bibnamefont
  {Rychlewski}}}\ (\bibinfo  {publisher} {Springer},\ \bibinfo {year} {2003})\
  pp.\ \bibinfo {pages} {3--90}\BibitemShut {NoStop}%
\bibitem [{\citenamefont {Roos}, \citenamefont {Taylor},\ and\ \citenamefont
  {Siegbahn}(1980)}]{Roos:1980wj}%
  \BibitemOpen
  \bibfield  {author} {\bibinfo {author} {\bibfnamefont {B.~O.}\ \bibnamefont
  {Roos}}, \bibinfo {author} {\bibfnamefont {P.~R.}\ \bibnamefont {Taylor}}, \
  and\ \bibinfo {author} {\bibfnamefont {P.~E.~M.}\ \bibnamefont {Siegbahn}},\
  }\href@noop {} {\bibfield  {journal} {\bibinfo  {journal} {Chem. Phys.}\
  }\textbf {\bibinfo {volume} {48}},\ \bibinfo {pages} {157} (\bibinfo {year}
  {1980})}\BibitemShut {NoStop}%
\bibitem [{\citenamefont {Laughlin}\ and\ \citenamefont
  {Pines}(2000)}]{Laughlin2000TheTheory}%
  \BibitemOpen
  \bibfield  {author} {\bibinfo {author} {\bibfnamefont {R.}~\bibnamefont
  {Laughlin}}\ and\ \bibinfo {author} {\bibfnamefont {D.}~\bibnamefont
  {Pines}},\ }\href@noop {} {\bibfield  {journal} {\bibinfo  {journal} {Proc.
  Natl. Acad. Sci. U.S.A}\ }\textbf {\bibinfo {volume} {97}},\ \bibinfo {pages}
  {28} (\bibinfo {year} {2000})}\BibitemShut {NoStop}%
\bibitem [{\citenamefont {Vogiatzis}\ \emph {et~al.}(2017)\citenamefont
  {Vogiatzis}, \citenamefont {Ma}, \citenamefont {Olsen}, \citenamefont
  {Gagliardi},\ and\ \citenamefont {De~Jong}}]{vogiatzis2017pushing}%
  \BibitemOpen
  \bibfield  {author} {\bibinfo {author} {\bibfnamefont {K.~D.}\ \bibnamefont
  {Vogiatzis}}, \bibinfo {author} {\bibfnamefont {D.}~\bibnamefont {Ma}},
  \bibinfo {author} {\bibfnamefont {J.}~\bibnamefont {Olsen}}, \bibinfo
  {author} {\bibfnamefont {L.}~\bibnamefont {Gagliardi}}, \ and\ \bibinfo
  {author} {\bibfnamefont {W.~A.}\ \bibnamefont {De~Jong}},\ }\href@noop {}
  {\bibfield  {journal} {\bibinfo  {journal} {J. Chem. Phys.}\ }\textbf
  {\bibinfo {volume} {147}},\ \bibinfo {pages} {184111} (\bibinfo {year}
  {2017})}\BibitemShut {NoStop}%
\bibitem [{\citenamefont {McArdle}\ \emph {et~al.}(2020)\citenamefont
  {McArdle}, \citenamefont {Endo}, \citenamefont {Aspuru-Guzik}, \citenamefont
  {Benjamin},\ and\ \citenamefont {Yuan}}]{mcardle2020quantum}%
  \BibitemOpen
  \bibfield  {author} {\bibinfo {author} {\bibfnamefont {S.}~\bibnamefont
  {McArdle}}, \bibinfo {author} {\bibfnamefont {S.}~\bibnamefont {Endo}},
  \bibinfo {author} {\bibfnamefont {A.}~\bibnamefont {Aspuru-Guzik}}, \bibinfo
  {author} {\bibfnamefont {S.~C.}\ \bibnamefont {Benjamin}}, \ and\ \bibinfo
  {author} {\bibfnamefont {X.}~\bibnamefont {Yuan}},\ }\href@noop {} {\bibfield
   {journal} {\bibinfo  {journal} {Rev. Mod. Phys.}\ }\textbf {\bibinfo
  {volume} {92}},\ \bibinfo {pages} {015003} (\bibinfo {year}
  {2020})}\BibitemShut {NoStop}%
\bibitem [{\citenamefont {Peruzzo}\ \emph {et~al.}(2014)\citenamefont
  {Peruzzo}, \citenamefont {McClean}, \citenamefont {Shadbolt}, \citenamefont
  {Yung}, \citenamefont {Zhou}, \citenamefont {Love}, \citenamefont
  {Aspuru-Guzik},\ and\ \citenamefont {O'Brien}}]{Peruzzo:2014kca}%
  \BibitemOpen
  \bibfield  {author} {\bibinfo {author} {\bibfnamefont {A.}~\bibnamefont
  {Peruzzo}}, \bibinfo {author} {\bibfnamefont {J.}~\bibnamefont {McClean}},
  \bibinfo {author} {\bibfnamefont {P.}~\bibnamefont {Shadbolt}}, \bibinfo
  {author} {\bibfnamefont {M.-H.}\ \bibnamefont {Yung}}, \bibinfo {author}
  {\bibfnamefont {X.-Q.}\ \bibnamefont {Zhou}}, \bibinfo {author}
  {\bibfnamefont {P.~J.}\ \bibnamefont {Love}}, \bibinfo {author}
  {\bibfnamefont {A.}~\bibnamefont {Aspuru-Guzik}}, \ and\ \bibinfo {author}
  {\bibfnamefont {J.~L.}\ \bibnamefont {O'Brien}},\ }\href@noop {} {\bibfield
  {journal} {\bibinfo  {journal} {Nat. Commun.}\ }\textbf {\bibinfo {volume}
  {5}},\ \bibinfo {pages} {4213} (\bibinfo {year} {2014})}\BibitemShut
  {NoStop}%
\bibitem [{\citenamefont {Yung}\ \emph {et~al.}(2014)\citenamefont {Yung},
  \citenamefont {Casanova}, \citenamefont {Mezzacapo}, \citenamefont {Mcclean},
  \citenamefont {Lamata}, \citenamefont {Aspuru-Guzik},\ and\ \citenamefont
  {Solano}}]{yung2014transistor}%
  \BibitemOpen
  \bibfield  {author} {\bibinfo {author} {\bibfnamefont {M.-H.}\ \bibnamefont
  {Yung}}, \bibinfo {author} {\bibfnamefont {J.}~\bibnamefont {Casanova}},
  \bibinfo {author} {\bibfnamefont {A.}~\bibnamefont {Mezzacapo}}, \bibinfo
  {author} {\bibfnamefont {J.}~\bibnamefont {Mcclean}}, \bibinfo {author}
  {\bibfnamefont {L.}~\bibnamefont {Lamata}}, \bibinfo {author} {\bibfnamefont
  {A.}~\bibnamefont {Aspuru-Guzik}}, \ and\ \bibinfo {author} {\bibfnamefont
  {E.}~\bibnamefont {Solano}},\ }\href@noop {} {\bibfield  {journal} {\bibinfo
  {journal} {Sci. Rep.}\ }\textbf {\bibinfo {volume} {4}},\ \bibinfo {pages}
  {3589} (\bibinfo {year} {2014})}\BibitemShut {NoStop}%
\bibitem [{\citenamefont {McClean}\ \emph {et~al.}(2016)\citenamefont
  {McClean}, \citenamefont {Romero}, \citenamefont {Babbush},\ and\
  \citenamefont {Aspuru-Guzik}}]{McClean:2016bs}%
  \BibitemOpen
  \bibfield  {author} {\bibinfo {author} {\bibfnamefont {J.~R.}\ \bibnamefont
  {McClean}}, \bibinfo {author} {\bibfnamefont {J.}~\bibnamefont {Romero}},
  \bibinfo {author} {\bibfnamefont {R.}~\bibnamefont {Babbush}}, \ and\
  \bibinfo {author} {\bibfnamefont {A.}~\bibnamefont {Aspuru-Guzik}},\
  }\href@noop {} {\bibfield  {journal} {\bibinfo  {journal} {New J. Phys.}\
  }\textbf {\bibinfo {volume} {18}},\ \bibinfo {pages} {023023} (\bibinfo
  {year} {2016})}\BibitemShut {NoStop}%
\bibitem [{\citenamefont {Grimsley}\ \emph {et~al.}(2019)\citenamefont
  {Grimsley}, \citenamefont {Economou}, \citenamefont {Barnes},\ and\
  \citenamefont {Mayhall}}]{grimsley2019adaptive}%
  \BibitemOpen
  \bibfield  {author} {\bibinfo {author} {\bibfnamefont {H.~R.}\ \bibnamefont
  {Grimsley}}, \bibinfo {author} {\bibfnamefont {S.~E.}\ \bibnamefont
  {Economou}}, \bibinfo {author} {\bibfnamefont {E.}~\bibnamefont {Barnes}}, \
  and\ \bibinfo {author} {\bibfnamefont {N.~J.}\ \bibnamefont {Mayhall}},\
  }\href@noop {} {\bibfield  {journal} {\bibinfo  {journal} {Nat. Commun.}\
  }\textbf {\bibinfo {volume} {10}},\ \bibinfo {pages} {1} (\bibinfo {year}
  {2019})}\BibitemShut {NoStop}%
\bibitem [{\citenamefont {McClean}\ \emph {et~al.}(2017)\citenamefont
  {McClean}, \citenamefont {Kimchi-Schwartz}, \citenamefont {Carter},\ and\
  \citenamefont {de~Jong}}]{mcclean2017hybrid}%
  \BibitemOpen
  \bibfield  {author} {\bibinfo {author} {\bibfnamefont {J.~R.}\ \bibnamefont
  {McClean}}, \bibinfo {author} {\bibfnamefont {M.~E.}\ \bibnamefont
  {Kimchi-Schwartz}}, \bibinfo {author} {\bibfnamefont {J.}~\bibnamefont
  {Carter}}, \ and\ \bibinfo {author} {\bibfnamefont {W.~A.}\ \bibnamefont
  {de~Jong}},\ }\href@noop {} {\bibfield  {journal} {\bibinfo  {journal} {Phys.
  Rev. A}\ }\textbf {\bibinfo {volume} {95}},\ \bibinfo {pages} {042308}
  (\bibinfo {year} {2017})}\BibitemShut {NoStop}%
\bibitem [{\citenamefont {Motta}\ \emph
  {et~al.}(2019{\natexlab{a}})\citenamefont {Motta}, \citenamefont {Sun},
  \citenamefont {Tan}, \citenamefont {O'Rourke}, \citenamefont {Ye},
  \citenamefont {Minnich}, \citenamefont {Brand{\~a}o},\ and\ \citenamefont
  {Chan}}]{motta2019determining}%
  \BibitemOpen
  \bibfield  {author} {\bibinfo {author} {\bibfnamefont {M.}~\bibnamefont
  {Motta}}, \bibinfo {author} {\bibfnamefont {C.}~\bibnamefont {Sun}}, \bibinfo
  {author} {\bibfnamefont {A.~T.}\ \bibnamefont {Tan}}, \bibinfo {author}
  {\bibfnamefont {M.~J.}\ \bibnamefont {O'Rourke}}, \bibinfo {author}
  {\bibfnamefont {E.}~\bibnamefont {Ye}}, \bibinfo {author} {\bibfnamefont
  {A.~J.}\ \bibnamefont {Minnich}}, \bibinfo {author} {\bibfnamefont {F.~G.}\
  \bibnamefont {Brand{\~a}o}}, \ and\ \bibinfo {author} {\bibfnamefont
  {G.~K.-L.}\ \bibnamefont {Chan}},\ }\href@noop {} {\bibfield  {journal}
  {\bibinfo  {journal} {Nat. Phys.}\ }\textbf {\bibinfo {volume} {16}},\
  \bibinfo {pages} {1} (\bibinfo {year} {2019}{\natexlab{a}})}\BibitemShut
  {NoStop}%
\bibitem [{\citenamefont {Parrish}\ and\ \citenamefont
  {McMahon}(2019)}]{Parrish:2019tc}%
  \BibitemOpen
  \bibfield  {author} {\bibinfo {author} {\bibfnamefont {R.~M.}\ \bibnamefont
  {Parrish}}\ and\ \bibinfo {author} {\bibfnamefont {P.~L.}\ \bibnamefont
  {McMahon}},\ }\href@noop {} {\bibfield  {journal} {\bibinfo  {journal}
  {e-print arXiv:1909.08925 [quant-ph]}\ } (\bibinfo {year}
  {2019})}\BibitemShut {NoStop}%
\bibitem [{\citenamefont {Stair}, \citenamefont {Huang},\ and\ \citenamefont
  {Evangelista}(2020)}]{Stair_2020}%
  \BibitemOpen
  \bibfield  {author} {\bibinfo {author} {\bibfnamefont {N.~H.}\ \bibnamefont
  {Stair}}, \bibinfo {author} {\bibfnamefont {R.}~\bibnamefont {Huang}}, \ and\
  \bibinfo {author} {\bibfnamefont {F.~A.}\ \bibnamefont {Evangelista}},\
  }\href@noop {} {\bibfield  {journal} {\bibinfo  {journal} {J. Chem. Theory
  Comput.}\ }\textbf {\bibinfo {volume} {16}},\ \bibinfo {pages} {2236}
  (\bibinfo {year} {2020})}\BibitemShut {NoStop}%
\bibitem [{\citenamefont {Huron}, \citenamefont {Malrieu},\ and\ \citenamefont
  {Rancurel}(1973)}]{Huron1973IterativePerturbation}%
  \BibitemOpen
  \bibfield  {author} {\bibinfo {author} {\bibfnamefont {B.}~\bibnamefont
  {Huron}}, \bibinfo {author} {\bibfnamefont {J.-P.}\ \bibnamefont {Malrieu}},
  \ and\ \bibinfo {author} {\bibfnamefont {P.}~\bibnamefont {Rancurel}},\
  }\href {\doibase 10.1063/1.1679199} {\bibfield  {journal} {\bibinfo
  {journal} {J. Chem. Phys.}\ }\textbf {\bibinfo {volume} {58}},\ \bibinfo
  {pages} {5745} (\bibinfo {year} {1973})}\BibitemShut {NoStop}%
\bibitem [{\citenamefont {Buenker}\ and\ \citenamefont
  {Peyerimhoff}(1974)}]{Buenker1974IndividualizedConfiguration}%
  \BibitemOpen
  \bibfield  {author} {\bibinfo {author} {\bibfnamefont {R.~J.}\ \bibnamefont
  {Buenker}}\ and\ \bibinfo {author} {\bibfnamefont {S.~D.}\ \bibnamefont
  {Peyerimhoff}},\ }\href {\doibase 10.1007/PL00020553} {\bibfield  {journal}
  {\bibinfo  {journal} {Theor. Chim. Acta}\ }\textbf {\bibinfo {volume} {35}},\
  \bibinfo {pages} {33} (\bibinfo {year} {1974})}\BibitemShut {NoStop}%
\bibitem [{\citenamefont {Buenker}\ and\ \citenamefont
  {Peyerimhoff}(1975)}]{Buenker1975EnergyExtrapolation}%
  \BibitemOpen
  \bibfield  {author} {\bibinfo {author} {\bibfnamefont {R.~J.}\ \bibnamefont
  {Buenker}}\ and\ \bibinfo {author} {\bibfnamefont {S.~D.}\ \bibnamefont
  {Peyerimhoff}},\ }\href {\doibase 10.1007/BF00555301} {\bibfield  {journal}
  {\bibinfo  {journal} {Theor. Chim. Acta}\ }\textbf {\bibinfo {volume} {39}},\
  \bibinfo {pages} {217} (\bibinfo {year} {1975})}\BibitemShut {NoStop}%
\bibitem [{\citenamefont {Evangelisti}, \citenamefont {Daudey},\ and\
  \citenamefont {Malrieu}(1983)}]{Evangelisti1983ConvergenceOf}%
  \BibitemOpen
  \bibfield  {author} {\bibinfo {author} {\bibfnamefont {S.}~\bibnamefont
  {Evangelisti}}, \bibinfo {author} {\bibfnamefont {J.-P.}\ \bibnamefont
  {Daudey}}, \ and\ \bibinfo {author} {\bibfnamefont {J.-P.}\ \bibnamefont
  {Malrieu}},\ }\href {\doibase https://doi.org/10.1016/0301-0104(83)85011-3}
  {\bibfield  {journal} {\bibinfo  {journal} {Chem. Phys.}\ }\textbf {\bibinfo
  {volume} {75}},\ \bibinfo {pages} {91 } (\bibinfo {year} {1983})}\BibitemShut
  {NoStop}%
\bibitem [{\citenamefont {Koch}\ and\ \citenamefont
  {Dalgaard}(1992)}]{Koch1992AVariational}%
  \BibitemOpen
  \bibfield  {author} {\bibinfo {author} {\bibfnamefont {H.}~\bibnamefont
  {Koch}}\ and\ \bibinfo {author} {\bibfnamefont {E.}~\bibnamefont
  {Dalgaard}},\ }\href {\doibase https://doi.org/10.1016/0009-2614(92)90048-R}
  {\bibfield  {journal} {\bibinfo  {journal} {Chem. Phys. Lett.}\ }\textbf
  {\bibinfo {volume} {198}},\ \bibinfo {pages} {51 } (\bibinfo {year}
  {1992})}\BibitemShut {NoStop}%
\bibitem [{\citenamefont {Taylor}(2013)}]{Taylor2013LosslessCompression}%
  \BibitemOpen
  \bibfield  {author} {\bibinfo {author} {\bibfnamefont {P.~R.}\ \bibnamefont
  {Taylor}},\ }\href {\doibase 10.1063/1.4816769} {\bibfield  {journal}
  {\bibinfo  {journal} {J. Chem. Phys.}\ }\textbf {\bibinfo {volume} {139}},\
  \bibinfo {pages} {074113} (\bibinfo {year} {2013})}\BibitemShut {NoStop}%
\bibitem [{\citenamefont {Fales}\ \emph {et~al.}(2018)\citenamefont {Fales},
  \citenamefont {Seritan}, \citenamefont {Settje}, \citenamefont {Levine},
  \citenamefont {Koch},\ and\ \citenamefont
  {Mart{\'\i}nez}}]{Fales2018LargeScale}%
  \BibitemOpen
  \bibfield  {author} {\bibinfo {author} {\bibfnamefont {B.~S.}\ \bibnamefont
  {Fales}}, \bibinfo {author} {\bibfnamefont {S.}~\bibnamefont {Seritan}},
  \bibinfo {author} {\bibfnamefont {N.~F.}\ \bibnamefont {Settje}}, \bibinfo
  {author} {\bibfnamefont {B.~G.}\ \bibnamefont {Levine}}, \bibinfo {author}
  {\bibfnamefont {H.}~\bibnamefont {Koch}}, \ and\ \bibinfo {author}
  {\bibfnamefont {T.~J.}\ \bibnamefont {Mart{\'\i}nez}},\ }\href@noop {}
  {\bibfield  {journal} {\bibinfo  {journal} {J. Chem. Theory Comput.}\
  }\textbf {\bibinfo {volume} {14}},\ \bibinfo {pages} {4139} (\bibinfo {year}
  {2018})}\BibitemShut {NoStop}%
\bibitem [{\citenamefont {White}(1992)}]{White1992DensityMatrix}%
  \BibitemOpen
  \bibfield  {author} {\bibinfo {author} {\bibfnamefont {S.~R.}\ \bibnamefont
  {White}},\ }\href {\doibase 10.1103/PhysRevLett.69.2863} {\bibfield
  {journal} {\bibinfo  {journal} {Phys. Rev. Lett.}\ }\textbf {\bibinfo
  {volume} {69}},\ \bibinfo {pages} {2863} (\bibinfo {year}
  {1992})}\BibitemShut {NoStop}%
\bibitem [{\citenamefont {Garc{\'\i}a}\ \emph {et~al.}(1995)\citenamefont
  {Garc{\'\i}a}, \citenamefont {Castell}, \citenamefont {Caballol},\ and\
  \citenamefont {Malrieu}}]{Garcia1995AnIterative}%
  \BibitemOpen
  \bibfield  {author} {\bibinfo {author} {\bibfnamefont {V.}~\bibnamefont
  {Garc{\'\i}a}}, \bibinfo {author} {\bibfnamefont {O.}~\bibnamefont
  {Castell}}, \bibinfo {author} {\bibfnamefont {R.}~\bibnamefont {Caballol}}, \
  and\ \bibinfo {author} {\bibfnamefont {J.-P.}\ \bibnamefont {Malrieu}},\
  }\href {\doibase https://doi.org/10.1016/0009-2614(95)00438-A} {\bibfield
  {journal} {\bibinfo  {journal} {Chem. Phys. Lett.}\ }\textbf {\bibinfo
  {volume} {238}},\ \bibinfo {pages} {222 } (\bibinfo {year}
  {1995})}\BibitemShut {NoStop}%
\bibitem [{\citenamefont {Neese}(2003)}]{Neese2003ASpectroscopy}%
  \BibitemOpen
  \bibfield  {author} {\bibinfo {author} {\bibfnamefont {F.}~\bibnamefont
  {Neese}},\ }\href@noop {} {\bibfield  {journal} {\bibinfo  {journal} {J.
  Chem. Phys.}\ }\textbf {\bibinfo {volume} {119}},\ \bibinfo {pages} {9428}
  (\bibinfo {year} {2003})}\BibitemShut {NoStop}%
\bibitem [{\citenamefont {Nakatsuji}\ and\ \citenamefont
  {Ehara}(2005)}]{Nakatsuji2005IterativeCI}%
  \BibitemOpen
  \bibfield  {author} {\bibinfo {author} {\bibfnamefont {H.}~\bibnamefont
  {Nakatsuji}}\ and\ \bibinfo {author} {\bibfnamefont {M.}~\bibnamefont
  {Ehara}},\ }\href@noop {} {\bibfield  {journal} {\bibinfo  {journal} {J.
  Chem. Phys.}\ }\textbf {\bibinfo {volume} {122}},\ \bibinfo {pages} {194108}
  (\bibinfo {year} {2005})}\BibitemShut {NoStop}%
\bibitem [{\citenamefont {Abrams}\ and\ \citenamefont
  {Sherrill}(2005)}]{Abrams2005ImportantConfigurations}%
  \BibitemOpen
  \bibfield  {author} {\bibinfo {author} {\bibfnamefont {M.~L.}\ \bibnamefont
  {Abrams}}\ and\ \bibinfo {author} {\bibfnamefont {C.~D.}\ \bibnamefont
  {Sherrill}},\ }\href {\doibase https://doi.org/10.1016/j.cplett.2005.06.107}
  {\bibfield  {journal} {\bibinfo  {journal} {Chem. Phys. Lett.}\ }\textbf
  {\bibinfo {volume} {412}},\ \bibinfo {pages} {121 } (\bibinfo {year}
  {2005})}\BibitemShut {NoStop}%
\bibitem [{\citenamefont {Bytautas}\ and\ \citenamefont
  {Ruedenberg}(2009)}]{Bytautas2009APriori}%
  \BibitemOpen
  \bibfield  {author} {\bibinfo {author} {\bibfnamefont {L.}~\bibnamefont
  {Bytautas}}\ and\ \bibinfo {author} {\bibfnamefont {K.}~\bibnamefont
  {Ruedenberg}},\ }\href {\doibase
  https://doi.org/10.1016/j.chemphys.2008.11.021} {\bibfield  {journal}
  {\bibinfo  {journal} {Chem. Phys.}\ }\textbf {\bibinfo {volume} {356}},\
  \bibinfo {pages} {64 } (\bibinfo {year} {2009})}\BibitemShut {NoStop}%
\bibitem [{\citenamefont {Roth}(2009)}]{Roth2009ImportanceTruncation}%
  \BibitemOpen
  \bibfield  {author} {\bibinfo {author} {\bibfnamefont {R.}~\bibnamefont
  {Roth}},\ }\href {\doibase 10.1103/PhysRevC.79.064324} {\bibfield  {journal}
  {\bibinfo  {journal} {Phys. Rev. C}\ }\textbf {\bibinfo {volume} {79}},\
  \bibinfo {pages} {064324} (\bibinfo {year} {2009})}\BibitemShut {NoStop}%
\bibitem [{\citenamefont {Evangelista}(2014)}]{Evangelista2014Adaptive}%
  \BibitemOpen
  \bibfield  {author} {\bibinfo {author} {\bibfnamefont {F.~A.}\ \bibnamefont
  {Evangelista}},\ }\href@noop {} {\bibfield  {journal} {\bibinfo  {journal}
  {J. Chem. Phys.}\ }\textbf {\bibinfo {volume} {140}},\ \bibinfo {pages}
  {124114} (\bibinfo {year} {2014})}\BibitemShut {NoStop}%
\bibitem [{\citenamefont {Knowles}(2015)}]{Knowles2015CompressiveSampling}%
  \BibitemOpen
  \bibfield  {author} {\bibinfo {author} {\bibfnamefont {P.~J.}\ \bibnamefont
  {Knowles}},\ }\href@noop {} {\bibfield  {journal} {\bibinfo  {journal} {Mol.
  Phys.}\ }\textbf {\bibinfo {volume} {113}},\ \bibinfo {pages} {1655}
  (\bibinfo {year} {2015})}\BibitemShut {NoStop}%
\bibitem [{\citenamefont {Liu}\ and\ \citenamefont
  {Hoffmann}(2016)}]{Liu2016iCI}%
  \BibitemOpen
  \bibfield  {author} {\bibinfo {author} {\bibfnamefont {W.}~\bibnamefont
  {Liu}}\ and\ \bibinfo {author} {\bibfnamefont {M.~R.}\ \bibnamefont
  {Hoffmann}},\ }\href@noop {} {\bibfield  {journal} {\bibinfo  {journal} {J.
  Chem. Theory Comput.}\ }\textbf {\bibinfo {volume} {12}},\ \bibinfo {pages}
  {1169} (\bibinfo {year} {2016})}\BibitemShut {NoStop}%
\bibitem [{\citenamefont {Schriber}\ and\ \citenamefont
  {Evangelista}(2016)}]{Schriber2016Adaptive}%
  \BibitemOpen
  \bibfield  {author} {\bibinfo {author} {\bibfnamefont {J.~B.}\ \bibnamefont
  {Schriber}}\ and\ \bibinfo {author} {\bibfnamefont {F.~A.}\ \bibnamefont
  {Evangelista}},\ }\href@noop {} {\bibfield  {journal} {\bibinfo  {journal}
  {J. Chem. Phys.}\ }\textbf {\bibinfo {volume} {144}},\ \bibinfo {pages}
  {161106} (\bibinfo {year} {2016})}\BibitemShut {NoStop}%
\bibitem [{\citenamefont {Holmes}, \citenamefont {Tubman},\ and\ \citenamefont
  {Umrigar}(2016)}]{Holmes2016HeatBath}%
  \BibitemOpen
  \bibfield  {author} {\bibinfo {author} {\bibfnamefont {A.~A.}\ \bibnamefont
  {Holmes}}, \bibinfo {author} {\bibfnamefont {N.~M.}\ \bibnamefont {Tubman}},
  \ and\ \bibinfo {author} {\bibfnamefont {C.~J.}\ \bibnamefont {Umrigar}},\
  }\href@noop {} {\bibfield  {journal} {\bibinfo  {journal} {J. Chem. Theory
  Comput.}\ }\textbf {\bibinfo {volume} {12}},\ \bibinfo {pages} {3674}
  (\bibinfo {year} {2016})}\BibitemShut {NoStop}%
\bibitem [{\citenamefont {Schriber}\ and\ \citenamefont
  {Evangelista}(2017)}]{Schriber2017Adaptive}%
  \BibitemOpen
  \bibfield  {author} {\bibinfo {author} {\bibfnamefont {J.~B.}\ \bibnamefont
  {Schriber}}\ and\ \bibinfo {author} {\bibfnamefont {F.~A.}\ \bibnamefont
  {Evangelista}},\ }\href@noop {} {\bibfield  {journal} {\bibinfo  {journal}
  {J. Chem. Theory Comput.}\ }\textbf {\bibinfo {volume} {13}},\ \bibinfo
  {pages} {5354} (\bibinfo {year} {2017})}\BibitemShut {NoStop}%
\bibitem [{\citenamefont {Greer}(1995)}]{Greer1995Estimating}%
  \BibitemOpen
  \bibfield  {author} {\bibinfo {author} {\bibfnamefont {J.~C.}\ \bibnamefont
  {Greer}},\ }\href {\doibase 10.1063/1.469756} {\bibfield  {journal} {\bibinfo
   {journal} {J. Chem. Phys.}\ }\textbf {\bibinfo {volume} {103}},\ \bibinfo
  {pages} {1821} (\bibinfo {year} {1995})}\BibitemShut {NoStop}%
\bibitem [{\citenamefont {Greer}(1998)}]{Greer1998MonteCarlo}%
  \BibitemOpen
  \bibfield  {author} {\bibinfo {author} {\bibfnamefont {J.~C.}\ \bibnamefont
  {Greer}},\ }\href {\doibase https://doi.org/10.1006/jcph.1998.5953}
  {\bibfield  {journal} {\bibinfo  {journal} {J. Comput. Phys.}\ }\textbf
  {\bibinfo {volume} {146}},\ \bibinfo {pages} {181 } (\bibinfo {year}
  {1998})}\BibitemShut {NoStop}%
\bibitem [{\citenamefont {Coe}, \citenamefont {Murphy},\ and\ \citenamefont
  {Paterson}(2014)}]{Coe2014ApplyingMonte}%
  \BibitemOpen
  \bibfield  {author} {\bibinfo {author} {\bibfnamefont {J.}~\bibnamefont
  {Coe}}, \bibinfo {author} {\bibfnamefont {P.}~\bibnamefont {Murphy}}, \ and\
  \bibinfo {author} {\bibfnamefont {M.}~\bibnamefont {Paterson}},\ }\href
  {\doibase https://doi.org/10.1016/j.cplett.2014.04.050} {\bibfield  {journal}
  {\bibinfo  {journal} {Chem. Phys. Lett.}\ }\textbf {\bibinfo {volume}
  {604}},\ \bibinfo {pages} {46 } (\bibinfo {year} {2014})}\BibitemShut
  {NoStop}%
\bibitem [{\citenamefont {Coe}\ and\ \citenamefont
  {Paterson}(2013)}]{Coe2013StateAveraged}%
  \BibitemOpen
  \bibfield  {author} {\bibinfo {author} {\bibfnamefont {J.~P.}\ \bibnamefont
  {Coe}}\ and\ \bibinfo {author} {\bibfnamefont {M.~J.}\ \bibnamefont
  {Paterson}},\ }\href {\doibase 10.1063/1.4824888} {\bibfield  {journal}
  {\bibinfo  {journal} {J. Chem. Phys.}\ }\textbf {\bibinfo {volume} {139}},\
  \bibinfo {pages} {154103} (\bibinfo {year} {2013})}\BibitemShut {NoStop}%
\bibitem [{\citenamefont {Coe}\ and\ \citenamefont
  {Paterson}(2012)}]{Coe2012DevelopmentOf}%
  \BibitemOpen
  \bibfield  {author} {\bibinfo {author} {\bibfnamefont {J.~P.}\ \bibnamefont
  {Coe}}\ and\ \bibinfo {author} {\bibfnamefont {M.~J.}\ \bibnamefont
  {Paterson}},\ }\href {\doibase 10.1063/1.4767436} {\bibfield  {journal}
  {\bibinfo  {journal} {J. Chem. Phys.}\ }\textbf {\bibinfo {volume} {137}},\
  \bibinfo {pages} {204108} (\bibinfo {year} {2012})}\BibitemShut {NoStop}%
\bibitem [{\citenamefont {Gy{\H o}rffy}, \citenamefont {Bartlett},\ and\
  \citenamefont {Greer}(2008)}]{Gyorffy2008MonteCarlo}%
  \BibitemOpen
  \bibfield  {author} {\bibinfo {author} {\bibfnamefont {W.}~\bibnamefont
  {Gy{\H o}rffy}}, \bibinfo {author} {\bibfnamefont {R.~J.}\ \bibnamefont
  {Bartlett}}, \ and\ \bibinfo {author} {\bibfnamefont {J.~C.}\ \bibnamefont
  {Greer}},\ }\href {\doibase 10.1063/1.2965529} {\bibfield  {journal}
  {\bibinfo  {journal} {J. Chem. Phys.}\ }\textbf {\bibinfo {volume} {129}},\
  \bibinfo {pages} {064103} (\bibinfo {year} {2008})}\BibitemShut {NoStop}%
\bibitem [{\citenamefont {Sharma}\ \emph {et~al.}(2017)\citenamefont {Sharma},
  \citenamefont {Holmes}, \citenamefont {Jeanmairet}, \citenamefont {Alavi},\
  and\ \citenamefont {Umrigar}}]{Sharma2017SemistochasticHeat}%
  \BibitemOpen
  \bibfield  {author} {\bibinfo {author} {\bibfnamefont {S.}~\bibnamefont
  {Sharma}}, \bibinfo {author} {\bibfnamefont {A.~A.}\ \bibnamefont {Holmes}},
  \bibinfo {author} {\bibfnamefont {G.}~\bibnamefont {Jeanmairet}}, \bibinfo
  {author} {\bibfnamefont {A.}~\bibnamefont {Alavi}}, \ and\ \bibinfo {author}
  {\bibfnamefont {C.~J.}\ \bibnamefont {Umrigar}},\ }\href {\doibase
  10.1021/acs.jctc.6b01028} {\bibfield  {journal} {\bibinfo  {journal} {J.
  Chem. Theory Comput.}\ }\textbf {\bibinfo {volume} {13}},\ \bibinfo {pages}
  {1595} (\bibinfo {year} {2017})}\BibitemShut {NoStop}%
\bibitem [{\citenamefont {Holmes}, \citenamefont {Umrigar},\ and\ \citenamefont
  {Sharma}(2017)}]{Holmes2017ExcitedStates}%
  \BibitemOpen
  \bibfield  {author} {\bibinfo {author} {\bibfnamefont {A.~A.}\ \bibnamefont
  {Holmes}}, \bibinfo {author} {\bibfnamefont {C.~J.}\ \bibnamefont {Umrigar}},
  \ and\ \bibinfo {author} {\bibfnamefont {S.}~\bibnamefont {Sharma}},\
  }\href@noop {} {\bibfield  {journal} {\bibinfo  {journal} {J. Chem. Phys.}\
  }\textbf {\bibinfo {volume} {147}},\ \bibinfo {pages} {164111} (\bibinfo
  {year} {2017})}\BibitemShut {NoStop}%
\bibitem [{\citenamefont {Chien}\ \emph {et~al.}(2018)\citenamefont {Chien},
  \citenamefont {Holmes}, \citenamefont {Otten}, \citenamefont {Umrigar},
  \citenamefont {Sharma},\ and\ \citenamefont
  {Zimmerman}}]{Chien2018ExcitedStates}%
  \BibitemOpen
  \bibfield  {author} {\bibinfo {author} {\bibfnamefont {A.~D.}\ \bibnamefont
  {Chien}}, \bibinfo {author} {\bibfnamefont {A.~A.}\ \bibnamefont {Holmes}},
  \bibinfo {author} {\bibfnamefont {M.}~\bibnamefont {Otten}}, \bibinfo
  {author} {\bibfnamefont {C.~J.}\ \bibnamefont {Umrigar}}, \bibinfo {author}
  {\bibfnamefont {S.}~\bibnamefont {Sharma}}, \ and\ \bibinfo {author}
  {\bibfnamefont {P.~M.}\ \bibnamefont {Zimmerman}},\ }\href {\doibase
  10.1021/acs.jpca.8b01554} {\bibfield  {journal} {\bibinfo  {journal} {J.
  Phys. Chem. A}\ }\textbf {\bibinfo {volume} {122}},\ \bibinfo {pages} {2714}
  (\bibinfo {year} {2018})}\BibitemShut {NoStop}%
\bibitem [{\citenamefont {Li}\ \emph {et~al.}(2018)\citenamefont {Li},
  \citenamefont {Otten}, \citenamefont {Holmes}, \citenamefont {Sharma},\ and\
  \citenamefont {Umrigar}}]{li2018fast}%
  \BibitemOpen
  \bibfield  {author} {\bibinfo {author} {\bibfnamefont {J.}~\bibnamefont
  {Li}}, \bibinfo {author} {\bibfnamefont {M.}~\bibnamefont {Otten}}, \bibinfo
  {author} {\bibfnamefont {A.~A.}\ \bibnamefont {Holmes}}, \bibinfo {author}
  {\bibfnamefont {S.}~\bibnamefont {Sharma}}, \ and\ \bibinfo {author}
  {\bibfnamefont {C.~J.}\ \bibnamefont {Umrigar}},\ }\href@noop {} {\bibfield
  {journal} {\bibinfo  {journal} {J. Chem. Phys.}\ }\textbf {\bibinfo {volume}
  {149}},\ \bibinfo {pages} {214110} (\bibinfo {year} {2018})}\BibitemShut
  {NoStop}%
\bibitem [{\citenamefont {Booth}, \citenamefont {Thom},\ and\ \citenamefont
  {Alavi}(2009)}]{Booth2009FermionMonte}%
  \BibitemOpen
  \bibfield  {author} {\bibinfo {author} {\bibfnamefont {G.~H.}\ \bibnamefont
  {Booth}}, \bibinfo {author} {\bibfnamefont {A.~J.~W.}\ \bibnamefont {Thom}},
  \ and\ \bibinfo {author} {\bibfnamefont {A.}~\bibnamefont {Alavi}},\ }\href
  {\doibase 10.1063/1.3193710} {\bibfield  {journal} {\bibinfo  {journal} {J.
  Chem. Phys.}\ }\textbf {\bibinfo {volume} {131}},\ \bibinfo {pages} {054106}
  (\bibinfo {year} {2009})}\BibitemShut {NoStop}%
\bibitem [{\citenamefont {Cleland}, \citenamefont {Booth},\ and\ \citenamefont
  {Alavi}(2010)}]{Cleland2010SurvivalOf}%
  \BibitemOpen
  \bibfield  {author} {\bibinfo {author} {\bibfnamefont {D.}~\bibnamefont
  {Cleland}}, \bibinfo {author} {\bibfnamefont {G.~H.}\ \bibnamefont {Booth}},
  \ and\ \bibinfo {author} {\bibfnamefont {A.}~\bibnamefont {Alavi}},\ }\href
  {\doibase 10.1063/1.3302277} {\bibfield  {journal} {\bibinfo  {journal} {J.
  Chem. Phys.}\ }\textbf {\bibinfo {volume} {132}},\ \bibinfo {pages} {041103}
  (\bibinfo {year} {2010})}\BibitemShut {NoStop}%
\bibitem [{\citenamefont {Booth}\ \emph {et~al.}(2011)\citenamefont {Booth},
  \citenamefont {Cleland}, \citenamefont {Thom},\ and\ \citenamefont
  {Alavi}}]{Booth2011BreakingThe}%
  \BibitemOpen
  \bibfield  {author} {\bibinfo {author} {\bibfnamefont {G.~H.}\ \bibnamefont
  {Booth}}, \bibinfo {author} {\bibfnamefont {D.}~\bibnamefont {Cleland}},
  \bibinfo {author} {\bibfnamefont {A.~J.~W.}\ \bibnamefont {Thom}}, \ and\
  \bibinfo {author} {\bibfnamefont {A.}~\bibnamefont {Alavi}},\ }\href
  {\doibase 10.1063/1.3624383} {\bibfield  {journal} {\bibinfo  {journal} {J.
  Chem. Phys.}\ }\textbf {\bibinfo {volume} {135}},\ \bibinfo {pages} {084104}
  (\bibinfo {year} {2011})}\BibitemShut {NoStop}%
\bibitem [{\citenamefont {Cleland}, \citenamefont {Booth},\ and\ \citenamefont
  {Alavi}(2011)}]{Cleland2011AStudy}%
  \BibitemOpen
  \bibfield  {author} {\bibinfo {author} {\bibfnamefont {D.}~\bibnamefont
  {Cleland}}, \bibinfo {author} {\bibfnamefont {G.~H.}\ \bibnamefont {Booth}},
  \ and\ \bibinfo {author} {\bibfnamefont {A.}~\bibnamefont {Alavi}},\ }\href
  {\doibase 10.1063/1.3525712} {\bibfield  {journal} {\bibinfo  {journal} {J.
  Chem. Phys.}\ }\textbf {\bibinfo {volume} {134}},\ \bibinfo {pages} {024112}
  (\bibinfo {year} {2011})}\BibitemShut {NoStop}%
\bibitem [{\citenamefont {Cleland}\ \emph {et~al.}(2012)\citenamefont
  {Cleland}, \citenamefont {Booth}, \citenamefont {Overy},\ and\ \citenamefont
  {Alavi}}]{Cleland2012TamingThe}%
  \BibitemOpen
  \bibfield  {author} {\bibinfo {author} {\bibfnamefont {D.}~\bibnamefont
  {Cleland}}, \bibinfo {author} {\bibfnamefont {G.~H.}\ \bibnamefont {Booth}},
  \bibinfo {author} {\bibfnamefont {C.}~\bibnamefont {Overy}}, \ and\ \bibinfo
  {author} {\bibfnamefont {A.}~\bibnamefont {Alavi}},\ }\href {\doibase
  10.1021/ct300504f} {\bibfield  {journal} {\bibinfo  {journal} {J. Chem.
  Theory Comput.}\ }\textbf {\bibinfo {volume} {8}},\ \bibinfo {pages} {4138}
  (\bibinfo {year} {2012})}\BibitemShut {NoStop}%
\bibitem [{\citenamefont {Thomas}\ \emph {et~al.}(2014)\citenamefont {Thomas},
  \citenamefont {Overy}, \citenamefont {Booth},\ and\ \citenamefont
  {Alavi}}]{Thomas2014SymmetryBreaking}%
  \BibitemOpen
  \bibfield  {author} {\bibinfo {author} {\bibfnamefont {R.~E.}\ \bibnamefont
  {Thomas}}, \bibinfo {author} {\bibfnamefont {C.}~\bibnamefont {Overy}},
  \bibinfo {author} {\bibfnamefont {G.~H.}\ \bibnamefont {Booth}}, \ and\
  \bibinfo {author} {\bibfnamefont {A.}~\bibnamefont {Alavi}},\ }\href
  {\doibase 10.1021/ct400835u} {\bibfield  {journal} {\bibinfo  {journal} {J.
  Chem. Theory Comput.}\ }\textbf {\bibinfo {volume} {10}},\ \bibinfo {pages}
  {1915} (\bibinfo {year} {2014})}\BibitemShut {NoStop}%
\bibitem [{\citenamefont {Booth}, \citenamefont {Smart},\ and\ \citenamefont
  {Alavi}(2014)}]{Booth2014LinearScaling}%
  \BibitemOpen
  \bibfield  {author} {\bibinfo {author} {\bibfnamefont {G.~H.}\ \bibnamefont
  {Booth}}, \bibinfo {author} {\bibfnamefont {S.~D.}\ \bibnamefont {Smart}}, \
  and\ \bibinfo {author} {\bibfnamefont {A.}~\bibnamefont {Alavi}},\ }\href
  {\doibase 10.1080/00268976.2013.877165} {\bibfield  {journal} {\bibinfo
  {journal} {Mol. Phys.}\ }\textbf {\bibinfo {volume} {112}},\ \bibinfo {pages}
  {1855} (\bibinfo {year} {2014})}\BibitemShut {NoStop}%
\bibitem [{\citenamefont {Li~Manni}, \citenamefont {Smart},\ and\ \citenamefont
  {Alavi}(2016)}]{Li2016CombiningThe}%
  \BibitemOpen
  \bibfield  {author} {\bibinfo {author} {\bibfnamefont {G.}~\bibnamefont
  {Li~Manni}}, \bibinfo {author} {\bibfnamefont {S.~D.}\ \bibnamefont {Smart}},
  \ and\ \bibinfo {author} {\bibfnamefont {A.}~\bibnamefont {Alavi}},\
  }\href@noop {} {\bibfield  {journal} {\bibinfo  {journal} {J. Chem. Theory
  Comput.}\ }\textbf {\bibinfo {volume} {12}},\ \bibinfo {pages} {1245}
  (\bibinfo {year} {2016})}\BibitemShut {NoStop}%
\bibitem [{\citenamefont {Kinoshita}, \citenamefont {Hino},\ and\ \citenamefont
  {Bartlett}(2003)}]{Kinoshita2003SingularValue}%
  \BibitemOpen
  \bibfield  {author} {\bibinfo {author} {\bibfnamefont {T.}~\bibnamefont
  {Kinoshita}}, \bibinfo {author} {\bibfnamefont {O.}~\bibnamefont {Hino}}, \
  and\ \bibinfo {author} {\bibfnamefont {R.~J.}\ \bibnamefont {Bartlett}},\
  }\href {\doibase 10.1063/1.1609442} {\bibfield  {journal} {\bibinfo
  {journal} {J. Chem. Phys.}\ }\textbf {\bibinfo {volume} {119}},\ \bibinfo
  {pages} {7756} (\bibinfo {year} {2003})}\BibitemShut {NoStop}%
\bibitem [{\citenamefont {Hino}, \citenamefont {Kinoshita},\ and\ \citenamefont
  {Bartlett}(2004)}]{Hino2004SingularValue}%
  \BibitemOpen
  \bibfield  {author} {\bibinfo {author} {\bibfnamefont {O.}~\bibnamefont
  {Hino}}, \bibinfo {author} {\bibfnamefont {T.}~\bibnamefont {Kinoshita}}, \
  and\ \bibinfo {author} {\bibfnamefont {R.~J.}\ \bibnamefont {Bartlett}},\
  }\href {\doibase 10.1063/1.1763575} {\bibfield  {journal} {\bibinfo
  {journal} {J. Chem. Phys.}\ }\textbf {\bibinfo {volume} {121}},\ \bibinfo
  {pages} {1206} (\bibinfo {year} {2004})}\BibitemShut {NoStop}%
\bibitem [{\citenamefont {Lewis}, \citenamefont {Calvin},\ and\ \citenamefont
  {Valeev}(2016)}]{Lewis2016ClusteredLow}%
  \BibitemOpen
  \bibfield  {author} {\bibinfo {author} {\bibfnamefont {C.~A.}\ \bibnamefont
  {Lewis}}, \bibinfo {author} {\bibfnamefont {J.~A.}\ \bibnamefont {Calvin}}, \
  and\ \bibinfo {author} {\bibfnamefont {E.~F.}\ \bibnamefont {Valeev}},\
  }\href {\doibase 10.1021/acs.jctc.6b00884} {\bibfield  {journal} {\bibinfo
  {journal} {J. Chem. Theory Comput.}\ }\textbf {\bibinfo {volume} {12}},\
  \bibinfo {pages} {5868} (\bibinfo {year} {2016})}\BibitemShut {NoStop}%
\bibitem [{\citenamefont {L\"owdin}\ and\ \citenamefont
  {Shull}(1956)}]{Lowdin1956NaturalOrbitals}%
  \BibitemOpen
  \bibfield  {author} {\bibinfo {author} {\bibfnamefont {P.-O.}\ \bibnamefont
  {L\"owdin}}\ and\ \bibinfo {author} {\bibfnamefont {H.}~\bibnamefont
  {Shull}},\ }\href {\doibase 10.1103/PhysRev.101.1730} {\bibfield  {journal}
  {\bibinfo  {journal} {Phys. Rev.}\ }\textbf {\bibinfo {volume} {101}},\
  \bibinfo {pages} {1730} (\bibinfo {year} {1956})}\BibitemShut {NoStop}%
\bibitem [{\citenamefont {Bischoff}\ and\ \citenamefont
  {Valeev}(2011)}]{Bischoff2011LowOrder}%
  \BibitemOpen
  \bibfield  {author} {\bibinfo {author} {\bibfnamefont {F.~A.}\ \bibnamefont
  {Bischoff}}\ and\ \bibinfo {author} {\bibfnamefont {E.~F.}\ \bibnamefont
  {Valeev}},\ }\href {\doibase 10.1063/1.3560091} {\bibfield  {journal}
  {\bibinfo  {journal} {J. Chem. Phys.}\ }\textbf {\bibinfo {volume} {134}},\
  \bibinfo {pages} {104104} (\bibinfo {year} {2011})}\BibitemShut {NoStop}%
\bibitem [{\citenamefont {Malmqvist}\ and\ \citenamefont
  {Veryazov}(2012)}]{Malmqvist2012TheBinatural}%
  \BibitemOpen
  \bibfield  {author} {\bibinfo {author} {\bibfnamefont {P.~{\AA}.}\
  \bibnamefont {Malmqvist}}\ and\ \bibinfo {author} {\bibfnamefont
  {V.}~\bibnamefont {Veryazov}},\ }\href {\doibase
  10.1080/00268976.2012.697587} {\bibfield  {journal} {\bibinfo  {journal}
  {Mol. Phys.}\ }\textbf {\bibinfo {volume} {110}},\ \bibinfo {pages} {2455}
  (\bibinfo {year} {2012})}\BibitemShut {NoStop}%
\bibitem [{\citenamefont {Beran}\ and\ \citenamefont
  {Head-Gordon}(2004)}]{Beran2004ExtracrtingPair}%
  \BibitemOpen
  \bibfield  {author} {\bibinfo {author} {\bibfnamefont {G.~J.~O.}\
  \bibnamefont {Beran}}\ and\ \bibinfo {author} {\bibfnamefont
  {M.}~\bibnamefont {Head-Gordon}},\ }\href {\doibase 10.1063/1.1756860}
  {\bibfield  {journal} {\bibinfo  {journal} {J. Chem. Phys.}\ }\textbf
  {\bibinfo {volume} {121}},\ \bibinfo {pages} {78} (\bibinfo {year}
  {2004})}\BibitemShut {NoStop}%
\bibitem [{\citenamefont {Mayer}(2007)}]{Mayer2007UsingSingular}%
  \BibitemOpen
  \bibfield  {author} {\bibinfo {author} {\bibfnamefont {I.}~\bibnamefont
  {Mayer}},\ }\href {\doibase https://doi.org/10.1016/j.cplett.2007.02.038}
  {\bibfield  {journal} {\bibinfo  {journal} {Chem. Phys. Lett.}\ }\textbf
  {\bibinfo {volume} {437}},\ \bibinfo {pages} {284 } (\bibinfo {year}
  {2007})}\BibitemShut {NoStop}%
\bibitem [{\citenamefont {Knowles}\ and\ \citenamefont
  {Handy}(1984)}]{Knowles:1984ud}%
  \BibitemOpen
  \bibfield  {author} {\bibinfo {author} {\bibfnamefont {P.~J.}\ \bibnamefont
  {Knowles}}\ and\ \bibinfo {author} {\bibfnamefont {N.~C.}\ \bibnamefont
  {Handy}},\ }\href@noop {} {\bibfield  {journal} {\bibinfo  {journal} {Chem.
  Phys. Lett.}\ }\textbf {\bibinfo {volume} {111}},\ \bibinfo {pages} {315}
  (\bibinfo {year} {1984})}\BibitemShut {NoStop}%
\bibitem [{\citenamefont {Weinstein}, \citenamefont {Auerbach},\ and\
  \citenamefont {Chandra}(2011)}]{Weinstein2011ReducingMemory}%
  \BibitemOpen
  \bibfield  {author} {\bibinfo {author} {\bibfnamefont {M.}~\bibnamefont
  {Weinstein}}, \bibinfo {author} {\bibfnamefont {A.}~\bibnamefont {Auerbach}},
  \ and\ \bibinfo {author} {\bibfnamefont {V.~R.}\ \bibnamefont {Chandra}},\
  }\href {\doibase 10.1103/PhysRevE.84.056701} {\bibfield  {journal} {\bibinfo
  {journal} {Phys. Rev. E}\ }\textbf {\bibinfo {volume} {84}},\ \bibinfo
  {pages} {056701} (\bibinfo {year} {2011})}\BibitemShut {NoStop}%
\bibitem [{\citenamefont {{\"O}stlund}\ and\ \citenamefont
  {Rommer}(1995)}]{ostlund1995thermodynamic}%
  \BibitemOpen
  \bibfield  {author} {\bibinfo {author} {\bibfnamefont {S.}~\bibnamefont
  {{\"O}stlund}}\ and\ \bibinfo {author} {\bibfnamefont {S.}~\bibnamefont
  {Rommer}},\ }\href@noop {} {\bibfield  {journal} {\bibinfo  {journal} {Phys.
  Rev. Lett.}\ }\textbf {\bibinfo {volume} {75}},\ \bibinfo {pages} {3537}
  (\bibinfo {year} {1995})}\BibitemShut {NoStop}%
\bibitem [{\citenamefont {Dukelsky}\ \emph {et~al.}(1998)\citenamefont
  {Dukelsky}, \citenamefont {Mart{\'\i}n-Delgado}, \citenamefont {Nishino},\
  and\ \citenamefont {Sierra}}]{dukelsky1998equivalence}%
  \BibitemOpen
  \bibfield  {author} {\bibinfo {author} {\bibfnamefont {J.}~\bibnamefont
  {Dukelsky}}, \bibinfo {author} {\bibfnamefont {M.~A.}\ \bibnamefont
  {Mart{\'\i}n-Delgado}}, \bibinfo {author} {\bibfnamefont {T.}~\bibnamefont
  {Nishino}}, \ and\ \bibinfo {author} {\bibfnamefont {G.}~\bibnamefont
  {Sierra}},\ }\href@noop {} {\bibfield  {journal} {\bibinfo  {journal} {EPL}\
  }\textbf {\bibinfo {volume} {43}},\ \bibinfo {pages} {457} (\bibinfo {year}
  {1998})}\BibitemShut {NoStop}%
\bibitem [{\citenamefont {Schollw{\"o}ck}(2011)}]{schollwock2011density}%
  \BibitemOpen
  \bibfield  {author} {\bibinfo {author} {\bibfnamefont {U.}~\bibnamefont
  {Schollw{\"o}ck}},\ }\href@noop {} {\bibfield  {journal} {\bibinfo  {journal}
  {Ann. Phys. (N. Y.)}\ }\textbf {\bibinfo {volume} {326}},\ \bibinfo {pages}
  {96} (\bibinfo {year} {2011})}\BibitemShut {NoStop}%
\bibitem [{\citenamefont {Eisert}, \citenamefont {Cramer},\ and\ \citenamefont
  {Plenio}(2009)}]{Eisert2009AreaLaws}%
  \BibitemOpen
  \bibfield  {author} {\bibinfo {author} {\bibfnamefont {J.}~\bibnamefont
  {Eisert}}, \bibinfo {author} {\bibfnamefont {M.}~\bibnamefont {Cramer}}, \
  and\ \bibinfo {author} {\bibfnamefont {M.}~\bibnamefont {Plenio}},\
  }\href@noop {} {\bibfield  {journal} {\bibinfo  {journal} {Rev. Mod. Phys}\
  }\textbf {\bibinfo {volume} {20}},\ \bibinfo {pages} {30} (\bibinfo {year}
  {2009})}\BibitemShut {NoStop}%
\bibitem [{\citenamefont {Evenbly}\ and\ \citenamefont
  {Vidal}(2011)}]{Evenbly2011TensorNetwork}%
  \BibitemOpen
  \bibfield  {author} {\bibinfo {author} {\bibfnamefont {G.}~\bibnamefont
  {Evenbly}}\ and\ \bibinfo {author} {\bibfnamefont {G.}~\bibnamefont
  {Vidal}},\ }\href {\doibase 10.1007/s10955-011-0237-4} {\bibfield  {journal}
  {\bibinfo  {journal} {J. Stat. Phys.}\ }\textbf {\bibinfo {volume} {145}},\
  \bibinfo {pages} {891} (\bibinfo {year} {2011})}\BibitemShut {NoStop}%
\bibitem [{\citenamefont {Murg}\ \emph {et~al.}(2010)\citenamefont {Murg},
  \citenamefont {Verstraete}, \citenamefont {Legeza},\ and\ \citenamefont
  {Noack}}]{Murg2010SimulatingStrongly}%
  \BibitemOpen
  \bibfield  {author} {\bibinfo {author} {\bibfnamefont {V.}~\bibnamefont
  {Murg}}, \bibinfo {author} {\bibfnamefont {F.}~\bibnamefont {Verstraete}},
  \bibinfo {author} {\bibfnamefont {O.}~\bibnamefont {Legeza}}, \ and\ \bibinfo
  {author} {\bibfnamefont {R.~M.}\ \bibnamefont {Noack}},\ }\href {\doibase
  10.1103/PhysRevB.82.205105} {\bibfield  {journal} {\bibinfo  {journal} {Phys.
  Rev. B}\ }\textbf {\bibinfo {volume} {82}},\ \bibinfo {pages} {205105}
  (\bibinfo {year} {2010})}\BibitemShut {NoStop}%
\bibitem [{\citenamefont {Nakatani}\ and\ \citenamefont
  {Chan}(2013)}]{Nakatani2013EfficientTree}%
  \BibitemOpen
  \bibfield  {author} {\bibinfo {author} {\bibfnamefont {N.}~\bibnamefont
  {Nakatani}}\ and\ \bibinfo {author} {\bibfnamefont {G.~K.-L.}\ \bibnamefont
  {Chan}},\ }\href {\doibase 10.1063/1.4798639} {\bibfield  {journal} {\bibinfo
   {journal} {J. Chem. Phys.}\ }\textbf {\bibinfo {volume} {138}},\ \bibinfo
  {pages} {134113} (\bibinfo {year} {2013})}\BibitemShut {NoStop}%
\bibitem [{\citenamefont {Chan}\ and\ \citenamefont
  {Head-Gordon}(2002)}]{Chan2002HighlyCorrelated}%
  \BibitemOpen
  \bibfield  {author} {\bibinfo {author} {\bibfnamefont {G.~K.-L.}\
  \bibnamefont {Chan}}\ and\ \bibinfo {author} {\bibfnamefont {M.}~\bibnamefont
  {Head-Gordon}},\ }\href {\doibase 10.1063/1.1449459} {\bibfield  {journal}
  {\bibinfo  {journal} {J. Chem. Phys.}\ }\textbf {\bibinfo {volume} {116}},\
  \bibinfo {pages} {4462} (\bibinfo {year} {2002})}\BibitemShut {NoStop}%
\bibitem [{\citenamefont {Chan}\ and\ \citenamefont
  {Sharma}(2011)}]{chan2011density}%
  \BibitemOpen
  \bibfield  {author} {\bibinfo {author} {\bibfnamefont {G.~K.-L.}\
  \bibnamefont {Chan}}\ and\ \bibinfo {author} {\bibfnamefont {S.}~\bibnamefont
  {Sharma}},\ }\href@noop {} {\bibfield  {journal} {\bibinfo  {journal} {Annu.
  Rev. Phys. Chem.}\ }\textbf {\bibinfo {volume} {62}},\ \bibinfo {pages} {465}
  (\bibinfo {year} {2011})}\BibitemShut {NoStop}%
\bibitem [{\citenamefont {Moritz}, \citenamefont {Wolf},\ and\ \citenamefont
  {Reiher}(2005)}]{Moritz2005RelitivisticDMRG}%
  \BibitemOpen
  \bibfield  {author} {\bibinfo {author} {\bibfnamefont {G.}~\bibnamefont
  {Moritz}}, \bibinfo {author} {\bibfnamefont {A.}~\bibnamefont {Wolf}}, \ and\
  \bibinfo {author} {\bibfnamefont {M.}~\bibnamefont {Reiher}},\ }\href
  {\doibase 10.1063/1.2104447} {\bibfield  {journal} {\bibinfo  {journal} {J.
  Chem. Phys.}\ }\textbf {\bibinfo {volume} {123}},\ \bibinfo {pages} {184105}
  (\bibinfo {year} {2005})}\BibitemShut {NoStop}%
\bibitem [{\citenamefont {Kurashige}\ and\ \citenamefont
  {Yanai}(2009)}]{Kurashige2009HighPerformance}%
  \BibitemOpen
  \bibfield  {author} {\bibinfo {author} {\bibfnamefont {Y.}~\bibnamefont
  {Kurashige}}\ and\ \bibinfo {author} {\bibfnamefont {T.}~\bibnamefont
  {Yanai}},\ }\href {\doibase 10.1063/1.3152576} {\bibfield  {journal}
  {\bibinfo  {journal} {J. Chem. Phys.}\ }\textbf {\bibinfo {volume} {130}},\
  \bibinfo {pages} {234114} (\bibinfo {year} {2009})}\BibitemShut {NoStop}%
\bibitem [{\citenamefont {Olivares-Amaya}\ \emph {et~al.}(2015)\citenamefont
  {Olivares-Amaya}, \citenamefont {Hu}, \citenamefont {Nakatani}, \citenamefont
  {Sharma}, \citenamefont {Yang},\ and\ \citenamefont
  {Chan}}]{Olivares2015TheAbinitio}%
  \BibitemOpen
  \bibfield  {author} {\bibinfo {author} {\bibfnamefont {R.}~\bibnamefont
  {Olivares-Amaya}}, \bibinfo {author} {\bibfnamefont {W.}~\bibnamefont {Hu}},
  \bibinfo {author} {\bibfnamefont {N.}~\bibnamefont {Nakatani}}, \bibinfo
  {author} {\bibfnamefont {S.}~\bibnamefont {Sharma}}, \bibinfo {author}
  {\bibfnamefont {J.}~\bibnamefont {Yang}}, \ and\ \bibinfo {author}
  {\bibfnamefont {G.~K.-L.}\ \bibnamefont {Chan}},\ }\href {\doibase
  10.1063/1.4905329} {\bibfield  {journal} {\bibinfo  {journal} {J. Chem.
  Phys.}\ }\textbf {\bibinfo {volume} {142}},\ \bibinfo {pages} {034102}
  (\bibinfo {year} {2015})}\BibitemShut {NoStop}%
\bibitem [{\citenamefont {Hachmann}, \citenamefont {Cardoen},\ and\
  \citenamefont {Chan}(2006)}]{Hachmann2006MultireferenceCorrelation}%
  \BibitemOpen
  \bibfield  {author} {\bibinfo {author} {\bibfnamefont {J.}~\bibnamefont
  {Hachmann}}, \bibinfo {author} {\bibfnamefont {W.}~\bibnamefont {Cardoen}}, \
  and\ \bibinfo {author} {\bibfnamefont {G.~K.-L.}\ \bibnamefont {Chan}},\
  }\href@noop {} {\bibfield  {journal} {\bibinfo  {journal} {J. Chem. Phys.}\
  }\textbf {\bibinfo {volume} {125}},\ \bibinfo {pages} {144101} (\bibinfo
  {year} {2006})}\BibitemShut {NoStop}%
\bibitem [{\citenamefont {Mitrushchenkov}\ \emph {et~al.}(2012)\citenamefont
  {Mitrushchenkov}, \citenamefont {Fano}, \citenamefont {Linguerri},\ and\
  \citenamefont {Palmieri}}]{Mitrushchenkov2012OnTheImportance}%
  \BibitemOpen
  \bibfield  {author} {\bibinfo {author} {\bibfnamefont {A.~O.}\ \bibnamefont
  {Mitrushchenkov}}, \bibinfo {author} {\bibfnamefont {G.}~\bibnamefont
  {Fano}}, \bibinfo {author} {\bibfnamefont {R.}~\bibnamefont {Linguerri}}, \
  and\ \bibinfo {author} {\bibfnamefont {P.}~\bibnamefont {Palmieri}},\
  }\href@noop {} {\bibfield  {journal} {\bibinfo  {journal} {Int. J. Quantum
  Chem.}\ }\textbf {\bibinfo {volume} {112}},\ \bibinfo {pages} {1606}
  (\bibinfo {year} {2012})}\BibitemShut {NoStop}%
\bibitem [{\citenamefont {Wouters}\ \emph {et~al.}(2012)\citenamefont
  {Wouters}, \citenamefont {Limacher}, \citenamefont {Van~Neck},\ and\
  \citenamefont {Ayers}}]{Wouters2012LongitudinalStatic}%
  \BibitemOpen
  \bibfield  {author} {\bibinfo {author} {\bibfnamefont {S.}~\bibnamefont
  {Wouters}}, \bibinfo {author} {\bibfnamefont {P.~A.}\ \bibnamefont
  {Limacher}}, \bibinfo {author} {\bibfnamefont {D.}~\bibnamefont {Van~Neck}},
  \ and\ \bibinfo {author} {\bibfnamefont {P.~W.}\ \bibnamefont {Ayers}},\
  }\href@noop {} {\bibfield  {journal} {\bibinfo  {journal} {J. Chem. Phys.}\
  }\textbf {\bibinfo {volume} {136}},\ \bibinfo {pages} {134110} (\bibinfo
  {year} {2012})}\BibitemShut {NoStop}%
\bibitem [{\citenamefont {Ma}\ and\ \citenamefont
  {Ma}(2013)}]{Ma2013AssessmentOf}%
  \BibitemOpen
  \bibfield  {author} {\bibinfo {author} {\bibfnamefont {Y.}~\bibnamefont
  {Ma}}\ and\ \bibinfo {author} {\bibfnamefont {H.}~\bibnamefont {Ma}},\
  }\href@noop {} {\bibfield  {journal} {\bibinfo  {journal} {J. Chem. Phys.}\
  }\textbf {\bibinfo {volume} {138}},\ \bibinfo {pages} {224105} (\bibinfo
  {year} {2013})}\BibitemShut {NoStop}%
\bibitem [{\citenamefont {Hachmann}\ \emph {et~al.}(2007)\citenamefont
  {Hachmann}, \citenamefont {Dorando}, \citenamefont {Avil{\'e}s},\ and\
  \citenamefont {Chan}}]{Hachmann2007TheRadical}%
  \BibitemOpen
  \bibfield  {author} {\bibinfo {author} {\bibfnamefont {J.}~\bibnamefont
  {Hachmann}}, \bibinfo {author} {\bibfnamefont {J.~J.}\ \bibnamefont
  {Dorando}}, \bibinfo {author} {\bibfnamefont {M.}~\bibnamefont {Avil{\'e}s}},
  \ and\ \bibinfo {author} {\bibfnamefont {G.~K.-L.}\ \bibnamefont {Chan}},\
  }\href {\doibase 10.1063/1.2768362} {\bibfield  {journal} {\bibinfo
  {journal} {J. Chem. Phys.}\ }\textbf {\bibinfo {volume} {127}},\ \bibinfo
  {pages} {134309} (\bibinfo {year} {2007})}\BibitemShut {NoStop}%
\bibitem [{\citenamefont {Raghu}, \citenamefont {Pati},\ and\ \citenamefont
  {Ramasesha}(2002)}]{Raghu2002StructuralAnd}%
  \BibitemOpen
  \bibfield  {author} {\bibinfo {author} {\bibfnamefont {C.}~\bibnamefont
  {Raghu}}, \bibinfo {author} {\bibfnamefont {Y.~A.}\ \bibnamefont {Pati}}, \
  and\ \bibinfo {author} {\bibfnamefont {S.}~\bibnamefont {Ramasesha}},\ }\href
  {\doibase 10.1103/PhysRevB.65.155204} {\bibfield  {journal} {\bibinfo
  {journal} {Phys. Rev. B}\ }\textbf {\bibinfo {volume} {65}},\ \bibinfo
  {pages} {155204} (\bibinfo {year} {2002})}\BibitemShut {NoStop}%
\bibitem [{\citenamefont {Raghu}, \citenamefont {Anusooya~Pati},\ and\
  \citenamefont {Ramasesha}(2002)}]{Raghu2002DensityMatrix}%
  \BibitemOpen
  \bibfield  {author} {\bibinfo {author} {\bibfnamefont {C.}~\bibnamefont
  {Raghu}}, \bibinfo {author} {\bibfnamefont {Y.}~\bibnamefont
  {Anusooya~Pati}}, \ and\ \bibinfo {author} {\bibfnamefont {S.}~\bibnamefont
  {Ramasesha}},\ }\href {\doibase 10.1103/PhysRevB.66.035116} {\bibfield
  {journal} {\bibinfo  {journal} {Phys. Rev. B}\ }\textbf {\bibinfo {volume}
  {66}},\ \bibinfo {pages} {035116} (\bibinfo {year} {2002})}\BibitemShut
  {NoStop}%
\bibitem [{\citenamefont {Kurashige}, \citenamefont {Chan},\ and\ \citenamefont
  {Yanai}(2013)}]{Kurashige2013EntangledQuantum}%
  \BibitemOpen
  \bibfield  {author} {\bibinfo {author} {\bibfnamefont {Y.}~\bibnamefont
  {Kurashige}}, \bibinfo {author} {\bibfnamefont {G.~K.-L.}\ \bibnamefont
  {Chan}}, \ and\ \bibinfo {author} {\bibfnamefont {T.}~\bibnamefont {Yanai}},\
  }\href {http://dx.doi.org/10.1038/nchem.1677} {\bibfield  {journal} {\bibinfo
   {journal} {Nat. Chem.}\ }\textbf {\bibinfo {volume} {5}},\ \bibinfo {pages}
  {660 } (\bibinfo {year} {2013})}\BibitemShut {NoStop}%
\bibitem [{\citenamefont {Sharma}\ \emph {et~al.}(2014)\citenamefont {Sharma},
  \citenamefont {Sivalingam}, \citenamefont {Neese},\ and\ \citenamefont
  {Chan}}]{Sharma2014LowEnergy}%
  \BibitemOpen
  \bibfield  {author} {\bibinfo {author} {\bibfnamefont {S.}~\bibnamefont
  {Sharma}}, \bibinfo {author} {\bibfnamefont {K.}~\bibnamefont {Sivalingam}},
  \bibinfo {author} {\bibfnamefont {F.}~\bibnamefont {Neese}}, \ and\ \bibinfo
  {author} {\bibfnamefont {G.~K.-L.}\ \bibnamefont {Chan}},\ }\href
  {http://dx.doi.org/10.1038/nchem.2041} {\bibfield  {journal} {\bibinfo
  {journal} {Nat. Chem.}\ }\textbf {\bibinfo {volume} {6}},\ \bibinfo {pages}
  {927} (\bibinfo {year} {2014})}\BibitemShut {NoStop}%
\bibitem [{\citenamefont {Sinitskiy}, \citenamefont {Greenman},\ and\
  \citenamefont {Mazziotti}(2010)}]{Sinitskiy2010StrongCorrelation}%
  \BibitemOpen
  \bibfield  {author} {\bibinfo {author} {\bibfnamefont {A.~V.}\ \bibnamefont
  {Sinitskiy}}, \bibinfo {author} {\bibfnamefont {L.}~\bibnamefont {Greenman}},
  \ and\ \bibinfo {author} {\bibfnamefont {D.~A.}\ \bibnamefont {Mazziotti}},\
  }\href@noop {} {\bibfield  {journal} {\bibinfo  {journal} {J. Chem. Phys.}\
  }\textbf {\bibinfo {volume} {133}},\ \bibinfo {pages} {014104} (\bibinfo
  {year} {2010})}\BibitemShut {NoStop}%
\bibitem [{\citenamefont {Motta}\ \emph {et~al.}(2017)\citenamefont {Motta},
  \citenamefont {Ceperley}, \citenamefont {Chan}, \citenamefont {Gomez},
  \citenamefont {Gull}, \citenamefont {Guo}, \citenamefont {Jim{\'e}nez-Hoyos},
  \citenamefont {Lan}, \citenamefont {Li}, \citenamefont {Ma} \emph
  {et~al.}}]{Motta2017TowardsThe}%
  \BibitemOpen
  \bibfield  {author} {\bibinfo {author} {\bibfnamefont {M.}~\bibnamefont
  {Motta}}, \bibinfo {author} {\bibfnamefont {D.~M.}\ \bibnamefont {Ceperley}},
  \bibinfo {author} {\bibfnamefont {G.~K.-L.}\ \bibnamefont {Chan}}, \bibinfo
  {author} {\bibfnamefont {J.~A.}\ \bibnamefont {Gomez}}, \bibinfo {author}
  {\bibfnamefont {E.}~\bibnamefont {Gull}}, \bibinfo {author} {\bibfnamefont
  {S.}~\bibnamefont {Guo}}, \bibinfo {author} {\bibfnamefont {C.~A.}\
  \bibnamefont {Jim{\'e}nez-Hoyos}}, \bibinfo {author} {\bibfnamefont {T.~N.}\
  \bibnamefont {Lan}}, \bibinfo {author} {\bibfnamefont {J.}~\bibnamefont
  {Li}}, \bibinfo {author} {\bibfnamefont {F.}~\bibnamefont {Ma}},  \emph
  {et~al.},\ }\href@noop {} {\bibfield  {journal} {\bibinfo  {journal} {Phys.
  Rev. X}\ }\textbf {\bibinfo {volume} {7}},\ \bibinfo {pages} {031059}
  (\bibinfo {year} {2017})}\BibitemShut {NoStop}%
\bibitem [{\citenamefont {Motta}\ \emph
  {et~al.}(2019{\natexlab{b}})\citenamefont {Motta}, \citenamefont {Genovese},
  \citenamefont {Ma}, \citenamefont {Cui}, \citenamefont {Sawaya},
  \citenamefont {Chan}, \citenamefont {Chepiga}, \citenamefont {Helms},
  \citenamefont {Jimenez-Hoyos}, \citenamefont {Millis} \emph
  {et~al.}}]{motta2019ground}%
  \BibitemOpen
  \bibfield  {author} {\bibinfo {author} {\bibfnamefont {M.}~\bibnamefont
  {Motta}}, \bibinfo {author} {\bibfnamefont {C.}~\bibnamefont {Genovese}},
  \bibinfo {author} {\bibfnamefont {F.}~\bibnamefont {Ma}}, \bibinfo {author}
  {\bibfnamefont {Z.-H.}\ \bibnamefont {Cui}}, \bibinfo {author} {\bibfnamefont
  {R.}~\bibnamefont {Sawaya}}, \bibinfo {author} {\bibfnamefont {G.~K.}\
  \bibnamefont {Chan}}, \bibinfo {author} {\bibfnamefont {N.}~\bibnamefont
  {Chepiga}}, \bibinfo {author} {\bibfnamefont {P.}~\bibnamefont {Helms}},
  \bibinfo {author} {\bibfnamefont {C.}~\bibnamefont {Jimenez-Hoyos}}, \bibinfo
  {author} {\bibfnamefont {A.~J.}\ \bibnamefont {Millis}},  \emph {et~al.},\
  }\href@noop {} {\bibfield  {journal} {\bibinfo  {journal} {e-print
  arXiv:1911.01618 [quant-ph]}\ } (\bibinfo {year}
  {2019}{\natexlab{b}})}\BibitemShut {NoStop}%
\bibitem [{\citenamefont {Yanai}\ \emph
  {et~al.}(2010{\natexlab{a}})\citenamefont {Yanai}, \citenamefont {Kurashige},
  \citenamefont {Neuscamman},\ and\ \citenamefont {Chan}}]{Yanai:2010kf}%
  \BibitemOpen
  \bibfield  {author} {\bibinfo {author} {\bibfnamefont {T.}~\bibnamefont
  {Yanai}}, \bibinfo {author} {\bibfnamefont {Y.}~\bibnamefont {Kurashige}},
  \bibinfo {author} {\bibfnamefont {E.}~\bibnamefont {Neuscamman}}, \ and\
  \bibinfo {author} {\bibfnamefont {G.~K.-L.}\ \bibnamefont {Chan}},\
  }\href@noop {} {\bibfield  {journal} {\bibinfo  {journal} {J. Chem. Phys.}\
  }\textbf {\bibinfo {volume} {132}},\ \bibinfo {pages} {024105} (\bibinfo
  {year} {2010}{\natexlab{a}})}\BibitemShut {NoStop}%
\bibitem [{\citenamefont {Saitow}, \citenamefont {Kurashige},\ and\
  \citenamefont {Yanai}(2013)}]{Saitow:2013ij}%
  \BibitemOpen
  \bibfield  {author} {\bibinfo {author} {\bibfnamefont {M.}~\bibnamefont
  {Saitow}}, \bibinfo {author} {\bibfnamefont {Y.}~\bibnamefont {Kurashige}}, \
  and\ \bibinfo {author} {\bibfnamefont {T.}~\bibnamefont {Yanai}},\
  }\href@noop {} {\bibfield  {journal} {\bibinfo  {journal} {J. Chem. Phys.}\
  }\textbf {\bibinfo {volume} {139}},\ \bibinfo {pages} {044118} (\bibinfo
  {year} {2013})}\BibitemShut {NoStop}%
\bibitem [{\citenamefont {Kurashige}\ \emph {et~al.}(2014)\citenamefont
  {Kurashige}, \citenamefont {Chalupsk{\'{y}}}, \citenamefont {Lan},\ and\
  \citenamefont {Yanai}}]{Kurashige:2014bq}%
  \BibitemOpen
  \bibfield  {author} {\bibinfo {author} {\bibfnamefont {Y.}~\bibnamefont
  {Kurashige}}, \bibinfo {author} {\bibfnamefont {J.}~\bibnamefont
  {Chalupsk{\'{y}}}}, \bibinfo {author} {\bibfnamefont {T.~N.}\ \bibnamefont
  {Lan}}, \ and\ \bibinfo {author} {\bibfnamefont {T.}~\bibnamefont {Yanai}},\
  }\href@noop {} {\bibfield  {journal} {\bibinfo  {journal} {J. Chem. Phys.}\
  }\textbf {\bibinfo {volume} {141}},\ \bibinfo {pages} {174111} (\bibinfo
  {year} {2014})}\BibitemShut {NoStop}%
\bibitem [{\citenamefont {Guo}\ \emph {et~al.}(2016)\citenamefont {Guo},
  \citenamefont {Watson}, \citenamefont {Hu}, \citenamefont {Sun},\ and\
  \citenamefont {Chan}}]{Guo:2016fu}%
  \BibitemOpen
  \bibfield  {author} {\bibinfo {author} {\bibfnamefont {S.}~\bibnamefont
  {Guo}}, \bibinfo {author} {\bibfnamefont {M.~A.}\ \bibnamefont {Watson}},
  \bibinfo {author} {\bibfnamefont {W.}~\bibnamefont {Hu}}, \bibinfo {author}
  {\bibfnamefont {Q.}~\bibnamefont {Sun}}, \ and\ \bibinfo {author}
  {\bibfnamefont {G.~K.-L.}\ \bibnamefont {Chan}},\ }\href@noop {} {\bibfield
  {journal} {\bibinfo  {journal} {J. Chem. Theory Comput.}\ }\textbf {\bibinfo
  {volume} {12}},\ \bibinfo {pages} {1583} (\bibinfo {year}
  {2016})}\BibitemShut {NoStop}%
\bibitem [{\citenamefont {Wouters}, \citenamefont {Van~Speybroeck},\ and\
  \citenamefont {Van~Neck}(2016)}]{Wouters:2016fb}%
  \BibitemOpen
  \bibfield  {author} {\bibinfo {author} {\bibfnamefont {S.}~\bibnamefont
  {Wouters}}, \bibinfo {author} {\bibfnamefont {V.}~\bibnamefont
  {Van~Speybroeck}}, \ and\ \bibinfo {author} {\bibfnamefont {D.}~\bibnamefont
  {Van~Neck}},\ }\href@noop {} {\bibfield  {journal} {\bibinfo  {journal} {J.
  Chem. Phys.}\ }\textbf {\bibinfo {volume} {145}},\ \bibinfo {pages} {054120}
  (\bibinfo {year} {2016})}\BibitemShut {NoStop}%
\bibitem [{\citenamefont {Schriber}\ \emph {et~al.}(2018)\citenamefont
  {Schriber}, \citenamefont {Hannon}, \citenamefont {Li},\ and\ \citenamefont
  {Evangelista}}]{schriber2018combined}%
  \BibitemOpen
  \bibfield  {author} {\bibinfo {author} {\bibfnamefont {J.~B.}\ \bibnamefont
  {Schriber}}, \bibinfo {author} {\bibfnamefont {K.~P.}\ \bibnamefont
  {Hannon}}, \bibinfo {author} {\bibfnamefont {C.}~\bibnamefont {Li}}, \ and\
  \bibinfo {author} {\bibfnamefont {F.~A.}\ \bibnamefont {Evangelista}},\
  }\href@noop {} {\bibfield  {journal} {\bibinfo  {journal} {J. Chem. Theory
  Comput.}\ }\textbf {\bibinfo {volume} {14}},\ \bibinfo {pages} {6295}
  (\bibinfo {year} {2018})}\BibitemShut {NoStop}%
\bibitem [{\citenamefont {Freed}(1974)}]{Freed:1974vo}%
  \BibitemOpen
  \bibfield  {author} {\bibinfo {author} {\bibfnamefont {K.~F.}\ \bibnamefont
  {Freed}},\ }\href@noop {} {\bibfield  {journal} {\bibinfo  {journal} {J.
  Chem. Phys.}\ }\textbf {\bibinfo {volume} {60}},\ \bibinfo {pages} {1765}
  (\bibinfo {year} {1974})}\BibitemShut {NoStop}%
\bibitem [{\citenamefont {Welden}, \citenamefont {Rusakov},\ and\ \citenamefont
  {Zgid}(2016)}]{welden2016exploring}%
  \BibitemOpen
  \bibfield  {author} {\bibinfo {author} {\bibfnamefont {A.~R.}\ \bibnamefont
  {Welden}}, \bibinfo {author} {\bibfnamefont {A.~A.}\ \bibnamefont {Rusakov}},
  \ and\ \bibinfo {author} {\bibfnamefont {D.}~\bibnamefont {Zgid}},\
  }\href@noop {} {\bibfield  {journal} {\bibinfo  {journal} {J. Chem. Phys.}\
  }\textbf {\bibinfo {volume} {145}},\ \bibinfo {pages} {204106} (\bibinfo
  {year} {2016})}\BibitemShut {NoStop}%
\bibitem [{\citenamefont {White}\ and\ \citenamefont
  {Chan}(2018)}]{white2018time}%
  \BibitemOpen
  \bibfield  {author} {\bibinfo {author} {\bibfnamefont {A.~F.}\ \bibnamefont
  {White}}\ and\ \bibinfo {author} {\bibfnamefont {G.~K.-L.}\ \bibnamefont
  {Chan}},\ }\href@noop {} {\bibfield  {journal} {\bibinfo  {journal} {J. Chem.
  Theory Comput.}\ }\textbf {\bibinfo {volume} {14}},\ \bibinfo {pages} {5690}
  (\bibinfo {year} {2018})}\BibitemShut {NoStop}%
\bibitem [{\citenamefont {Harsha}, \citenamefont {Henderson},\ and\
  \citenamefont {Scuseria}(2019)}]{harsha2019thermofield}%
  \BibitemOpen
  \bibfield  {author} {\bibinfo {author} {\bibfnamefont {G.}~\bibnamefont
  {Harsha}}, \bibinfo {author} {\bibfnamefont {T.~M.}\ \bibnamefont
  {Henderson}}, \ and\ \bibinfo {author} {\bibfnamefont {G.~E.}\ \bibnamefont
  {Scuseria}},\ }\href@noop {} {\bibfield  {journal} {\bibinfo  {journal} {J.
  Chem. Theory Comput.}\ }\textbf {\bibinfo {volume} {15}},\ \bibinfo {pages}
  {6127} (\bibinfo {year} {2019})}\BibitemShut {NoStop}%
\bibitem [{\citenamefont {Purvis~III}\ and\ \citenamefont
  {Bartlett}(1982)}]{Purvis1982FullCoupled}%
  \BibitemOpen
  \bibfield  {author} {\bibinfo {author} {\bibfnamefont {G.~D.}\ \bibnamefont
  {Purvis~III}}\ and\ \bibinfo {author} {\bibfnamefont {R.~J.}\ \bibnamefont
  {Bartlett}},\ }\href@noop {} {\bibfield  {journal} {\bibinfo  {journal} {J.
  Chem. Phys.}\ }\textbf {\bibinfo {volume} {76}},\ \bibinfo {pages} {1910}
  (\bibinfo {year} {1982})}\BibitemShut {NoStop}%
\bibitem [{\citenamefont {Raghavachari}\ \emph {et~al.}(1989)\citenamefont
  {Raghavachari}, \citenamefont {Trucks}, \citenamefont {Pople},\ and\
  \citenamefont {Head-Gordon}}]{Raghavachari1989AFifth}%
  \BibitemOpen
  \bibfield  {author} {\bibinfo {author} {\bibfnamefont {K.}~\bibnamefont
  {Raghavachari}}, \bibinfo {author} {\bibfnamefont {G.~W.}\ \bibnamefont
  {Trucks}}, \bibinfo {author} {\bibfnamefont {J.~A.}\ \bibnamefont {Pople}}, \
  and\ \bibinfo {author} {\bibfnamefont {M.}~\bibnamefont {Head-Gordon}},\
  }\href@noop {} {\bibfield  {journal} {\bibinfo  {journal} {Chem. Phys.
  Lett.}\ }\textbf {\bibinfo {volume} {157}},\ \bibinfo {pages} {479} (\bibinfo
  {year} {1989})}\BibitemShut {NoStop}%
\bibitem [{\citenamefont {Piecuch}\ and\ \citenamefont
  {W{\l}och}(2005)}]{piecuch2005renormalized}%
  \BibitemOpen
  \bibfield  {author} {\bibinfo {author} {\bibfnamefont {P.}~\bibnamefont
  {Piecuch}}\ and\ \bibinfo {author} {\bibfnamefont {M.}~\bibnamefont
  {W{\l}och}},\ }\href@noop {} {\bibfield  {journal} {\bibinfo  {journal} {J.
  Chem. Phys.}\ }\textbf {\bibinfo {volume} {123}},\ \bibinfo {pages} {224105}
  (\bibinfo {year} {2005})}\BibitemShut {NoStop}%
\bibitem [{\citenamefont {Colmenero}\ and\ \citenamefont
  {Valdemoro}(1993)}]{Colmenero1993ApproximatingQorder}%
  \BibitemOpen
  \bibfield  {author} {\bibinfo {author} {\bibfnamefont {F.}~\bibnamefont
  {Colmenero}}\ and\ \bibinfo {author} {\bibfnamefont {C.}~\bibnamefont
  {Valdemoro}},\ }\href@noop {} {\bibfield  {journal} {\bibinfo  {journal}
  {Phys. Rev. A}\ }\textbf {\bibinfo {volume} {47}},\ \bibinfo {pages} {979}
  (\bibinfo {year} {1993})}\BibitemShut {NoStop}%
\bibitem [{\citenamefont {Nakatsuji}\ and\ \citenamefont
  {Yasuda}(1996)}]{Nakatsuji1996DirectDetermination}%
  \BibitemOpen
  \bibfield  {author} {\bibinfo {author} {\bibfnamefont {H.}~\bibnamefont
  {Nakatsuji}}\ and\ \bibinfo {author} {\bibfnamefont {K.}~\bibnamefont
  {Yasuda}},\ }\href@noop {} {\bibfield  {journal} {\bibinfo  {journal} {Phys.
  Rev. Lett.}\ }\textbf {\bibinfo {volume} {76}},\ \bibinfo {pages} {1039}
  (\bibinfo {year} {1996})}\BibitemShut {NoStop}%
\bibitem [{\citenamefont {Mazziotti}(1998)}]{Mazziotti1998ContractedSchro}%
  \BibitemOpen
  \bibfield  {author} {\bibinfo {author} {\bibfnamefont {D.~A.}\ \bibnamefont
  {Mazziotti}},\ }\href@noop {} {\bibfield  {journal} {\bibinfo  {journal}
  {Phys. Rev. A}\ }\textbf {\bibinfo {volume} {57}},\ \bibinfo {pages} {4219}
  (\bibinfo {year} {1998})}\BibitemShut {NoStop}%
\bibitem [{\citenamefont {Mazziotti}(2011)}]{Mazziotti2011TwoElectron}%
  \BibitemOpen
  \bibfield  {author} {\bibinfo {author} {\bibfnamefont {D.~A.}\ \bibnamefont
  {Mazziotti}},\ }\href@noop {} {\bibfield  {journal} {\bibinfo  {journal}
  {Chem. Rev.}\ }\textbf {\bibinfo {volume} {112}},\ \bibinfo {pages} {244}
  (\bibinfo {year} {2011})}\BibitemShut {NoStop}%
\bibitem [{\citenamefont {Fosso-Tande}\ \emph {et~al.}(2016)\citenamefont
  {Fosso-Tande}, \citenamefont {Nguyen}, \citenamefont {Gidofalvi},\ and\
  \citenamefont {DePrince~III}}]{Fosso2016LargeScale}%
  \BibitemOpen
  \bibfield  {author} {\bibinfo {author} {\bibfnamefont {J.}~\bibnamefont
  {Fosso-Tande}}, \bibinfo {author} {\bibfnamefont {T.-S.}\ \bibnamefont
  {Nguyen}}, \bibinfo {author} {\bibfnamefont {G.}~\bibnamefont {Gidofalvi}}, \
  and\ \bibinfo {author} {\bibfnamefont {A.~E.}\ \bibnamefont {DePrince~III}},\
  }\href@noop {} {\bibfield  {journal} {\bibinfo  {journal} {J. Chem. Theory
  Comput.}\ }\textbf {\bibinfo {volume} {12}},\ \bibinfo {pages} {2260}
  (\bibinfo {year} {2016})}\BibitemShut {NoStop}%
\bibitem [{\citenamefont {Evangelista}\ and\ \citenamefont
  {Stair}(2020)}]{HstudyRepo2020}%
  \BibitemOpen
  \bibfield  {author} {\bibinfo {author} {\bibfnamefont {F.~A.}\ \bibnamefont
  {Evangelista}}\ and\ \bibinfo {author} {\bibfnamefont {N.~H.}\ \bibnamefont
  {Stair}},\ }\href@noop {} {\enquote {\bibinfo {title} {Github repository:
  https://github.com/evangelistalab/hydrogen-models-data},}\ } (\bibinfo {year}
  {2020})\BibitemShut {NoStop}%
\bibitem [{\citenamefont {Handy}(1980)}]{Handy1980MultiRoot}%
  \BibitemOpen
  \bibfield  {author} {\bibinfo {author} {\bibfnamefont {N.~C.}\ \bibnamefont
  {Handy}},\ }\href@noop {} {\bibfield  {journal} {\bibinfo  {journal} {Chem.
  Phys. Lett.}\ }\textbf {\bibinfo {volume} {74}},\ \bibinfo {pages} {280}
  (\bibinfo {year} {1980})}\BibitemShut {NoStop}%
\bibitem [{\citenamefont {Wouters}\ and\ \citenamefont
  {Van~Neck}(2014)}]{Wouters2014TheDensity}%
  \BibitemOpen
  \bibfield  {author} {\bibinfo {author} {\bibfnamefont {S.}~\bibnamefont
  {Wouters}}\ and\ \bibinfo {author} {\bibfnamefont {D.}~\bibnamefont
  {Van~Neck}},\ }\href@noop {} {\bibfield  {journal} {\bibinfo  {journal} {Eur.
  Phys. J. D}\ }\textbf {\bibinfo {volume} {68}},\ \bibinfo {pages} {272}
  (\bibinfo {year} {2014})}\BibitemShut {NoStop}%
\bibitem [{\citenamefont {Legeza}, \citenamefont {R{\"o}der},\ and\
  \citenamefont {Hess}(2003)}]{legeza2003controlling}%
  \BibitemOpen
  \bibfield  {author} {\bibinfo {author} {\bibfnamefont {{\"O}.}~\bibnamefont
  {Legeza}}, \bibinfo {author} {\bibfnamefont {J.}~\bibnamefont {R{\"o}der}}, \
  and\ \bibinfo {author} {\bibfnamefont {B.}~\bibnamefont {Hess}},\ }\href@noop
  {} {\bibfield  {journal} {\bibinfo  {journal} {Phys. Rev. B}\ }\textbf
  {\bibinfo {volume} {67}},\ \bibinfo {pages} {125114} (\bibinfo {year}
  {2003})}\BibitemShut {NoStop}%
\bibitem [{\citenamefont {Moritz}, \citenamefont {Hess},\ and\ \citenamefont
  {Reiher}(2005)}]{moritz2005convergence}%
  \BibitemOpen
  \bibfield  {author} {\bibinfo {author} {\bibfnamefont {G.}~\bibnamefont
  {Moritz}}, \bibinfo {author} {\bibfnamefont {B.~A.}\ \bibnamefont {Hess}}, \
  and\ \bibinfo {author} {\bibfnamefont {M.}~\bibnamefont {Reiher}},\
  }\href@noop {} {\bibfield  {journal} {\bibinfo  {journal} {J. Chem. Phys.}\
  }\textbf {\bibinfo {volume} {122}},\ \bibinfo {pages} {024107} (\bibinfo
  {year} {2005})}\BibitemShut {NoStop}%
\bibitem [{\citenamefont {Legeza}\ and\ \citenamefont
  {S\'olyom}(2003)}]{Legeza2003OptemizingThe}%
  \BibitemOpen
  \bibfield  {author} {\bibinfo {author} {\bibfnamefont {O.}~\bibnamefont
  {Legeza}}\ and\ \bibinfo {author} {\bibfnamefont {J.}~\bibnamefont
  {S\'olyom}},\ }\href {\doibase 10.1103/PhysRevB.68.195116} {\bibfield
  {journal} {\bibinfo  {journal} {Phys. Rev. B}\ }\textbf {\bibinfo {volume}
  {68}},\ \bibinfo {pages} {195116} (\bibinfo {year} {2003})}\BibitemShut
  {NoStop}%
\bibitem [{\citenamefont {Rissler}, \citenamefont {Noack},\ and\ \citenamefont
  {White}(2006)}]{Rissler2006MeasuringOrbital}%
  \BibitemOpen
  \bibfield  {author} {\bibinfo {author} {\bibfnamefont {J.}~\bibnamefont
  {Rissler}}, \bibinfo {author} {\bibfnamefont {R.~M.}\ \bibnamefont {Noack}},
  \ and\ \bibinfo {author} {\bibfnamefont {S.~R.}\ \bibnamefont {White}},\
  }\href@noop {} {\bibfield  {journal} {\bibinfo  {journal} {Chem. Phys.}\
  }\textbf {\bibinfo {volume} {323}},\ \bibinfo {pages} {519} (\bibinfo {year}
  {2006})}\BibitemShut {NoStop}%
\bibitem [{\citenamefont {L{\"o}wdin}(1958)}]{lowdin1958correlation}%
  \BibitemOpen
  \bibfield  {author} {\bibinfo {author} {\bibfnamefont {P.-O.}\ \bibnamefont
  {L{\"o}wdin}},\ }\href@noop {} {\bibfield  {journal} {\bibinfo  {journal}
  {Adv. Chem. Phys.}\ }\textbf {\bibinfo {volume} {2}},\ \bibinfo {pages} {207}
  (\bibinfo {year} {1958})}\BibitemShut {NoStop}%
\bibitem [{\citenamefont {Sinanoglu}\ and\ \citenamefont
  {Fu-Tai~Tuan}(1964)}]{Sinanoglu1964QuantumTheory}%
  \BibitemOpen
  \bibfield  {author} {\bibinfo {author} {\bibfnamefont {O.}~\bibnamefont
  {Sinanoglu}}\ and\ \bibinfo {author} {\bibfnamefont {D.}~\bibnamefont
  {Fu-Tai~Tuan}},\ }\href@noop {} {\bibfield  {journal} {\bibinfo  {journal}
  {Annu. Rev. Phys. Chem.}\ }\textbf {\bibinfo {volume} {15}},\ \bibinfo
  {pages} {251} (\bibinfo {year} {1964})}\BibitemShut {NoStop}%
\bibitem [{\citenamefont {Bartlett}\ and\ \citenamefont
  {Stanton}(1994)}]{bartlett1994applications}%
  \BibitemOpen
  \bibfield  {author} {\bibinfo {author} {\bibfnamefont {R.~J.}\ \bibnamefont
  {Bartlett}}\ and\ \bibinfo {author} {\bibfnamefont {J.~F.}\ \bibnamefont
  {Stanton}}\ }(\bibinfo  {publisher} {Wiley Online Library},\ \bibinfo {year}
  {1994})\ pp.\ \bibinfo {pages} {65--169}\BibitemShut {NoStop}%
\bibitem [{\citenamefont {Lee}\ and\ \citenamefont
  {Taylor}(1989)}]{lee1989diagnostic}%
  \BibitemOpen
  \bibfield  {author} {\bibinfo {author} {\bibfnamefont {T.~J.}\ \bibnamefont
  {Lee}}\ and\ \bibinfo {author} {\bibfnamefont {P.~R.}\ \bibnamefont
  {Taylor}},\ }\href@noop {} {\bibfield  {journal} {\bibinfo  {journal} {Int.
  J. Quantum Chem.}\ }\textbf {\bibinfo {volume} {36}},\ \bibinfo {pages} {199}
  (\bibinfo {year} {1989})}\BibitemShut {NoStop}%
\bibitem [{\citenamefont {Nielsen}\ and\ \citenamefont
  {Janssen}(1999)}]{nielsen1999double}%
  \BibitemOpen
  \bibfield  {author} {\bibinfo {author} {\bibfnamefont {I.~M.~B.}\
  \bibnamefont {Nielsen}}\ and\ \bibinfo {author} {\bibfnamefont {C.~L.}\
  \bibnamefont {Janssen}},\ }\href@noop {} {\bibfield  {journal} {\bibinfo
  {journal} {Chem. Phys. Lett.}\ }\textbf {\bibinfo {volume} {310}},\ \bibinfo
  {pages} {568} (\bibinfo {year} {1999})}\BibitemShut {NoStop}%
\bibitem [{\citenamefont {Luzanov}\ and\ \citenamefont
  {Prezhdo}(2005)}]{Luzanov2005IrreducibleCharge}%
  \BibitemOpen
  \bibfield  {author} {\bibinfo {author} {\bibfnamefont {A.~V.}\ \bibnamefont
  {Luzanov}}\ and\ \bibinfo {author} {\bibfnamefont {O.~V.}\ \bibnamefont
  {Prezhdo}},\ }\href {\doibase 10.1002/qua.20438} {\bibfield  {journal}
  {\bibinfo  {journal} {Int. J. Quantum Chem.}\ }\textbf {\bibinfo {volume}
  {102}},\ \bibinfo {pages} {582} (\bibinfo {year} {2005})}\BibitemShut
  {NoStop}%
\bibitem [{\citenamefont {Huang}, \citenamefont {Wang},\ and\ \citenamefont
  {Kais}(2006)}]{Huang2006Entanglement}%
  \BibitemOpen
  \bibfield  {author} {\bibinfo {author} {\bibfnamefont {Z.}~\bibnamefont
  {Huang}}, \bibinfo {author} {\bibfnamefont {H.}~\bibnamefont {Wang}}, \ and\
  \bibinfo {author} {\bibfnamefont {S.}~\bibnamefont {Kais}},\ }\href {\doibase
  10.1080/09500340600955674} {\bibfield  {journal} {\bibinfo  {journal} {J.
  Mod. Opt.}\ }\textbf {\bibinfo {volume} {53}},\ \bibinfo {pages} {2543}
  (\bibinfo {year} {2006})}\BibitemShut {NoStop}%
\bibitem [{\citenamefont {Juh{\'a}sz}\ and\ \citenamefont
  {Mazziotti}(2006)}]{Juhasz2006TheCumulant}%
  \BibitemOpen
  \bibfield  {author} {\bibinfo {author} {\bibfnamefont {T.}~\bibnamefont
  {Juh{\'a}sz}}\ and\ \bibinfo {author} {\bibfnamefont {D.~A.}\ \bibnamefont
  {Mazziotti}},\ }\href {\doibase 10.1063/1.2378768} {\bibfield  {journal}
  {\bibinfo  {journal} {J. Chem. Phys.}\ }\textbf {\bibinfo {volume} {125}},\
  \bibinfo {pages} {174105} (\bibinfo {year} {2006})}\BibitemShut {NoStop}%
\bibitem [{\citenamefont {Luzanov}\ and\ \citenamefont
  {Prezhdo}(2007)}]{Luzanova2007HighOrder}%
  \BibitemOpen
  \bibfield  {author} {\bibinfo {author} {\bibfnamefont {A.~V.}\ \bibnamefont
  {Luzanov}}\ and\ \bibinfo {author} {\bibfnamefont {O.}~\bibnamefont
  {Prezhdo}},\ }\href {\doibase 10.1080/00268970701725039} {\bibfield
  {journal} {\bibinfo  {journal} {Mol. Phys.}\ }\textbf {\bibinfo {volume}
  {105}},\ \bibinfo {pages} {2879} (\bibinfo {year} {2007})}\BibitemShut
  {NoStop}%
\bibitem [{\citenamefont {Alcoba}\ \emph {et~al.}(2010)\citenamefont {Alcoba},
  \citenamefont {Bochicchio}, \citenamefont {Lain},\ and\ \citenamefont
  {Torre}}]{Alcoba2010OnThe}%
  \BibitemOpen
  \bibfield  {author} {\bibinfo {author} {\bibfnamefont {D.~R.}\ \bibnamefont
  {Alcoba}}, \bibinfo {author} {\bibfnamefont {R.~C.}\ \bibnamefont
  {Bochicchio}}, \bibinfo {author} {\bibfnamefont {L.}~\bibnamefont {Lain}}, \
  and\ \bibinfo {author} {\bibfnamefont {A.}~\bibnamefont {Torre}},\ }\href
  {\doibase 10.1063/1.3503766} {\bibfield  {journal} {\bibinfo  {journal} {J.
  Chem. Phys.}\ }\textbf {\bibinfo {volume} {133}},\ \bibinfo {pages} {144104}
  (\bibinfo {year} {2010})}\BibitemShut {NoStop}%
\bibitem [{\citenamefont {Bochicchio}(1998)}]{Bochicchio1998OnSpin}%
  \BibitemOpen
  \bibfield  {author} {\bibinfo {author} {\bibfnamefont {R.~C.}\ \bibnamefont
  {Bochicchio}},\ }\href {\doibase
  https://doi.org/10.1016/S0166-1280(97)00357-6} {\bibfield  {journal}
  {\bibinfo  {journal} {J. Mol. Struc.-Theochem.}\ }\textbf {\bibinfo {volume}
  {429}},\ \bibinfo {pages} {229 } (\bibinfo {year} {1998})}\BibitemShut
  {NoStop}%
\bibitem [{\citenamefont {Lain}\ \emph {et~al.}(2009)\citenamefont {Lain},
  \citenamefont {Torre}, \citenamefont {Alcoba},\ and\ \citenamefont
  {Bochicchio}}]{Lain2009ADecomposition}%
  \BibitemOpen
  \bibfield  {author} {\bibinfo {author} {\bibfnamefont {L.}~\bibnamefont
  {Lain}}, \bibinfo {author} {\bibfnamefont {A.}~\bibnamefont {Torre}},
  \bibinfo {author} {\bibfnamefont {D.~R.}\ \bibnamefont {Alcoba}}, \ and\
  \bibinfo {author} {\bibfnamefont {R.~C.}\ \bibnamefont {Bochicchio}},\ }\href
  {\doibase https://doi.org/10.1016/j.cplett.2009.05.071} {\bibfield  {journal}
  {\bibinfo  {journal} {Chem. Phys. Lett.}\ }\textbf {\bibinfo {volume}
  {476}},\ \bibinfo {pages} {101 } (\bibinfo {year} {2009})}\BibitemShut
  {NoStop}%
\bibitem [{\citenamefont {Alcoba}\ \emph {et~al.}(2006)\citenamefont {Alcoba},
  \citenamefont {Bochicchio}, \citenamefont {Lain},\ and\ \citenamefont
  {Torre}}]{Alcoba2006OnTheDefinition}%
  \BibitemOpen
  \bibfield  {author} {\bibinfo {author} {\bibfnamefont {D.~R.}\ \bibnamefont
  {Alcoba}}, \bibinfo {author} {\bibfnamefont {R.~C.}\ \bibnamefont
  {Bochicchio}}, \bibinfo {author} {\bibfnamefont {L.}~\bibnamefont {Lain}}, \
  and\ \bibinfo {author} {\bibfnamefont {A.}~\bibnamefont {Torre}},\ }\href
  {\doibase https://doi.org/10.1016/j.cplett.2006.07.068} {\bibfield  {journal}
  {\bibinfo  {journal} {Chem. Phys. Lett.}\ }\textbf {\bibinfo {volume}
  {429}},\ \bibinfo {pages} {286 } (\bibinfo {year} {2006})}\BibitemShut
  {NoStop}%
\bibitem [{\citenamefont {Pipek}\ and\ \citenamefont
  {Mezey}(1989)}]{Pipek1989FastIntrinsic}%
  \BibitemOpen
  \bibfield  {author} {\bibinfo {author} {\bibfnamefont {J.}~\bibnamefont
  {Pipek}}\ and\ \bibinfo {author} {\bibfnamefont {P.~G.}\ \bibnamefont
  {Mezey}},\ }\href@noop {} {\bibfield  {journal} {\bibinfo  {journal} {J.
  Chem. Phys.}\ }\textbf {\bibinfo {volume} {90}},\ \bibinfo {pages} {4916}
  (\bibinfo {year} {1989})}\BibitemShut {NoStop}%
\bibitem [{\citenamefont {Jim{\'e}nez-Hoyos}, \citenamefont
  {Rodr{\'\i}guez-Guzm{\'a}n},\ and\ \citenamefont
  {Scuseria}(2014)}]{Hoyos2014PolyradicalCharacter}%
  \BibitemOpen
  \bibfield  {author} {\bibinfo {author} {\bibfnamefont {C.~A.}\ \bibnamefont
  {Jim{\'e}nez-Hoyos}}, \bibinfo {author} {\bibfnamefont {R.}~\bibnamefont
  {Rodr{\'\i}guez-Guzm{\'a}n}}, \ and\ \bibinfo {author} {\bibfnamefont
  {G.~E.}\ \bibnamefont {Scuseria}},\ }\href {\doibase 10.1021/jp508383z}
  {\bibfield  {journal} {\bibinfo  {journal} {J. Phys. Chem. A}\ }\textbf
  {\bibinfo {volume} {118}},\ \bibinfo {pages} {9925} (\bibinfo {year}
  {2014})}\BibitemShut {NoStop}%
\bibitem [{\citenamefont {Boguslawski}\ and\ \citenamefont
  {Tecmer}(2015)}]{Boguslawski2015OrbtitalEntanglement}%
  \BibitemOpen
  \bibfield  {author} {\bibinfo {author} {\bibfnamefont {K.}~\bibnamefont
  {Boguslawski}}\ and\ \bibinfo {author} {\bibfnamefont {P.}~\bibnamefont
  {Tecmer}},\ }\href {\doibase 10.1002/qua.24832} {\bibfield  {journal}
  {\bibinfo  {journal} {Int. J. Quantum Chem.}\ }\textbf {\bibinfo {volume}
  {115}},\ \bibinfo {pages} {1289} (\bibinfo {year} {2015})}\BibitemShut
  {NoStop}%
\bibitem [{\citenamefont {Boguslawski}\ \emph {et~al.}(2013)\citenamefont
  {Boguslawski}, \citenamefont {Tecmer}, \citenamefont {Barcza}, \citenamefont
  {Legeza},\ and\ \citenamefont {Reiher}}]{Boguslawski2013OrbitalEntanglement}%
  \BibitemOpen
  \bibfield  {author} {\bibinfo {author} {\bibfnamefont {K.}~\bibnamefont
  {Boguslawski}}, \bibinfo {author} {\bibfnamefont {P.}~\bibnamefont {Tecmer}},
  \bibinfo {author} {\bibfnamefont {G.}~\bibnamefont {Barcza}}, \bibinfo
  {author} {\bibfnamefont {{\"O}.}~\bibnamefont {Legeza}}, \ and\ \bibinfo
  {author} {\bibfnamefont {M.}~\bibnamefont {Reiher}},\ }\href {\doibase
  10.1021/ct400247p} {\bibfield  {journal} {\bibinfo  {journal} {J. Chem.
  Theory Comput.}\ }\textbf {\bibinfo {volume} {9}},\ \bibinfo {pages} {2959}
  (\bibinfo {year} {2013})}\BibitemShut {NoStop}%
\bibitem [{\citenamefont {Stein}\ and\ \citenamefont
  {Reiher}(2017)}]{stein2017automated}%
  \BibitemOpen
  \bibfield  {author} {\bibinfo {author} {\bibfnamefont {C.~J.}\ \bibnamefont
  {Stein}}\ and\ \bibinfo {author} {\bibfnamefont {M.}~\bibnamefont {Reiher}},\
  }\href@noop {} {\bibfield  {journal} {\bibinfo  {journal} {Chimia}\ }\textbf
  {\bibinfo {volume} {71}},\ \bibinfo {pages} {170} (\bibinfo {year}
  {2017})}\BibitemShut {NoStop}%
\bibitem [{\citenamefont {Fertitta}\ \emph {et~al.}(2014)\citenamefont
  {Fertitta}, \citenamefont {Paulus}, \citenamefont {Barcza},\ and\
  \citenamefont {Legeza}}]{fertitta2014investigation}%
  \BibitemOpen
  \bibfield  {author} {\bibinfo {author} {\bibfnamefont {E.}~\bibnamefont
  {Fertitta}}, \bibinfo {author} {\bibfnamefont {B.}~\bibnamefont {Paulus}},
  \bibinfo {author} {\bibfnamefont {G.}~\bibnamefont {Barcza}}, \ and\ \bibinfo
  {author} {\bibfnamefont {{\"O}.}~\bibnamefont {Legeza}},\ }\href@noop {}
  {\bibfield  {journal} {\bibinfo  {journal} {Phys. Rev. B}\ }\textbf {\bibinfo
  {volume} {90}},\ \bibinfo {pages} {245129} (\bibinfo {year}
  {2014})}\BibitemShut {NoStop}%
\bibitem [{\citenamefont {Murg}\ \emph {et~al.}(2015)\citenamefont {Murg},
  \citenamefont {Verstraete}, \citenamefont {Schneider}, \citenamefont {Nagy},\
  and\ \citenamefont {Legeza}}]{murg2015tree}%
  \BibitemOpen
  \bibfield  {author} {\bibinfo {author} {\bibfnamefont {V.}~\bibnamefont
  {Murg}}, \bibinfo {author} {\bibfnamefont {F.}~\bibnamefont {Verstraete}},
  \bibinfo {author} {\bibfnamefont {R.}~\bibnamefont {Schneider}}, \bibinfo
  {author} {\bibfnamefont {P.~R.}\ \bibnamefont {Nagy}}, \ and\ \bibinfo
  {author} {\bibfnamefont {O.}~\bibnamefont {Legeza}},\ }\href@noop {}
  {\bibfield  {journal} {\bibinfo  {journal} {J. Chem. Theory Comput.}\
  }\textbf {\bibinfo {volume} {11}},\ \bibinfo {pages} {1027} (\bibinfo {year}
  {2015})}\BibitemShut {NoStop}%
\bibitem [{\citenamefont {Parrish}\ \emph {et~al.}(2017)\citenamefont
  {Parrish}, \citenamefont {Burns}, \citenamefont {Smith}, \citenamefont
  {Simmonett}, \citenamefont {DePrince~III}, \citenamefont {Hohenstein},
  \citenamefont {Bozkaya}, \citenamefont {Sokolov}, \citenamefont {Di~Remigio},
  \citenamefont {Richard} \emph {et~al.}}]{Parrish2017Psi4}%
  \BibitemOpen
  \bibfield  {author} {\bibinfo {author} {\bibfnamefont {R.~M.}\ \bibnamefont
  {Parrish}}, \bibinfo {author} {\bibfnamefont {L.~A.}\ \bibnamefont {Burns}},
  \bibinfo {author} {\bibfnamefont {D.~G.}\ \bibnamefont {Smith}}, \bibinfo
  {author} {\bibfnamefont {A.~C.}\ \bibnamefont {Simmonett}}, \bibinfo {author}
  {\bibfnamefont {A.~E.}\ \bibnamefont {DePrince~III}}, \bibinfo {author}
  {\bibfnamefont {E.~G.}\ \bibnamefont {Hohenstein}}, \bibinfo {author}
  {\bibfnamefont {U.}~\bibnamefont {Bozkaya}}, \bibinfo {author} {\bibfnamefont
  {A.~Y.}\ \bibnamefont {Sokolov}}, \bibinfo {author} {\bibfnamefont
  {R.}~\bibnamefont {Di~Remigio}}, \bibinfo {author} {\bibfnamefont {R.~M.}\
  \bibnamefont {Richard}},  \emph {et~al.},\ }\href@noop {} {\bibfield
  {journal} {\bibinfo  {journal} {J. Chem. Theory Comput.}\ }\textbf {\bibinfo
  {volume} {13}},\ \bibinfo {pages} {3185} (\bibinfo {year}
  {2017})}\BibitemShut {NoStop}%
\bibitem [{\citenamefont {Smith}\ \emph {et~al.}(2020)\citenamefont {Smith},
  \citenamefont {Burns}, \citenamefont {Simmonett}, \citenamefont {Parrish},
  \citenamefont {Schieber}, \citenamefont {Galvelis}, \citenamefont {Kraus},
  \citenamefont {Kruse}, \citenamefont {Di~Remigio}, \citenamefont {Alenaizan}
  \emph {et~al.}}]{smith2020psi4}%
  \BibitemOpen
  \bibfield  {author} {\bibinfo {author} {\bibfnamefont {D.~G.}\ \bibnamefont
  {Smith}}, \bibinfo {author} {\bibfnamefont {L.~A.}\ \bibnamefont {Burns}},
  \bibinfo {author} {\bibfnamefont {A.~C.}\ \bibnamefont {Simmonett}}, \bibinfo
  {author} {\bibfnamefont {R.~M.}\ \bibnamefont {Parrish}}, \bibinfo {author}
  {\bibfnamefont {M.~C.}\ \bibnamefont {Schieber}}, \bibinfo {author}
  {\bibfnamefont {R.}~\bibnamefont {Galvelis}}, \bibinfo {author}
  {\bibfnamefont {P.}~\bibnamefont {Kraus}}, \bibinfo {author} {\bibfnamefont
  {H.}~\bibnamefont {Kruse}}, \bibinfo {author} {\bibfnamefont
  {R.}~\bibnamefont {Di~Remigio}}, \bibinfo {author} {\bibfnamefont
  {A.}~\bibnamefont {Alenaizan}},  \emph {et~al.},\ }\href@noop {} {\bibfield
  {journal} {\bibinfo  {journal} {J. Chem. Phys.}\ }\textbf {\bibinfo {volume}
  {152}},\ \bibinfo {pages} {184108} (\bibinfo {year} {2020})}\BibitemShut
  {NoStop}%
\bibitem [{\citenamefont {Hehre}, \citenamefont {Stewart},\ and\ \citenamefont
  {Pople}(1969)}]{Hehre1969ASelf}%
  \BibitemOpen
  \bibfield  {author} {\bibinfo {author} {\bibfnamefont {W.~J.}\ \bibnamefont
  {Hehre}}, \bibinfo {author} {\bibfnamefont {R.~F.}\ \bibnamefont {Stewart}},
  \ and\ \bibinfo {author} {\bibfnamefont {J.~A.}\ \bibnamefont {Pople}},\
  }\href@noop {} {\bibfield  {journal} {\bibinfo  {journal} {J. Chem. Phys.}\
  }\textbf {\bibinfo {volume} {51}},\ \bibinfo {pages} {2657} (\bibinfo {year}
  {1969})}\BibitemShut {NoStop}%
\bibitem [{\citenamefont {Barca}\ \emph {et~al.}(2020)\citenamefont {Barca},
  \citenamefont {Bertoni}, \citenamefont {Carrington}, \citenamefont {Datta},
  \citenamefont {De~Silva}, \citenamefont {Deustua}, \citenamefont {Fedorov},
  \citenamefont {Gour}, \citenamefont {Gunina}, \citenamefont {Guidez} \emph
  {et~al.}}]{barca2020recent}%
  \BibitemOpen
  \bibfield  {author} {\bibinfo {author} {\bibfnamefont {G.~M.}\ \bibnamefont
  {Barca}}, \bibinfo {author} {\bibfnamefont {C.}~\bibnamefont {Bertoni}},
  \bibinfo {author} {\bibfnamefont {L.}~\bibnamefont {Carrington}}, \bibinfo
  {author} {\bibfnamefont {D.}~\bibnamefont {Datta}}, \bibinfo {author}
  {\bibfnamefont {N.}~\bibnamefont {De~Silva}}, \bibinfo {author}
  {\bibfnamefont {J.~E.}\ \bibnamefont {Deustua}}, \bibinfo {author}
  {\bibfnamefont {D.~G.}\ \bibnamefont {Fedorov}}, \bibinfo {author}
  {\bibfnamefont {J.~R.}\ \bibnamefont {Gour}}, \bibinfo {author}
  {\bibfnamefont {A.~O.}\ \bibnamefont {Gunina}}, \bibinfo {author}
  {\bibfnamefont {E.}~\bibnamefont {Guidez}},  \emph {et~al.},\ }\href@noop {}
  {\bibfield  {journal} {\bibinfo  {journal} {J. Chem. Phys.}\ }\textbf
  {\bibinfo {volume} {152}},\ \bibinfo {pages} {154102} (\bibinfo {year}
  {2020})}\BibitemShut {NoStop}%
\bibitem [{\citenamefont {Evangelista}(2020)}]{Evangelista2019Forte}%
  \BibitemOpen
  \bibfield  {author} {\bibinfo {author} {\bibfnamefont {F.~A.}\ \bibnamefont
  {Evangelista}},\ }\href@noop {} {\enquote {\bibinfo {title} {Forte: an open
  source plugin for strongly correlated electronic systems},}\ } (\bibinfo
  {year} {2020})\BibitemShut {NoStop}%
\bibitem [{\citenamefont {Wouters}\ \emph {et~al.}(2014)\citenamefont
  {Wouters}, \citenamefont {Poelmans}, \citenamefont {Ayers},\ and\
  \citenamefont {Van~Neck}}]{Wouters2014Chemps2}%
  \BibitemOpen
  \bibfield  {author} {\bibinfo {author} {\bibfnamefont {S.}~\bibnamefont
  {Wouters}}, \bibinfo {author} {\bibfnamefont {W.}~\bibnamefont {Poelmans}},
  \bibinfo {author} {\bibfnamefont {P.~W.}\ \bibnamefont {Ayers}}, \ and\
  \bibinfo {author} {\bibfnamefont {D.}~\bibnamefont {Van~Neck}},\ }\href@noop
  {} {\bibfield  {journal} {\bibinfo  {journal} {Comput. Phys. Commun.}\
  }\textbf {\bibinfo {volume} {185}},\ \bibinfo {pages} {1501} (\bibinfo {year}
  {2014})}\BibitemShut {NoStop}%
\bibitem [{\citenamefont {Moritz}\ and\ \citenamefont
  {Reiher}(2006)}]{Moritz2006ConstructionOf}%
  \BibitemOpen
  \bibfield  {author} {\bibinfo {author} {\bibfnamefont {G.}~\bibnamefont
  {Moritz}}\ and\ \bibinfo {author} {\bibfnamefont {M.}~\bibnamefont
  {Reiher}},\ }\href@noop {} {\bibfield  {journal} {\bibinfo  {journal} {J.
  Chem. Phys.}\ }\textbf {\bibinfo {volume} {124}},\ \bibinfo {pages} {034103}
  (\bibinfo {year} {2006})}\BibitemShut {NoStop}%
\bibitem [{\citenamefont {Yanai}\ \emph
  {et~al.}(2010{\natexlab{b}})\citenamefont {Yanai}, \citenamefont {Kurashige},
  \citenamefont {Neuscamman},\ and\ \citenamefont
  {Chan}}]{Yanai2010MultireferenceQuantum}%
  \BibitemOpen
  \bibfield  {author} {\bibinfo {author} {\bibfnamefont {T.}~\bibnamefont
  {Yanai}}, \bibinfo {author} {\bibfnamefont {Y.}~\bibnamefont {Kurashige}},
  \bibinfo {author} {\bibfnamefont {E.}~\bibnamefont {Neuscamman}}, \ and\
  \bibinfo {author} {\bibfnamefont {G.~K.-L.}\ \bibnamefont {Chan}},\
  }\href@noop {} {\bibfield  {journal} {\bibinfo  {journal} {J. Chem. Phys.}\
  }\textbf {\bibinfo {volume} {132}},\ \bibinfo {pages} {024105} (\bibinfo
  {year} {2010}{\natexlab{b}})}\BibitemShut {NoStop}%
\bibitem [{\citenamefont {Barcza}\ \emph {et~al.}(2011)\citenamefont {Barcza},
  \citenamefont {Legeza}, \citenamefont {Marti},\ and\ \citenamefont
  {Reiher}}]{Barcza2011QuantumInfromation}%
  \BibitemOpen
  \bibfield  {author} {\bibinfo {author} {\bibfnamefont {G.}~\bibnamefont
  {Barcza}}, \bibinfo {author} {\bibfnamefont {{\"O}.}~\bibnamefont {Legeza}},
  \bibinfo {author} {\bibfnamefont {K.~H.}\ \bibnamefont {Marti}}, \ and\
  \bibinfo {author} {\bibfnamefont {M.}~\bibnamefont {Reiher}},\ }\href@noop {}
  {\bibfield  {journal} {\bibinfo  {journal} {Phys. Rev. A}\ }\textbf {\bibinfo
  {volume} {83}},\ \bibinfo {pages} {012508} (\bibinfo {year}
  {2011})}\BibitemShut {NoStop}%
\bibitem [{\citenamefont {Boguslawski}\ \emph {et~al.}(2012)\citenamefont
  {Boguslawski}, \citenamefont {Tecmer}, \citenamefont {Legeza},\ and\
  \citenamefont {Reiher}}]{Boguslawski2012EntanglementMeasures}%
  \BibitemOpen
  \bibfield  {author} {\bibinfo {author} {\bibfnamefont {K.}~\bibnamefont
  {Boguslawski}}, \bibinfo {author} {\bibfnamefont {P.}~\bibnamefont {Tecmer}},
  \bibinfo {author} {\bibfnamefont {{\"O}.}~\bibnamefont {Legeza}}, \ and\
  \bibinfo {author} {\bibfnamefont {M.}~\bibnamefont {Reiher}},\ }\href
  {\doibase 10.1021/jz301319v} {\bibfield  {journal} {\bibinfo  {journal} {J.
  Phys. Chem. Let.}\ }\textbf {\bibinfo {volume} {3}},\ \bibinfo {pages} {3129}
  (\bibinfo {year} {2012})}\BibitemShut {NoStop}%
\bibitem [{\citenamefont {Baker}\ \emph {et~al.}(2012)\citenamefont {Baker},
  \citenamefont {Timco}, \citenamefont {Piligkos}, \citenamefont {Mathieson},
  \citenamefont {Mutka}, \citenamefont {Tuna}, \citenamefont {Koz{\l}owski},
  \citenamefont {Antkowiak}, \citenamefont {Guidi}, \citenamefont {Gupta} \emph
  {et~al.}}]{Baker2012ClassificationOf}%
  \BibitemOpen
  \bibfield  {author} {\bibinfo {author} {\bibfnamefont {M.~L.}\ \bibnamefont
  {Baker}}, \bibinfo {author} {\bibfnamefont {G.~A.}\ \bibnamefont {Timco}},
  \bibinfo {author} {\bibfnamefont {S.}~\bibnamefont {Piligkos}}, \bibinfo
  {author} {\bibfnamefont {J.~S.}\ \bibnamefont {Mathieson}}, \bibinfo {author}
  {\bibfnamefont {H.}~\bibnamefont {Mutka}}, \bibinfo {author} {\bibfnamefont
  {F.}~\bibnamefont {Tuna}}, \bibinfo {author} {\bibfnamefont {P.}~\bibnamefont
  {Koz{\l}owski}}, \bibinfo {author} {\bibfnamefont {M.}~\bibnamefont
  {Antkowiak}}, \bibinfo {author} {\bibfnamefont {T.}~\bibnamefont {Guidi}},
  \bibinfo {author} {\bibfnamefont {T.}~\bibnamefont {Gupta}},  \emph
  {et~al.},\ }\href@noop {} {\bibfield  {journal} {\bibinfo  {journal} {Proc.
  Natl. Acad. Sci. U.S.A}\ }\textbf {\bibinfo {volume} {109}},\ \bibinfo
  {pages} {19113} (\bibinfo {year} {2012})}\BibitemShut {NoStop}%
\bibitem [{\citenamefont {Limacher}\ \emph {et~al.}(2013)\citenamefont
  {Limacher}, \citenamefont {Ayers}, \citenamefont {Johnson}, \citenamefont
  {De~Baerdemacker}, \citenamefont {Van~Neck},\ and\ \citenamefont
  {Bultinck}}]{Limacher2013NewMean}%
  \BibitemOpen
  \bibfield  {author} {\bibinfo {author} {\bibfnamefont {P.~A.}\ \bibnamefont
  {Limacher}}, \bibinfo {author} {\bibfnamefont {P.~W.}\ \bibnamefont {Ayers}},
  \bibinfo {author} {\bibfnamefont {P.~A.}\ \bibnamefont {Johnson}}, \bibinfo
  {author} {\bibfnamefont {S.}~\bibnamefont {De~Baerdemacker}}, \bibinfo
  {author} {\bibfnamefont {D.}~\bibnamefont {Van~Neck}}, \ and\ \bibinfo
  {author} {\bibfnamefont {P.}~\bibnamefont {Bultinck}},\ }\href@noop {}
  {\bibfield  {journal} {\bibinfo  {journal} {J. Chem. Theory Comput.}\
  }\textbf {\bibinfo {volume} {9}},\ \bibinfo {pages} {1394} (\bibinfo {year}
  {2013})}\BibitemShut {NoStop}%
\bibitem [{\citenamefont {Bulik}, \citenamefont {Henderson},\ and\
  \citenamefont {Scuseria}(2015)}]{Bulik2015CanSingle}%
  \BibitemOpen
  \bibfield  {author} {\bibinfo {author} {\bibfnamefont {I.~W.}\ \bibnamefont
  {Bulik}}, \bibinfo {author} {\bibfnamefont {T.~M.}\ \bibnamefont
  {Henderson}}, \ and\ \bibinfo {author} {\bibfnamefont {G.~E.}\ \bibnamefont
  {Scuseria}},\ }\href@noop {} {\bibfield  {journal} {\bibinfo  {journal} {J.
  Chem. Theory Comput.}\ }\textbf {\bibinfo {volume} {11}},\ \bibinfo {pages}
  {3171} (\bibinfo {year} {2015})}\BibitemShut {NoStop}%
\bibitem [{\citenamefont {Hastings}(2004)}]{hastings2004lieb}%
  \BibitemOpen
  \bibfield  {author} {\bibinfo {author} {\bibfnamefont {M.~B.}\ \bibnamefont
  {Hastings}},\ }\href@noop {} {\bibfield  {journal} {\bibinfo  {journal}
  {Phys. Rev. B}\ }\textbf {\bibinfo {volume} {69}},\ \bibinfo {pages} {104431}
  (\bibinfo {year} {2004})}\BibitemShut {NoStop}%
\bibitem [{\citenamefont {Stoudenmire}\ and\ \citenamefont
  {White}(2012)}]{stoudenmire2012studying}%
  \BibitemOpen
  \bibfield  {author} {\bibinfo {author} {\bibfnamefont {E.~M.}\ \bibnamefont
  {Stoudenmire}}\ and\ \bibinfo {author} {\bibfnamefont {S.~R.}\ \bibnamefont
  {White}},\ }\href@noop {} {\bibfield  {journal} {\bibinfo  {journal} {Annu.
  Rev. Condens. Matter Phys.}\ }\textbf {\bibinfo {volume} {3}},\ \bibinfo
  {pages} {111} (\bibinfo {year} {2012})}\BibitemShut {NoStop}%
\bibitem [{\citenamefont {Verstraete}\ and\ \citenamefont
  {Cirac}(2004)}]{verstraete2004renormalization}%
  \BibitemOpen
  \bibfield  {author} {\bibinfo {author} {\bibfnamefont {F.}~\bibnamefont
  {Verstraete}}\ and\ \bibinfo {author} {\bibfnamefont {J.~I.}\ \bibnamefont
  {Cirac}},\ }\href@noop {} {\bibfield  {journal} {\bibinfo  {journal} {e-print
  arXiv:0407066 [cond-mat.str-el]}\ } (\bibinfo {year} {2004})}\BibitemShut
  {NoStop}%
\bibitem [{\citenamefont {Vidal}(2007)}]{vidal2007entanglement}%
  \BibitemOpen
  \bibfield  {author} {\bibinfo {author} {\bibfnamefont {G.}~\bibnamefont
  {Vidal}},\ }\href@noop {} {\bibfield  {journal} {\bibinfo  {journal} {Phys.
  Rev. Lett.}\ }\textbf {\bibinfo {volume} {99}},\ \bibinfo {pages} {220405}
  (\bibinfo {year} {2007})}\BibitemShut {NoStop}%
\bibitem [{\citenamefont {Needs}\ \emph {et~al.}(2009)\citenamefont {Needs},
  \citenamefont {Towler}, \citenamefont {Drummond},\ and\ \citenamefont
  {R{\'\i}os}}]{needs2009continuum}%
  \BibitemOpen
  \bibfield  {author} {\bibinfo {author} {\bibfnamefont {R.~J.}\ \bibnamefont
  {Needs}}, \bibinfo {author} {\bibfnamefont {M.~D.}\ \bibnamefont {Towler}},
  \bibinfo {author} {\bibfnamefont {N.~D.}\ \bibnamefont {Drummond}}, \ and\
  \bibinfo {author} {\bibfnamefont {P.~L.}\ \bibnamefont {R{\'\i}os}},\
  }\href@noop {} {\bibfield  {journal} {\bibinfo  {journal} {J. Phys.: Condens.
  Matter}\ }\textbf {\bibinfo {volume} {22}},\ \bibinfo {pages} {023201}
  (\bibinfo {year} {2009})}\BibitemShut {NoStop}%
\bibitem [{\citenamefont {Motta}\ and\ \citenamefont
  {Zhang}(2018)}]{motta2018ab}%
  \BibitemOpen
  \bibfield  {author} {\bibinfo {author} {\bibfnamefont {M.}~\bibnamefont
  {Motta}}\ and\ \bibinfo {author} {\bibfnamefont {S.}~\bibnamefont {Zhang}},\
  }\href@noop {} {\bibfield  {journal} {\bibinfo  {journal} {Wiley Interdiscip.
  Rev. Comput. Mol. Sci.}\ }\textbf {\bibinfo {volume} {8}},\ \bibinfo {pages}
  {1364} (\bibinfo {year} {2018})}\BibitemShut {NoStop}%
\bibitem [{\citenamefont {Abrams}\ and\ \citenamefont
  {Lloyd}(1997)}]{Abrams:1997ha}%
  \BibitemOpen
  \bibfield  {author} {\bibinfo {author} {\bibfnamefont {D.~S.}\ \bibnamefont
  {Abrams}}\ and\ \bibinfo {author} {\bibfnamefont {S.}~\bibnamefont {Lloyd}},\
  }\href@noop {} {\bibfield  {journal} {\bibinfo  {journal} {Phys. Rev. Lett.}\
  }\textbf {\bibinfo {volume} {79}},\ \bibinfo {pages} {2586} (\bibinfo {year}
  {1997})}\BibitemShut {NoStop}%
\bibitem [{\citenamefont {Abrams}\ and\ \citenamefont
  {Lloyd}(1999)}]{Abrams:1999ur}%
  \BibitemOpen
  \bibfield  {author} {\bibinfo {author} {\bibfnamefont {D.~S.}\ \bibnamefont
  {Abrams}}\ and\ \bibinfo {author} {\bibfnamefont {S.}~\bibnamefont {Lloyd}},\
  }\href@noop {} {\bibfield  {journal} {\bibinfo  {journal} {Phys. Rev. Lett.}\
  }\textbf {\bibinfo {volume} {83}},\ \bibinfo {pages} {5162} (\bibinfo {year}
  {1999})}\BibitemShut {NoStop}%
\bibitem [{\citenamefont {Babbush}\ \emph {et~al.}(2018)\citenamefont
  {Babbush}, \citenamefont {Wiebe}, \citenamefont {McClean}, \citenamefont
  {McClain}, \citenamefont {Neven},\ and\ \citenamefont
  {Chan}}]{babbush2018low}%
  \BibitemOpen
  \bibfield  {author} {\bibinfo {author} {\bibfnamefont {R.}~\bibnamefont
  {Babbush}}, \bibinfo {author} {\bibfnamefont {N.}~\bibnamefont {Wiebe}},
  \bibinfo {author} {\bibfnamefont {J.~R.}\ \bibnamefont {McClean}}, \bibinfo
  {author} {\bibfnamefont {J.}~\bibnamefont {McClain}}, \bibinfo {author}
  {\bibfnamefont {H.}~\bibnamefont {Neven}}, \ and\ \bibinfo {author}
  {\bibfnamefont {G.~K.-L.}\ \bibnamefont {Chan}},\ }\href@noop {} {\bibfield
  {journal} {\bibinfo  {journal} {Phys. Rev. X}\ }\textbf {\bibinfo {volume}
  {8}},\ \bibinfo {pages} {011044} (\bibinfo {year} {2018})}\BibitemShut
  {NoStop}%
\bibitem [{\citenamefont {McClean}\ \emph {et~al.}(2019)\citenamefont
  {McClean}, \citenamefont {Faulstich}, \citenamefont {Zhu}, \citenamefont
  {O'Gorman}, \citenamefont {Qiu}, \citenamefont {White}, \citenamefont
  {Babbush},\ and\ \citenamefont {Lin}}]{mcclean2019discontinuous}%
  \BibitemOpen
  \bibfield  {author} {\bibinfo {author} {\bibfnamefont {J.~R.}\ \bibnamefont
  {McClean}}, \bibinfo {author} {\bibfnamefont {F.~M.}\ \bibnamefont
  {Faulstich}}, \bibinfo {author} {\bibfnamefont {Q.}~\bibnamefont {Zhu}},
  \bibinfo {author} {\bibfnamefont {B.}~\bibnamefont {O'Gorman}}, \bibinfo
  {author} {\bibfnamefont {Y.}~\bibnamefont {Qiu}}, \bibinfo {author}
  {\bibfnamefont {S.~R.}\ \bibnamefont {White}}, \bibinfo {author}
  {\bibfnamefont {R.}~\bibnamefont {Babbush}}, \ and\ \bibinfo {author}
  {\bibfnamefont {L.}~\bibnamefont {Lin}},\ }\href@noop {} {\bibfield
  {journal} {\bibinfo  {journal} {e-print arXiv:1909.00028 [quant-ph]}\ }
  (\bibinfo {year} {2019})}\BibitemShut {NoStop}%
\end{thebibliography}%

\end{document}